\RequirePackage{pdf14}
\documentclass[11pt,a4paper]{article}
\usepackage[a4paper]{geometry}
\usepackage{setspace}
\usepackage{array}
\usepackage[centertags,reqno]{amsmath}
\usepackage{graphics,subfigure}
\usepackage{hyperref}
\usepackage{appendix}
\usepackage{threeparttable}
\usepackage{rotating}
\usepackage{dcolumn}
\usepackage{natbib}
\usepackage{hyperref}
\usepackage{longtable}
\usepackage{caption}
\usepackage{pdflscape}
\usepackage{xcolor}
\usepackage{authblk}
\usepackage{amscd,latexsym,amsfonts}
\usepackage{verbatim,color}
\usepackage{booktabs}
\usepackage{epsfig}
\usepackage{epstopdf}
\usepackage{adjustbox}
\usepackage{multirow}
\usepackage{float,lscape}
\usepackage{bigstrut}
\usepackage{array}
\usepackage{tabularx}
\usepackage{fancyhdr}
\usepackage{lipsum}
\usepackage{bm}
\usepackage{amssymb}
\usepackage{ragged2e}
\usepackage{pgfplots}
\pgfplotsset{compat=newest,xlabel near ticks,ylabel near ticks,width=15cm}
\usepackage{tikz}
\usetikzlibrary{plotmarks}
\usepackage{etex}
\usepackage{filecontents}
\usetikzlibrary{calc}
\usepackage{environ}
\usepackage{xstring}

\usepackage{xcolor}
\usepackage{bchart}

\newcolumntype{d}[1]{D{.}{.}{#1}}
\newcolumntype{p}[1]{D{(}{(}{-1}}

\newcolumntype{L}[1]{>{\raggedright\let\newline\\\arraybackslash\hspace{0pt}}m{#1}}
\newcolumntype{C}[1]{>{\centering\let\newline\\\arraybackslash\hspace{0pt}}m{#1}}
\newcolumntype{R}[1]{>{\raggedleft\let\newline\\\arraybackslash\hspace{0pt}}m{#1}}

\begin{document}
	\title{Bridging the Covid-19 Data and the Epidemiological Model using Time-Varying Parameter SIRD Model   
	}
	
	\author[,1]{ Cem \c{C}akmakl{\i} \thanks{
			Correspondence to: Cem \c{C}akmakl{\i}, Ko\c{c} University, Rumelifeneri Yolu 34450 Sarıyer Istanbul Turkey, 
			e--mail: ccakmakli@ku.edu.tr. }
	}
	\author[,2]{Yasin \c{S}im\c{s}ek \thanks{e--mail: yasin.simsek@duke.edu}}
	\affil[1]{\emph{Ko\c{c} University}}
	\affil[2]{\emph{Duke University}}

	\maketitle
	\justifying
	\begin{abstract} \noindent
This paper extends the canonical model of epidemiology, the SIRD model, to allow for time-varying parameters for real-time measurement and prediction of the trajectory of the Covid-19 pandemic. Time variation in model parameters is captured using the generalized autoregressive score modeling structure designed for the typical daily count data related to the pandemic. The resulting specification permits a flexible yet parsimonious model with a low computational cost. The model is extended to allow for unreported cases as well. Results suggest that these cases' effects on the parameter estimates diminish with the increasing number of testing. Full sample results show that the flexible framework captures the successive waves of the pandemic accurately. A real-time exercise indicates that the proposed structure delivers timely and precise information on the pandemic's current stance. This superior performance, in turn, transforms into accurate predictions of the confirmed cases. 
		\end{abstract}

	\newpage
	
	\setcounter{page}{1}
\setstretch{1.5}
	
\section{Introduction}
The outbreak of the new coronavirus, Covid-19, pandemic is one of the most severe health crises the world has encountered in the last decades. Since the onset of the pandemic in early January of 2020, it exhibits a varying pattern for many reasons. First, countries have repeatedly taken various measures to reduce the transmission rate of the virus. These measures involve complete lockdown such as full closure of business and curfew, and partial lockdown that implies a partial closure of daily routines. These seem to mitigate the spread of the virus at various stages of the pandemic. Second, mutations of the virus might lead to changes in its main characteristics. Furthermore, with the increasing medical knowledge and ongoing research on the virus, the death rate seems to be partially lowered. The recovery rate has increased compared to the initial phases of the pandemic. 

While the pandemic evolves rapidly with successive waves of infections, efficient and timely monitoring of the pandemic is crucial. Making prompt and effective decisions of imposing or relaxation of lockdown measures for policymakers and taking timely precautions for individuals critically rely on the knowledge about the pandemic. Therefore, epidemiological models for estimating, and perhaps even more crucially, for predicting the pandemic's trajectory come to the forefront. Conventional statistical models of epidemiology mostly involve structural parameters that remain constant throughout the pandemic. However, suppose these interventions turn out to be effective and cause changes in the pandemic's natural course. In that case, this implies that the parameters of the epidemiological models alter to comply with this changing trajectory. Besides, the virus might undergo some mutations which alter the contagiousness and fatality of the virus. Hence, these mutations translate into changes in the key structural parameters of the model. These observations are the departure point of this paper. Specifically, we develop a computationally simple and statistically coherent model that allows for time variation in the epidemiological model parameters. 

We start our analysis by confronting a simple version of the workhorse epidemiological model with the existing data. From the perspective of econometrics, we specify a counting process for modeling the course of the Covid-19 pandemic for a selected set of countries based on the SIRD model. The SIRD model identifies the pandemic's four states as Susceptible, Infected, Recovered, and Death. It depicts these states' evolution depending on the total number of infected individuals, see \cite{kermack1927, Allen-2008, avery2020economist}. Using these four variables of daily counts of susceptible, infected, recovered, and death cases, the model is well-identified conditional on the infected cases' initial value. The pandemic's course is determined by the contestation of these forces, i.e. the structural parameters governing infection and resolution (either recovery or death) rates. As a result, if the infection rate is larger than the resolution rate, the number of infections evolves according to a nonstationary process representing the virus's increasing spread. In contrast, the opposite case results in a stationary process. Therefore, we opt for a Bayesian estimation strategy of the parameters for computing reliable credible intervals for inference conditional on the available data rather than utilizing asymptotic analysis.     
Equipped with these tools, we extend the econometric model by allowing for time variation in the structural parameters resorting to the Generalized Autoregressive Score (henceforth GAS) modeling framework that is a class of observation-driven models. The proposed model permits a flexible yet feasible framework to track structural parameters' evolution timely and accurately. A relevant aspect of our specification is its relatively low computational cost. The computational cost might be crucial, especially at the beginning of the pandemic, when the data is scarce and the uncertainty is overwhelming. Finally, we extend the model taking undocumented cases (as these infected individuals do not show symptoms) into consideration by exploiting the information on testing for infection following \cite{Grewelle-2020}. 

We construct a set of countries with distinct experiences related to the pandemic to demonstrate the proposed framework's efficacy. These include countries that can mitigate the pandemic but with differential momentum, countries where the pandemic starts relatively late, and countries experiencing the second wave of the pandemic fiercely. These distinct experiences provide us with a testing ground with various patterns to examine the proposed model's efficacy. Our results indicate that for a majority of countries, the structural parameters change over time. The infection rate typically starts at a high level at the onset of the pandemic, but it decreases at differential degrees depending on the success in containing the virus's transmission. In some countries, we observe recurring increases in the infection rate that boost the pandemic's second wave. Contrarily, the recovery rate starts at a low level, and it gets stabilized after an increase from these low values reflecting the increasing performance of health systems in handling the active cases. Consequently, the reproduction rate starts at high levels, often exceeding the value of 5, but it diminishes at differential rates. For many countries, the reproduction rate exhibits an increasing pattern towards the end of the sample, reflecting the pandemic's second wave in line with the stylized facts. Two crucial findings emerge from the outcomes of the proposed model with time-varying parameters. First, the US still cannot tackle the pandemic as the reproduction rates have almost never fallen below the value of 2. Second, Germany and Italy experience the second wave of the pandemic intensively, with the reproduction rates well above 1. Results are very similar when we take unobserved cases also into account.

We further examine the model performance in real-time by conducting a real-time recursive estimation and forecasting exercise. Results indicate that the proposed model with time-varying parameters provides timely information on the pandemic's current stance ahead of the competing models. Moreover, it yields superior forecasting performance up to 2 weeks horizon. Therefore, it provides crucial information, especially for decision-makers planning potential lockdown measures to contain the pandemic and efficient allocation of the resources to combat the pandemic. 

The literature on estimating the SIRD model (with fixed parameters) and variants to evaluate the current stance of the Covid-19 pandemic has exploded since the outbreak of the pandemic. Relatively earlier analysis include \cite{Read-etal-2020}, and \cite{lourenco2020fundamental} who estimate a SIRD based model with the data from China for the former and from the UK and Italy for the latter using a likelihood-based inference strategy. \cite{wu2020nowcasting} blend data related to the Covid-19 for China with mobility data and estimate the epidemiological model using Bayesian inference to predict the spread of the infection domestically and internationally. \cite{li2020substantial} conduct a similar analysis employing a modified SIR model together with a network structure and mobility data to uncover the size of the undocumented cases, see also \cite{hortaccsu2020estimating}. \cite{zhang2020prediction} extend the standard SIR model with many additional compartments and estimate a part of parameters using Bayesian inference. The identification of the model parameters in these models hinges upon the data availability for each compartment. Otherwise, parameter values are set based on the pandemic's stylized facts, see \cite{manski2020estimating, Atkeson-2020, KOROLEV2020}.

Several factors might lead to the time variation in the parameters of the epidemiological model. On the one hand, lockdown measures implemented by the policymakers isolate the infected from the susceptible individuals. Therefore, the parameter governing the infection rate, that is, the average number of contacts of an individual, is likely to alter with lockdown conduct, see \cite{hale2020variation} for example. On the other hand, advancements in the fight against Covid-19 such as the recovery of drugs and vaccination could effectively mitigate the course of the disease. In addition, the installment or the lack of the medical equipment such as ventilators might alter the rate of recovery or in other words, the duration of the state of being infected, see for example \cite{Greenhalgh-Day-2017} on time variation in recovery rates. Accordingly, \cite{anastassopoulou2020data} use a least-squares based approach on a rolling window of daily observations. They document the time variation of parameters in the SIRD based model using Chinese data. \cite{tan2020} also employ a similar but more articulated rolling window strategy to capture the time variation in the model parameters. Other frameworks with time-varying model parameters almost exclusively allow for the time variation only in the infection rate. An application before the Covid-19 outbreak includes, for example, \cite{xu2016bayesian} among others, who utilize a Gaussian process prior for the incidence rate involving the rate of infection using a Bayesian nonparametric structure. In the context of the Covid-19 pandemic, \cite{Kucharski-etal-2020} estimate a modified SIR model using a parameter-driven model framework allowing the infection rate to follow a geometric random walk with the remaining parameters kept as constant; see \cite{kucinskas2020tracking} for a similar approach. Similarly, \cite{yang2020short}, and \cite{fernandez2020estimating} allow for time variation in the rate of infection, keeping the remaining parameters constant. \cite{LIU-2020} provides an econometric specification where the growth rate of infections follow an autoregressive process around a deterministic trend with a structural break.

In this paper, we propose an alternative modeling strategy to capture the time variation in the structural parameters of the SIRD model. On the one hand, our modeling framework is statistically consistent with the typical count data structure related to the pandemic, unlike the models that either employ least-squares or likelihood-based inference using Gaussian distribution, that is, the Kalman filter. On the other hand, our framework is computationally inexpensive, unlike the models that are statistically consistent but computationally costly such as the particle filter. This computational efficiency might be crucial, most notably, when the data is scarce and uncertainty about the pandemic is abounding at the start of the pandemic.  Our framework belongs to the class of observation-driven models, specifically to the GAS models proposed by \cite{creal2013}. GAS models involve many of the celebrated econometric models like the Generalized Autoregressive Heteroskedasticity (GARCH) model and various variants as a specific case, and thus, they proved to be useful in both model fitting as well as prediction. \cite{koopman2016predicting} provide a comprehensive analysis of these models' predictive power compared to parameter-driven models in many settings, including models with count data.  

Observation-driven models for count data are considered in many different cases independent of the analysis of the Covid-19 pandemic. \cite{Davis-etal-2003} provide a comprehensive analysis of observation-driven models with a particular focus on data with (conditional) Poisson distributions. \cite{Ferland-etal-2006} derive an integer-valued analog of the GARCH model (IN-GARCH) using Poisson distribution instead of Gaussian distribution. \cite{Fokianos-etal-2009} consider the Poisson autoregression of linear and nonlinear form like the IN-GARCH model as a specific case. \cite{Chen-etal-2016} extend the Poisson autoregression to allow for smooth regime switches in parameters. Our framework naturally extends these approaches to the epidemiological model framework for each of the core compartments of the SIRD model using a multivariate structure.         

The remainder of the paper is organized as follows. Section~\ref{sec:Model specification} describes the canonical SIRD model and introduce the SIRD model with time-varying parameters. In Section~\ref{sec:Econometric inference}, we discuss econometric issues, including identification of model parameters and how to account for sample selection. In this section, we further elaborate on our estimation strategy and the resulting simulation scheme. Section~\ref{sec:Empirical results} presents estimation results using full sample data from various countries. In Section~\ref{sec:Real-TimePerformance}, we evaluate the real-time performance of our model framework in capturing the current stance of the pandemic and in short-term forecasting compared to frequently used competitors. Finally, we conclude in Section~\ref{sec:Conclusion}. 

\section{Model specification}
\label{sec:Model specification}

\subsection{The canonical model of pandemic, SIRD model}
	
We start our analysis by discussing the epidemiological model denoted as the SIRD model of \cite{kermack1927}. Specifically, the SIRD model categorizes a population into four classes of individuals representing four distinct states of the pandemic as Susceptible ($S(t)$), Infected ($I(t)$), Recovered ($Rc(t)$) and Death ($D(t)$) in period $t$. The susceptible group does not yet have immunity to disease, and individuals in this group have the possibility of getting infected. On the other hand, the recovered group consists of individuals who are immune to the disease, and finally, $D(t)$ represents individuals who have succumbed to the disease.  The Susceptible-Infected-Recovered-Death (SIRD) model builds on the principle that a fraction of the infected individuals in the population, $\frac{I(t)}{N}$, can transmit the disease to susceptible ones, $S(t)$, with a (structural) infection rate of $\beta$ by assuming a quadratic matching in the spirit of gravity law, see \cite{Acemoglu-etal-2020} for details on alternative matching structures. Therefore, the number of newly infected individuals in the current period is $\beta S(t) \frac{I(t)}{N}$. The newly infected individuals, that is, confirmed cases, $C(t)$, should be deducted from the susceptible individuals in the current period. Meanwhile, in each period, a fraction $\gamma$  of the infected people recover from the disease, which reduces the number of actively infected individuals. Similarly, a fraction $\nu$ of the infected people have succumbed to the disease, further reducing the number of actively infected individuals. Hence, a fraction $\gamma + \nu$ of the infections are `resolved' in total.
This leads to the following sets of equations: \vspace{-0.2cm}
\begin{equation}  \label{eq:SIR}
	\begin{array}{rcl}
\dot{S}(t)  & = & -\beta S(t) \frac{I(t)}{N} \\
\dot{Rc}(t) & = & \gamma I(t) \\
\dot{D}(t)  & = & \nu    I(t) \\
\dot{I}(t)  & = & \dot{S}(t) + \dot{Rc}(t) + \dot{D}(t)
	\end{array}	
\end{equation}
where $\dot{x}$ corresponds to $dx/dt$ and we assume that the population remains constant.\footnote{In fact, the number of  deaths reduces the total population. We assume that the total number of deaths is negligible compared to the population for the sake of tractability of the resulting SIRD model.} 

\subsection{Econometric analysis of the SIRD model with fixed parameters}
\label{subsec:FixedSIRDmodel}
The parameters of interest are the structural parameters $\beta$, $\gamma$, and $\nu$ that provide information on the transmission and resolution rates of the Covid-19 pandemic. A central metric that characterizes the course of the pandemic is the reproduction number, $R_0$. The reproduction number refers to the speed of the diffusion which can be computed by the ratio of newly confirmed cases, denoted as $\dot{C}(t)$, to the resolved cases, that is $\dot{C}(t) /(\dot{Rc}(t) + \dot{D}(t))$. Therefore, it serves as a threshold parameter of many epidemiological models for disease extinction or spread. Assuming that $S(t)/N \approx 1$, $R_{0}$ can be approximated by  $\beta/ (\gamma+\nu)$ in \eqref{eq:SIR} and it holds exactly when $t=0$, referred as the basic reproduction rate. In this sense, a value of $R_0$ being less than unity indicates that the pandemic is contained, and if it exceeds unity, this implies that the spread of the pandemic continues. Our main motivation for employing the model from the econometrics perspective is to conform to this canonical epidemiological model with the existing datasets and pinpoint the pandemic's stance timely. For that purpose, we first discretize \eqref{eq:SIR} as the typical Covid-19 dataset involves daily observations on the counts of individuals belonging to these states of health. Motivated by this, we specify a counting process for the states using the Poisson distribution conditional on past cases of active infections implying a nonhomogenous Poisson process for all the counts see for example \cite{Allen-2008, Yan-2008, Rizoiu-2018} for earlier examples and \cite{li2020substantial} in the Covid-19 context for a similar approach. We specify the following for the stochastic evolution of the counts of these states;
	\begin{equation} \label{eq:FPSIR}
	\begin{array}{rcl}
	\Delta C_{t}|\Omega_{t-1} &\sim&  Poisson( \beta \frac{S_{t-1}}{N}I_{t-1}) \\
	\Delta Rc_{t}|\Omega_{t-1} &\sim&  Poisson( \gamma I_{t-1})    \\
	\Delta D_{t}|\Omega_{t-1} &\sim&  Poisson( \nu I_{t-1})    \\
    \Delta I_{t}   &=&  \Delta C_t - \Delta Rc_{t} - \Delta D_{t},
	\end{array}	
	\end{equation}	
where $\Omega_{t}$ stands for information set that is available up to time $t$. We assume that $\Delta C_t$, $\Delta Rc_t$ and $\Delta D_t$, representing the daily counts of the pandemic states, are independent conditional on $\Omega_{t-1}$. Considering the fact that $\frac{S_{t-1}}{N} \approx 1$ the final identity leads an autoregressive process for the number of active infections, $I_t$. The resulting distribution for the number of active infections is a Skellam distribution (conditional on $\Omega_{t-1}$) with the mean $\pi I_{t-1}$, where $\pi = (1 +  \beta(1 -R_0^{-1}))$ and the variance as $\beta(1+R_0^{-1})I_{t-1}$. Here, we use the identity in the last equation of \eqref{eq:FPSIR} together with the definition of $R_0$. Therefore, stationarity of the resulting process depends on whether $R_0<1$ or $R_0\ge 1$, i.e., whether the pandemic is taken under control or not. In addition, the first and second unconditional moments are as follows, 
	\begin{equation} \label{eq:FPSIR-UnconMom}
	\begin{array}{rcl}
E[I_t] &=& \pi^t I_0  \\
Var(I_t) &=& \beta(1+R_0^{-1})\frac{\pi^{t-1}(1-\pi^{t})}{1-\pi} I_0,
	\end{array}	
	\end{equation}	
where we assume that the initial condition, $I_0$, is known. If the initial condition is considered a parameter to be estimated, then the variance is further amplified with a factor in terms of the initial condition's variance. Accordingly, the unconditional moments of the pandemic states are linear functions of these unconditional moments of $I_t$. We refer to \ref{app:sec:UncMomSIRD} for details.

\subsection{SIRD model with time-varying parameters - the TVP-SIRD model}
\label{subsec:TVPSIRDmodel}
In this section, we put forward the SIRD model with time-varying parameters. We use the framework of the Generalized Autoregressive Score model for modeling the time variation in parameters. This framework encompasses a wide range of celebrated models in econometrics, including the GARCH model and its variants. Briefly, the GAS model relies on the intuitive principle of modeling the time variation in key parameters in an autoregressive manner which evolves in the direction implied by the score function and thereby improving the (local) likelihood; see \cite{creal2013} for a detailed analysis of the GAS model. As in the case of the GARCH model, it effectively captures the time dependence in long lags in a parsimonious yet quite flexible structure. Perhaps more importantly, since it admits a recursive deterministic structure, the resulting observation-driven time variation in parameters is computationally inexpensive to estimate. This computational efficiency might be crucial, given that these flexible models are evaluated throughout the pandemic when data is often scarce, especially at its onset. Consider the SIRD model with time-varying parameters as $\beta_t$, $\gamma_t$ and $\nu_t$. We first transform these parameters into logarithmic terms to ensure the positivity of these parameters and, thereby, the positivity of the predicted counts every period. Let the parameter with a tilde denote the logarithmic transformations as $\tilde{\beta}_t = ln(\beta_t)$, $\tilde{\gamma}_t = ln(\gamma_t)$ and $\tilde{\nu}_t = ln(\nu_t)$. The resulting Time-Varying Parameters - SIRD (TVP-SIRD) model is as follows
\begin{equation}
\label{eq:TVP-SIRD}
    \begin{array}{rcl}
    \Delta C_{t}|\Omega_{t-1} &\sim&  Poisson( \beta_t \frac{S_{t-1}}{N}I_{t-1}) \\[-0.2em]
	\Delta Rc_{t}|\Omega_{t-1} &\sim&  Poisson( \gamma_t I_{t-1})    \\[-0.2em]
	\Delta D_{t}|\Omega_{t-1} &\sim&  Poisson( \nu_t I_{t-1})    \\[0.4em]
      \tilde{\beta_t}   & = &  \alpha_0 + \alpha_1 \tilde{\beta}_{t-1} + \alpha_2 s_{1,t}\\[-0.2em]
       \tilde{\gamma_t}  &= & \phi_0 + \phi_1 \tilde{\gamma}_{t-1} + \phi_2 s_{2,t} \\[-0.2em]
     \tilde{\nu_t}  &= & \psi_0 + \psi_1 \tilde{\nu}_{t-1} + \psi_2 s_{3,t} \\[0.4em]
    \Delta I_{t}   &=&  \Delta C_t - \Delta Rc_{t} - \Delta D_{t}, \\
    \end{array}
\end{equation}
where $s_{1,t}$, $s_{2,t}$ and $s_{3,t}$ are the (scaled) score functions of the joint likelihood. Since the SIRD model's likelihood function is constituted by the (conditionally) independent Poisson processes, each score function is derived using the corresponding compartment. Specifically, let   $\nabla_{1,t}=\frac{\partial L(\Delta C_{t}; \tilde{\beta}_t)}{\partial \tilde{\beta}_t}$, $\nabla_{2,t}=\frac{\partial L(\Delta Rc_{t}; \tilde{\gamma_t})}{\partial  \tilde{\gamma_t}}$ and $\nabla_{3,t}=\frac{\partial L(\Delta D_{t}; \tilde{\nu_t})}{\partial  \tilde{\nu_t}}$ denote the score functions for period $t$ observation. We specify $s_{i,t}$ such that the score functions are scaled by their variance as $s_{i,t}=\frac{\nabla_{i,t}}{\text{Var}(\nabla_{i,t})}$ for $i=1,2,3$.\footnote{Alternative approaches for scaling the score function include the standard deviation rather than the variance and the score function without scaling. Our findings suggest that using the variance as the scaling function leads to smoother and more robust evolution of parameters over time.}  In the specific case of the SIRD model, this modeling strategy leads to the following specification for the (scaled score functions) in terms of the logarithmic link function
	\begin{equation}
	\label{eq:ScoreFunctions}
        \begin{array}{rcl}
	    s_{1,t} &=& \frac{\Delta C_{t-1} - \lambda_{1,t-1}}{\lambda_{1,t-1}}  \\
	    s_{2,t} &=& \frac{\Delta Rc_{t-1} - \lambda_{2,t-1}}{\lambda_{2,t-1}} \\
	    s_{3,t} &=& \frac{\Delta D_{t-1} - \lambda_{3,t-1}}{\lambda_{3,t-1}}
    	\end{array}
     \end{equation} 
where $\lambda_{1,t} = \beta_t \frac{I_{t-1}S_{t-1}}{N}$, $\lambda_{2,t} = \gamma_t I_{t-1}$ and $\lambda_{3,t} = \nu_t I_{t-1}$. The resulting specification implies an intuitive updating rule because the parameters (in the logarithmic form) are updated using a combination of the previous parameter value and the previous percentage deviation from the mean. We refer to \ref{app:sec:UpdatingRules} for the details on derivation of \eqref{eq:ScoreFunctions}. The specification in \eqref{eq:TVP-SIRD} together with \eqref{eq:ScoreFunctions} lead to quite rich dynamics both in terms of mean and the variance of the resulting process. These rich dynamics enable us to accurately capture the pandemic's evolution reflected as the timely and prompt response of the parameters to the pandemic states' data changes. An appealing feature of the TVP-SIRD model is that it encompasses the SIRD model with fixed parameters. For example, when $\alpha_1=1$ together with  $\alpha_0$ and $\alpha_2$ to be zero, then the rate of infection, $\beta_t$, remains fixed throughout the pandemic. 
These restrictions would indicate that the lockdown measures are ineffective as it does not lead to a systematic change in the infection rate. Similar reasoning also applies to $\gamma_t$ and $\nu_t$. Therefore, it provides a solid framework for statistically testing the efficiency of measures to take the pandemic under control. 

To elaborate further, we discuss the implied moments of the resulting process. Consider the general form of equations in \eqref{eq:TVP-SIRD} as $Y_t|\Omega_{t-1}  \sim Poisson( \lambda_{t})$ together with $\lambda_{t}=\Psi_{t}K_{t-1}$  where $\Psi_t = \beta_t$, $\gamma_t$, and $\nu_t$, $K_t = \frac{I_{t-1}S_{t-1}}{N}, I_{t-1}$ and $I_{t-1}$ for $Y_t=\Delta C_t$, $\Delta R_t$ and $\Delta D_t$, respectively. The following heavily draws on the demonstration in \cite{Davis-etal-2003}. We consider $W_t=\log(\lambda_{t})=\log(\Psi_{t})+\log(K_{t-1})=\theta_t +X_{t-1}$. Considering the updating rules in \eqref{eq:TVP-SIRD} and \eqref{eq:ScoreFunctions}, we can describe the evolution of the model parameters as
\vspace{-0.2cm}
\begin{equation}
\begin{array}{rcl}
   \theta_{t} & = & \omega_0 + \omega_1 \theta_{t-1} + \omega_2 e_{t-1},
\end{array}    \vspace{-0.2cm} 
\end{equation}
where $e_{t-1}=\frac{Y_{t-1} - \lambda_{t-1}}{\lambda_{t-1}}$ and $\omega_i = \alpha_i, \phi_i$ and $\psi_i$ for $i=0,1,2$. Assuming stationarity conditions, $\theta_{t}$ can be solved as
\begin{equation}
\begin{array}{rcl} \label{app:eq:teta_t}
   \theta_{t} & = & \frac{\omega_0}{1-\omega_1}  + \sum_{i=1}^{t-1} \omega_1^i \omega_2 e_{t-i} =   \zeta_0  + \sum_{i=1}^{\infty} \zeta_i e_{t-i}.
\end{array}    
\end{equation}
For $e_t$, given the initial conditions $e_s=0$ for $s \le 0 $, the mean of $e_t$
\begin{equation}
\begin{array}{rcl}
   E[e_t|\Omega_{t-1}] & = & 0 \text{ for } s > 0, \text{ hence } \\
   E[e_t] &=& E[E[e_t|\Omega_{t-1}]] = 0,
\end{array}    
\end{equation}
and the variance could be computed as,
\begin{equation}
     E[e^2_t]  =  E[E[e^2_t| \lambda_{t}]] = E[\lambda_{t}^{-1}] \text{ for } s > 0.
\end{equation}
Moreover, $Cov(e_t, e_s) = 0$ for $t \neq s$. This suggests that $e_t$ series can be considered as innovations that are uncorrelated. Therefore, unconditional moments of $\theta_{t}$ follow as
\begin{equation}
\label{eq:UncExpTheta}
\begin{array}{rcl}
E[\theta_t] &=&  \zeta_0 = \frac{\omega_0}{1-\omega_1},
\end{array}    
\end{equation}
and 
\begin{equation}
\begin{array}{rcl}
    Var(\theta_t)  & = & \sum_{i=1}^{\infty} \zeta_i^2 E[\lambda_{t-i}^{-1}] \\
        Cov(\theta_t, \theta_{t-h}) & = & \sum_{i=1}^{\infty} \zeta_i \zeta_{i+h} E[\lambda_{t-i}^{-1}]. 
\end{array}
\end{equation}
Considering the evolution of the active infected cases, we can write the following
\begin{equation}
\label{eq:EvolI_t_TVP}
\begin{array}{rcl}
    I_{t} &=& (1 +  \beta_t - \gamma_t - \nu_t )I_{t-1} = \pi_t I_{t-1}. 
\end{array}
\end{equation}
Therefore, using \eqref{eq:UncExpTheta} unconditional moments can be written as
\begin{equation}
\label{eq:UncMom_I_t}
\begin{array}{rcl}
    E[I_{t}] &=& \left(1 + \frac{\alpha_0}{1-\alpha_1} - \frac{\phi_0}{1-\phi_1}- \frac{\psi_0}{1-\psi_1} \right)^t \\          &=& \bar{\pi}^t I_{0} \\
    Var(I_t) &=& \left(\frac{\alpha_0}{1-\alpha_1} + \frac{\phi_0}{1-\phi_1} + \frac{\psi_0}{1-\psi_1} \right) \frac{\bar{\pi}^{t-1}(1-\bar{\pi}^{t})}{1-\bar{\pi}} I_0,
\end{array}
\end{equation}
Accordingly, we can write $W_t = \zeta_0  + \sum_{i=1}^{\infty} \zeta_i e_{t-i} + X_{t-1}$ and the mean of $W_t$
\begin{equation}
\begin{array}{rcl}
  E[W_t] & = & \zeta_0 + E[X_{t-1}] \\
         & = & \zeta_0 + E[\log(I_{t-1})] \\
         & \approx & \zeta_0 + \log(E[I_{t-1}]) \\
\end{array}    
\end{equation}
For the variance term, we can write
\begin{equation}
\label{app:eq:VarCov(W)}
\begin{array}{rcl}
  Var(W_t) & = &  \sum_{i=1}^{\infty} \zeta^2_i E[\lambda_{t-i}^{-1}] + Var(X_{t-1})  \\
\end{array}    
\end{equation}
Following \cite{Davis-etal-2003}, for deriving the expectations of the variables representing the pandemic states, we use an approximation to Gaussian distribution, and we can write
\begin{equation}
\label{eq:UE_Y}
\begin{array}{rcl}
  E[Y_t]&=& E[E[Y_t|\lambda_t]]  = E[\lambda_t] = E[\exp(W_t)] \\ 
  & \approx &  \exp\left(\zeta_0 + E[X_{t-1}] + \frac{1}{2} \left(\sum_{i=1}^{\infty} \zeta^2_i E[\lambda_{t-i}^{-1}]  + Var(X_{t-1}) \right)  \right) 
\end{array}    
\end{equation}
For the variance of $Y_t$, using the law of total variance, it follows that
\begin{equation}
\begin{array}{rcl}
     Var(Y_t) &=& E[\lambda_t] + Var(\lambda_t) \\
     &\approx&  E[\lambda_t] +  Var(\exp(W_t)) \\
    & = & E[\lambda_t] + \exp\left(  \sum_{i=1}^{\infty} \zeta^2_i E[\lambda_{t-i}^{-1}] + Var(X_{t-1})-1 \right)\times \\
    &&\exp\left( 2(\zeta_0 + E[X_{t-1}] ) + \left(\sum_{i=1}^{\infty} \zeta^2_i E[\lambda_{t-i}^{-1}]  + Var(X_{t-1}) \right)\right))
\end{array}    
\end{equation}
The expectation in \eqref{eq:UE_Y} demonstrates the efficacy of the underlying process. Specifically, the expectation of $Y_t$ is driven by two forces. First, as in the SIRD model with fixed parameters, one of the main drivers is the $E[X_{t-1}]$. Second, unlike the SIRD model with fixed parameters, it is characterized by the past score functions' variances with a diminishing effect for longer lags. This structure is also valid for the variance process. This flexible dependence structure for both the first and second moments provides considerably rich dynamics for capturing the pandemic's evolution, reflected as the timely and prompt response of the parameters to the data changes, which we discuss in the next sections.



\section{Econometric inference}
\label{sec:Econometric inference}

\subsection{Identification of model parameters}
\label{subsec:Identification}
The pandemic's course in terms of the evolution of active cases depends on the structural parameters, $\beta, \gamma$ and $\nu$ that are used to construct $\pi$ in \eqref{eq:FPSIR-UnconMom}. Additionally, the initial condition  $I_0$ is required because the process might be nonstationary if the pandemic is not contained. We estimate the models starting from the period when the number of cumulative confirmed cases exceeds 1000, and we use this first observation in our sample as the initial condition.\footnote{Different starting points (such as the periods when the number of cumulative confirmed cases exceeds 10000) yield very similar results. Results are available upon request by the authors.} The structural parameters are representing the compartments of the SIRD model. Therefore, if the specific compartments' data, including the counts of recovery and deaths, are available, these structural parameters are well identified. Still, extensions of the model with additional compartments such as cases that are `exposed to the virus' or `quarantined' require additional parameters to be estimated, see \cite{lourenco2020fundamental} for example. Therefore, while these additional compartments provide further refinements to the SIRD model, these refinements plague identification of the structural parameters if additional data is missing, see, for example, the broader discussion in \cite{manski2020estimating} independent of the SIRD model. \cite{Atkeson-2020} discusses the identification of the structural parameters regarding the SIRD model. He demonstrates how different parameter setups might result in very similar initial phases of the pandemic but result in divergent patterns in the long-run absence of the data on the model's compartments. Similarly, \cite{KOROLEV2020} discusses identification problems of the structural parameters regarding the SIRD model with an additional compartment of `exposed' case. 

\subsection{Accounting for sample selection}
	\label{subsec:Asymptomatics}
A fundamental underlying assumption of the model specification in previous sections is that the variables of the infected, recovered, and succumbed individuals represent the aggregate numbers. However, one of the stylized facts related to the Covid-19 pandemic is the presence of the infected individuals who do not have any symptoms, denoted as asymptomatic. These hard to detect cases complicate the analysis as it leads to a selection bias in econometric inference, among other factors, see \cite{manski2020estimating}. These unreported infection cases prohibit the tests to be randomly assigned plaguing econometric inference. This section provides results on the total number of individuals based on some assumptions on the model structure to capture asymptomatic infected individuals.
Specifically, let $P_t$ denote an indicator function that takes the value one if an individual is infected in period $t$, and 0 otherwise. Further, let $T_t$ denote another indicator function, which takes the value one if an individual is tested for the infection in period $t$ and 0 otherwise. Using the Bayes rule, we can show that 
\begin{equation}
    P(P_t=1) = \frac{P(P_t=1|T_t=1)P(T_t=1)}{P(T_t=1|P_t=1)},
\end{equation}
see also \cite{stock2020}. In case the assignment for testing of an individual is carried out randomly, then $P(T_t=1)=P(T_t=1|P_t=1)$ and there is no identification problem due to sample selection. Neither $P(P_t=1)$ nor $P(T_t=1|P_t=1)$ are observed. Nevertheless, $P(T_t=1)$ could be computed as the fraction of tested individuals in the population in period $t$. Furthermore, $P(P_t=1|T_t=1)$ could be considered as the daily positive test rate. Equipped with these, identification of $P(T_t=1|P_t=1)$ boils down to identification of $P(P_t=1)$, the true prevalence of the infection including asymptomatic cases. Departing from \cite{Grewelle-2020}, we make use of a parametric identification strategy for approximation of $P(T_t=1|P_t=1)$,
\begin{equation}
\label{eq:Approx}
    P(T_t=1|P_t=1) = \exp(-k\rho_t)
\end{equation}
where $\rho_t$ is the fraction of positives in the tested individuals in period $t$, and $k$ is a positive constant. Briefly, the underlying idea stems from the fact that detecting infections, including asymptomatic individuals, would improve with the increasing number of testing. In that sense, the fraction of tested individuals among the population should be related to the ratio of reported infections to the total number of infections. With the increasing number of testing on the population, this fraction approaches 1. On the other hand, if testing is concentrated only on symptomatic individuals, this fraction approaches a lower bound, captured by the parameter $\exp(-k)$, where the functional form admits exponential decay.
Let $I^*_{t}$ be the number of infected individuals involving both asymptomatic and symptomatic cases and let $S^*_t$ and $Rc^*_t$ denote the total number of susceptible and recovered individuals, respectively. Let $\delta_t = 1-\exp(-k\rho_t)$ denotes the fraction of the unreported infection cases among all infection cases. Using \eqref{eq:Approx}, these could be computed as
\begin{equation}
\begin{array}{rcl}
    \Delta C_t^* &=&  \frac{\Delta C_t}{1-\delta_t} \\
   \Delta Rc_t^*  &=& \frac{\Delta Rc_t}{1-\delta_t}.
\end{array}
 \end{equation}
Finally, the TVP-SIRD model in terms of the total numbers can be written as
\begin{equation}
\label{eq:TVP-SIRD-A}
    \begin{array}{rcl}
     \Delta C^*_{t}|\Omega_{t-1} &\sim&  Poisson( \beta_t \frac{S^*_{t-1}}{N} I^*_{t-1}) \\[-0.2em]
     \Delta Rc^*_{t}|\Omega_{t-1} &\sim&  Poisson( \gamma_t  I^*_{t-1} )    \\[-0.2em]
   	   \Delta D_{t}|\Omega_{t-1} &\sim&  Poisson( \nu_t I^*_{t-1})    \\[0.4em]
      \tilde{\beta_t}   & = &  \alpha_0 + \alpha_1 \tilde{\beta}_{t-1} + \alpha_2 s_{1,t}\\[-0.2em]
       \tilde{\gamma_t}  &= & \phi_0 + \phi_1 \tilde{\gamma}_{t-1} + \phi_2 s_{2,t} \\[-0.2em]
     \tilde{\nu_t}  &= & \psi_0 + \psi_1 \tilde{\nu}_{t-1} + \psi_2 s_{3,t} \\[0.4em]
    \Delta C^*_t = -\Delta S^*_t    &=&  \Delta I^*_{t} + \Delta Rc^*_{t} + \Delta D_{t} \\
    \end{array}
\end{equation}
The new set of equations exploits two sources of information for dealing with potential sample selection bias. First, the fact that the observed number of deaths covers the total number facilitates identifying the infection's true prevalence. Second, the information on testing further provides a distinct source of information for the prevalence.

\subsection{Estimation strategy and the simulation algorithm} 
\subsubsection{Bayesian inference}
We use simulation-based Bayesian estimation techniques for inference on model parameters. Bayesian inference involves updating the prior distributions of model parameters with the data likelihood to form the parameters' posterior distributions. 
Considering the SIRD model, Bayesian inference is especially appealing, since the inference is conditional on the data at hand and does not require relying on asymptotic analysis. Therefore, we can compute the credible intervals when the underlying process of the number of infected cases, $I_t$, is nonstationary. This property is especially reassuring in our case since obviously, nonstationarity, or in other words, basic reproduction rate being larger than 1, $R_{0 t}>1$, is an inevitable feature of the pandemic. 

Here we demonstrate the likelihood function and prior specifications for only the TVP-SIRD model, because the SIRD model with fixed model boils down to a special case of this model. The likelihood function is based on the model in \eqref{eq:TVP-SIRD} where we specify conditionally independent Poisson distributions for each of the components of the SIRD model. Let $y_t=(\Delta C_t, \Delta Rc_t, \Delta D_t)'$ be the vector of observations. Notice that $I_t =I_{t-1} + \Delta C_t -\Delta Rc_t - \Delta D_t$, and thus the number of active infections can be computed using the information set $\Omega_t = (y_t', I_{t-1})'$. Accordingly, we have the following likelihood function
\begin{equation}             
\label{eq:LikelihoodFunction}
       \begin{array}{rcl}
    f(y_t|\Omega_{t-1}) = \frac{\lambda_{1,t}^{\Delta C_t} \exp(-\lambda_{1,t})}{\Gamma(\Delta C_t + 1)} ~ \frac{\lambda_{2,t}^{\Delta Rc_t} \exp(-\lambda_{2,t})}{\Gamma(\Delta Rc_t + 1)} ~\frac{\lambda_{3,t}^{\Delta D_t} \exp(-\lambda_{3,t})}{\Gamma(\Delta D_t + 1)},
 	\end{array}
 \end{equation}
 where $\lambda_{1,t} = \beta_t \frac{S_{t-1}I_{t-1}}{N}$, $\lambda_{2,t} = \gamma_t I_{t-1}$  and $\lambda_{3,t} = \nu_t I_{t-1}$ and $\Gamma(.)$ is the Gamma function. Note that the time varying parameters follow the recursion as
 \begin{equation}
 \label{eq:TVPrecursion}
    \begin{array}{rcl}
     \tilde{\beta}_t &=& \alpha_0 + \alpha_1 \tilde{\beta}_{t-1} + \alpha_2 \left(\frac{\Delta C_{t-1} - \lambda_{1,t-1}}{\lambda_{1,t-1}} \right) \\
     \tilde{\gamma}_t &=& \phi_0 + \phi_1 \tilde{\gamma}_{t-1} + \phi_2 \left(\frac{\Delta Rc_{t-1} - \lambda_{2,t-1}}{\lambda_{2,t-1}} \right) \\
    \tilde{\nu}_t &=& \psi_0 + \psi_1 \tilde{\nu}_{t-1} + \psi_2 \left( \frac{\Delta D_{t-1} - \lambda_{3,t-1}}{\lambda_{3,t-1}} \right) 
	\end{array}
 \end{equation}
 where $\tilde{\beta}_t=\log(\beta_t)$, $\tilde{\gamma}_t=\log(\gamma_t)$ and $\tilde{\nu}_t=\log(\nu_t)$. 
 
We would like to obtain posterior results that are driven by the data rather than by the prior distributions. Therefore, we impose rather diffuse prior specifications for the model parameters. Let $\omega=(\omega_0, \omega_1, \omega_2)'$ for $\omega_i = \alpha_i, \phi_i$ and $\psi_i$ for $i=0,1,2$, we specify the following improper prior specifications
\begin{equation}
\label{app:eq:Priors}
     \begin{array}{rcl}
    f(\omega) \propto 1
 	\end{array}
 \end{equation}
 when the process described in \eqref{eq:TVPrecursion} is stationary and 0 otherwise. 
 
 \subsubsection{Simulation scheme}
 For the SIRD model with fixed parameters, the likelihood with Poisson distributions as in \eqref{eq:LikelihoodFunction} with fixed parameters and noninformative or conjugate priors in the form of Gamma distribution lead to a Gamma distribution for the posterior distributions of the model parameters. Therefore, these can be sampled using plain Gibbs sampler. For the TVP-SIRD model, the fact that we have time-varying parameters with deterministic recursions leads to posterior distributions that are nonstandard. Therefore, we cannot use standard distributions that we can easily simulate from for the inference, as is the case for the Gibbs sampler. Instead, we resort to the (adaptive) random walk Metropolis-Hastings (MH) algorithm
 within the Gibbs sampler, see \cite{robert2013monte} for details. The algorithm is as follows
\begin{enumerate}
    \item Sample $\alpha=(\alpha_0, \alpha_1,\alpha_2)'$ from $f(\alpha|S^T, I^T)$ using MH step
    \item Sample $\phi=(\phi_0, \phi_1,\phi_2)'$ from $f(\phi|Rc^T, I^T)$ using MH step
    \item Sample $\psi=(\psi_0, \psi_1,\psi_2)'$ from $f(\psi|D^T, I^T)$ using MH step
\end{enumerate}
Here $Y^T= (Y_1,Y_2,\dots, Y_T)'$ for $Y=S,I,Rc,D$ indicating the full sample of the count data regarding to the states of the pandemic, respectively. For the MH steps, the candidate generating density is constructed using the random walk specification as
\begin{equation}
\label{eq:RWMH}
\begin{array}{rcl}
\omega_m& = &\omega_{m-1} + \Sigma_\omega^{1/2} \epsilon_m
\end{array}    
\end{equation}
where $\omega_m = \alpha_m, \phi_m$ and $\psi_m$ depending on the step at the iteration $m$ and $\epsilon_m$ follows a standard multivariate $t-$distribution with degrees of freedom 15. For the initial value of parameters, $\omega_0$, and for the covariance matrix $\Sigma_\omega$, we use the maximum likelihood estimate of the model parameters. Therefore, we use the mode and the inverse Hessian of the likelihood function at the mode in \eqref{eq:RWMH}. For improving the performance of the sampler, we  follow the adaptive scheme described in \cite{haario2001adaptive}. This involves replacing $\Sigma_\omega$ with $\tau S_m + \epsilon I$, once we obtain a sufficient number of draws to replace the inverse Hessian of the likelihood function with the simulated curvature of the posterior distribution. Here $S_m$ corresponds to the empirical covariance matrix computed using the draws up to step $M$, $I$ indicates identity matrix and $\epsilon$ is a small number. $\epsilon I$ ensures a nonsingular empirical covariance matrix. In addition, we use $\tau$ for optimizing the sampler's performance for the candidate generating density to be efficient enough to cover the tails of the posterior distribution. Let $\omega^{cand} \sim q(\omega^{cand}|\omega_{m-1})$ be a draw from the candidate generating density in iteration $m$ of the sampler, the candidate is accepted with probability
\begin{equation}
\label{eq:AccProb}
\begin{array}{rcl}
\alpha(\theta_{m-1}|\omega^{cand}) &=& \min \left\{ 1, \frac{q(\omega_{m-1}|\omega^{cand}) p(\omega^{cand}|Y^T)}{q(\omega^{cand}|\omega_{m-1})p(\omega_{m-1}|Y^T) }   \right\}. 
\end{array}    
\end{equation}
Here $p(\omega_{m-1}|Y^T)$ refers to the posterior distribution. Note that, due to the symmetry of the random walk specification, $q(\omega_{m-1}|\omega^{cand})$ and $q(\omega^{cand}|\omega_{m-1})$ are equivalent, hence they are of no use in \eqref{eq:AccProb}.

\section{Empirical results}
\label{sec:Empirical results}
\subsection{Dataset}
\begin{figure}[H]
    \centering
    \caption{The evolution of the cumulative active cases in selected countries}
\begin{tabular}{cccc}
 Brazil & Germany & India    \\[-0.2em]
  \includegraphics[trim = 0mm 6mm 0mm 0mm, clip,width=0.33\textwidth]{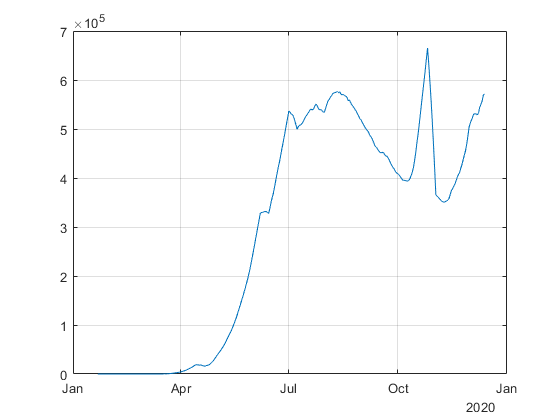}  
  & \includegraphics[trim = 0mm 6mm 0mm 0mm , clip, width=0.33\textwidth]{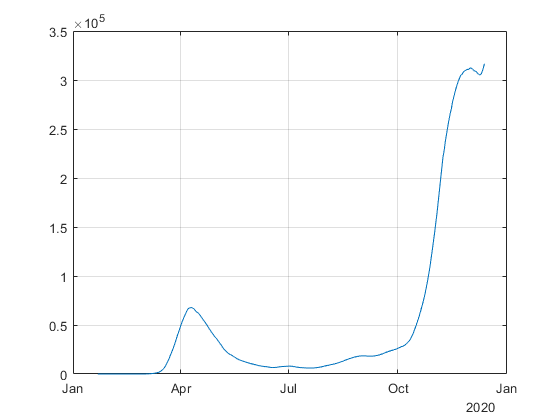} 
  &\includegraphics[trim = 0mm 6mm 0mm 0mm, clip,width=0.33\textwidth]{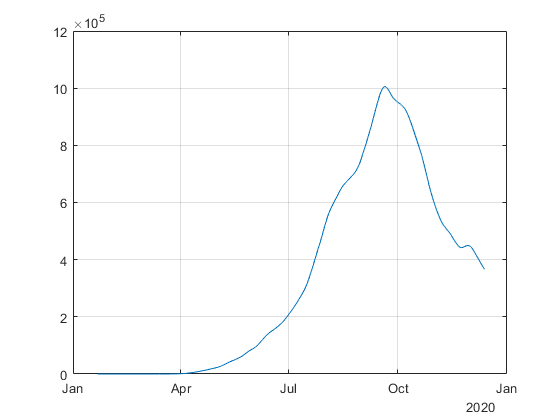}  \\[0.2em]
  Italy & S. Korea & US  \\[-0.2em]
    \includegraphics[trim = 0mm 6mm 0mm 0mm, clip,width=0.33\textwidth]{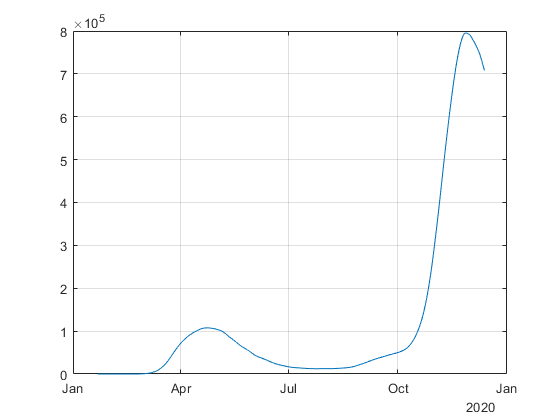}
  &  \includegraphics[trim = 0mm 6mm 0mm 0mm, clip,width=0.33\textwidth]{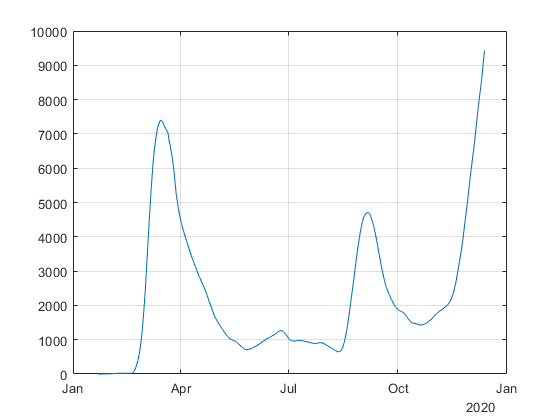}
  &  \includegraphics[trim = 0mm 6mm 0mm 0mm, clip,width=0.33\textwidth]{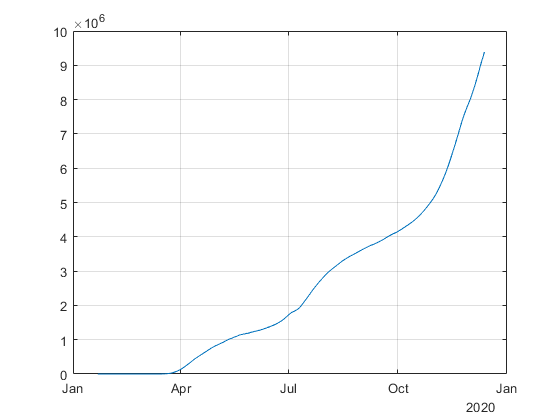}
  \end{tabular}
    \label{fig:ActiveCases}
\end{figure}

We use the data for a selected set of countries starting from the early days of the pandemic until the mid of December. Since daily data involves noise due to problems in timely announcements and many other reasons, we use a 7-day moving average of daily data following convention, see \cite{fernandez2020estimating} for example. We display the details on the dataset in Table~\ref{tab:dataset}. We, further, display the evolution of the cumulative active cases in these countries in Figure~\ref{fig:ActiveCases}.
\begin{center}
[Insert  Table~\ref{tab:dataset} and Figure~\ref{fig:ActiveCases} about here]
\end{center} 

These countries exhibit extensive heterogeneity in terms of their experience related to the pandemic, as shown in Figure~\ref{fig:ActiveCases}. Some countries in this set impose strict measures of full lockdown and successful testing and tracing policies at the onset of the pandemic, including South Korea. In contrast, other countries, including Italy, have imposed these immense measures after a certain threshold regarding the number of infected individuals. Some others opt for imposing mixed strategies involving partial lockdowns and voluntary quarantine, such as the US. We also include other interesting cases such as Brazil and India. These countries could not contain the pandemic below 300,000 active cases so far, and they have been experiencing differential phases with large swings since the start of the pandemic. Hence, this relatively rich and heterogeneous dataset involving countries with all sorts of pandemic experience enables us to examine the econometric model's success in tracking the changes in the structural parameters as a response to policy implementations and changes in the virus's characteristics. 

\subsection{Full sample results}
This section discusses the full sample estimates of the parameters for the models with fixed and time-varying parameters as described in \eqref{eq:FPSIR} and \eqref{eq:TVP-SIRD}, respectively. We further evaluate the model's parameter estimates with time-varying parameters when asymptomatic cases are explicitly considered, as shown in \eqref{eq:TVP-SIRD-A}.  
We start our analysis with the full sample estimates of the model parameters for the SIRD model with fixed parameters described in \eqref{eq:FPSIR}. These are displayed in Table~\ref{tab:FP-SIR-Parameters}. \begin{center}
[Insert  Table~\ref{tab:FP-SIR-Parameters} about here]
\end{center} 
The basic reproduction rate, $R_0$, as the main summary statistics on the course of the pandemic, is displayed in the last column of Table~\ref{tab:FP-SIR-Parameters}. For all countries, $R_0$ exceeds the threshold of 1. When the full sample of several days is taken into account, estimation results show that the pandemic could not be taken under control in these countries. For the US, which has been experiencing relatively prolonged earlier phases of the pandemic, we estimate an $R_0$ exceeding 2 departing from the rest of the countries in the sample. This large reproduction rate is due to the almost exceptional low recovery rate for the US. For Italy and Germany, $R_0$ is around 1.5, reflecting sizable swings with many infections when successive waves of the pandemic hit these countries. On the contrary, these waves are interrupted with periods of a much lower number of active cases. From a time-series perspective, these large swings lead to an explosive process with the $R_0$ well above the critical threshold of 1. This pattern also applies to South Korea, albeit to a milder extent. As for the structural parameters, for Germany, the rate of infection, $\beta$ is larger than those for Italy and South Korea. This larger rate of infection might be related to tighter lockdown restrictions imposed by these countries relative to Germany.\footnote{The stringency index reflecting governments' responses for containing pandemic was 79.63 for Italy versus 67.59 for Germany as of December 13, 2020. Larger values imply relatively tighter measures. See \url{https://ourworldindata.org/grapher/covid-stringency-index?stackMode=absolute&time=2020-12-13&region=World} for details.} Finally, India and Brazil constitute another cluster with lower $R_0$,  nevertheless still larger than 1. These results are driven by the fact that these countries exhibit little variation during the pandemic, but it does not imply that they successfully could contain the pandemic. On the contrary, while India experienced a single wave that could barely be contained, Brazil had only a minor relief between the pandemic waves. 

The estimation results in Table~\ref{tab:FP-SIR-Parameters} indicate that the SIRD model with fixed parameters might yield $R_0$ estimates that might not match with the stylized facts of countries with different experiences on the pandemic. These inconsistencies might stem from the rapid pace of infectiousness, captured by $\beta$, at the onset of the pandemic, brought under control due to swiftly imposed measures. Moreover, increasing knowledge about the SARS-Cov-2 virus, availability of the medical care facilities such as ICU's and more effective treatment of the infection potentially lead to changes in the recovery rate $\gamma$ and the death rate $\nu$.\footnote{We note the difference between the term \textit{death rate} and the terms  \textit{case fatality ratio} or \textit{mortality rate}. While the case fatality ratio refers to the ratio of (cumulative) number of the deaths to the (cumulative) number of the infected individuals, the \textit{mortality rate} measures the proportion of deaths due to a certain disease  among the entire population for a given period. On the other hand, the death rate, $\nu_t$, measures the portion of actively infected population who succumbed to the Covid-19 for a given period.} Besides, mutations of the virus might also lead to changes in the structural parameters reflecting the virus's key features. Therefore, it is crucial to efficiently model this time variation to monitor the pandemic's stance timely. 

Next, we display the parameter estimates of the TVP-SIRD model as described in \eqref{eq:TVP-SIRD} in Table~\ref{tab:TVP-SIRD-par}. 
\begin{center}
[Insert  Table~\ref{tab:TVP-SIRD-par} about here]
\end{center} 
These parameters govern the time variation in the structural parameters of the SIRD model. Therefore, it provides a testing tool for the potential time variation of these structural parameters. The parameter estimates in Table~\ref{tab:TVP-SIRD-par} indicate that the structural parameters governing the diffusion of the infection indeed exhibit time-varying behavior. The 95\% credible intervals constructed using posterior joint distribution exclude the set of $\alpha_0=\alpha_2=0$ together with $\alpha_1=1$, implying that $\beta_t$ indeed varies over time for all countries. Similar conclusions can be drawn for the recovery and death rates, $\gamma_t$ and $\nu_t$, respectively. 

We display the evolution of the underlying structural parameters,  $\beta_t$, $\gamma_t$, $\nu_t$ and the resulting $R_{0, t}$ over time in Figure~\ref{fig:TVParameters}. 
\begin{center}
[Insert  Figure~\ref{fig:TVParameters} about here]
\end{center} 
We first consider $\beta_t$, which is the rate of infection. We observe a generic spike at the onset of the pandemic, which dampens rapidly. Several measures imposed by these countries to contain the pandemic have weakened the virus's transmission throughout society. For most countries, we observe increases in the transmission rate, reflecting the spread's successive waves. These waves seem to be quite fierce for countries including Germany, Italy, and South Korea, while Brazil, India, and the US follow a milder pattern of fluctuations. On the one hand, the first group of countries exhibits sizable waves interrupted by calm periods with relatively fewer infections. On the other hand, the second group of countries could not contain the pandemic successfully, and thus, they have a relatively stable path of active infections with large numbers. The results show that the flexible model's dynamic structure could capture many types of infection rate patterns affected by the countries' containment measures. 

The recovery rate exhibits some variation for a majority of countries, albeit at a much-limited scale compared to the fluctuations of infection rate. We observe a typical path for the recovery rate, which usually starts with low levels as still the first wave of recoveries is limited. Over time it gets stabilized. For most countries, the mean recovery rate is around 0.07, indicating 14 days, on average, of recovery time from the infection in line with the WHO's official statistics. For Italy and the US, the recovery rate is particularly low, around values of 0.02-0.03. These low rates might be related to misreporting of the cases or recoveries. As a matter of fact, when asymptomatic cases or cases with relatively milder symptoms are excluded from the sample of infected cases, then only the recoveries of the severe cases are observed. These cases take much longer than the mild cases to recover: 2-8 weeks for severe cases according to the Report of the WHO-China Joint Mission on Coronavirus Disease 2019 (\cite{who2020}). Other underlying reasons include the demographic differences across countries and the health system's capacity leading to bottlenecks during the brunt of the pandemic. The pace of the recovery rate in India is rather unusual. It exhibits a gradual increase reflecting the gradual but constant improvement in the disease's treatment throughout the year, see also \cite{lancet2020covid} on this.  

When we consider the death rate, an impressive structure emerges. For all countries, except for South Korea, the value of death rate is smoothly stabilized around a fixed value. While this fixed value is 0.001 for most countries, it is lower for South Korea, reflecting the varying capability of these countries' health systems in coping with the pandemic. For South Korea, the pandemic's death rate is relatively low, and the seemingly volatile nature of the death rate might be due to these minuscule rates prone to fluctuations over time. 

The course of the reproduction rate, $R_{0, t}$ is of central importance for tracing the efficacy of the pandemic's containment efforts. The last column of the Figure~\ref{fig:TVParameters}  displays the evolution of the pandemic throughout 2020. The evolution of the reproduction rate closely follows the trace of the infection rate. First, for all countries, very high values of the reproduction rate at the onset of the pandemic are replaced by values around 2 for Brazil, the US, and India following the first wave. On the contrary, South Korea and Germany reduced the reproduction rate below 1 swiftly, thanks to intensive measures containing the pandemic. Second, the successive waves of increases in reproduction rates come into view for almost all countries except Brazil and India. For the US, the reproduction rate rose to values around 4 in July, but South Korea and Italy had experienced a spike in August and September. Although there are sizeable fluctuations in the US's reproduction rate, the rates could be reduced to only values around 2. Hence, the pandemic is far from being contained, and it continues to diffuse rapidly. Finally, for almost all countries, increasing reproduction rates in fall and winter 2020 can be traced quite explicitly except Brazil and India. The persistence pattern with many active infections and limited volatility make these countries suffer from reproduction rates continually fluctuating around threshold 1.

\subsection{Accounting for unreported cases}
\begin{figure}[H]
    \centering
    \caption{The evolution of $\delta_t$, $\beta_t$, $\gamma_t$, $\nu_t$ and $R_{0,t}$ when asymptomatic infected individuals are also considered in the sample}
\begin{tabular}{cccccc}
   & Italy  & S. Korea   &   US  \\[-0.2em]
    $\delta_t$  \hspace{-0.4cm}    &
  \includegraphics[trim = 10mm 6mm 10mm 0mm, clip,width=0.22\textwidth]{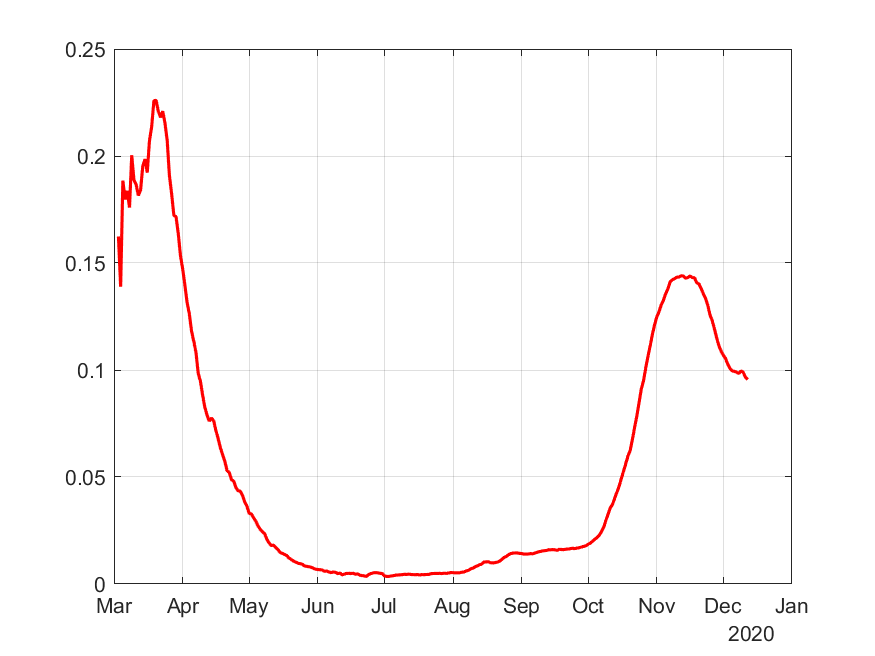} &
  \includegraphics[trim = 10mm 6mm 10mm 0mm, clip,width=0.22\textwidth]{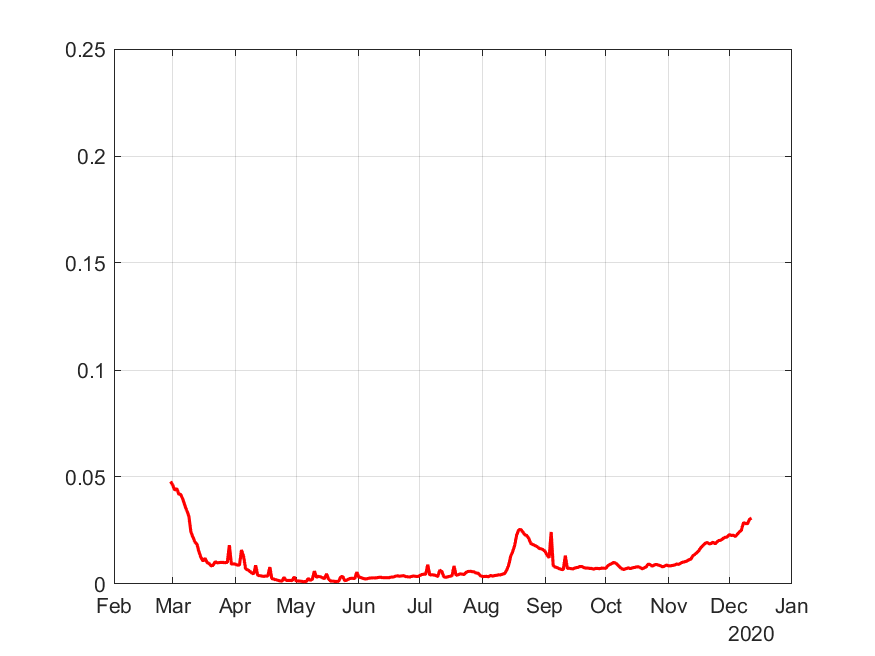} &
  \includegraphics[trim = 10mm 6mm 10mm 0mm, clip,width=0.22\textwidth]{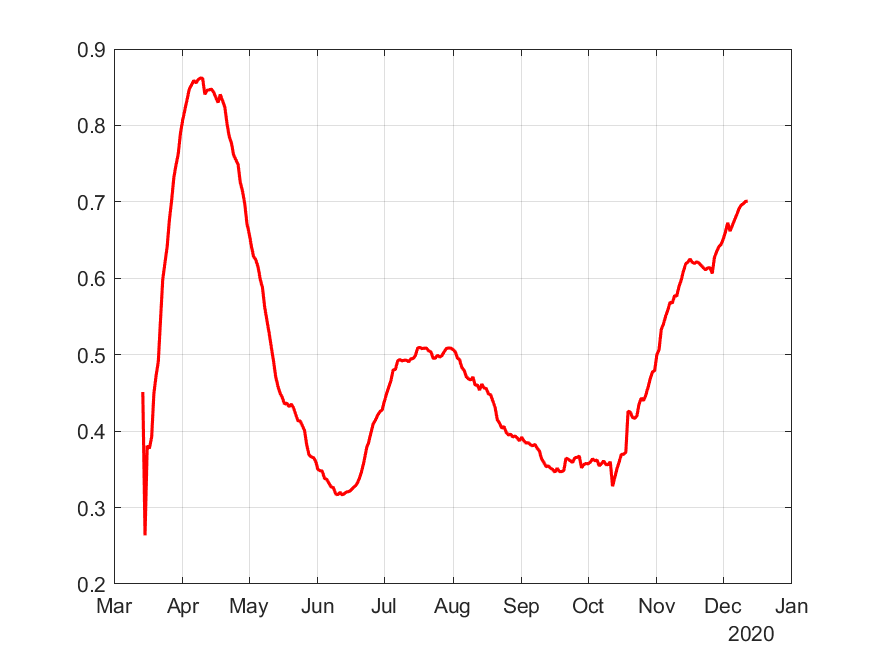}\\[-0.3em] 
  $\beta_t$  \hspace{-0.4cm} &
  \includegraphics[trim = 10mm 6mm 10mm 0mm, clip,width=0.22\textwidth]{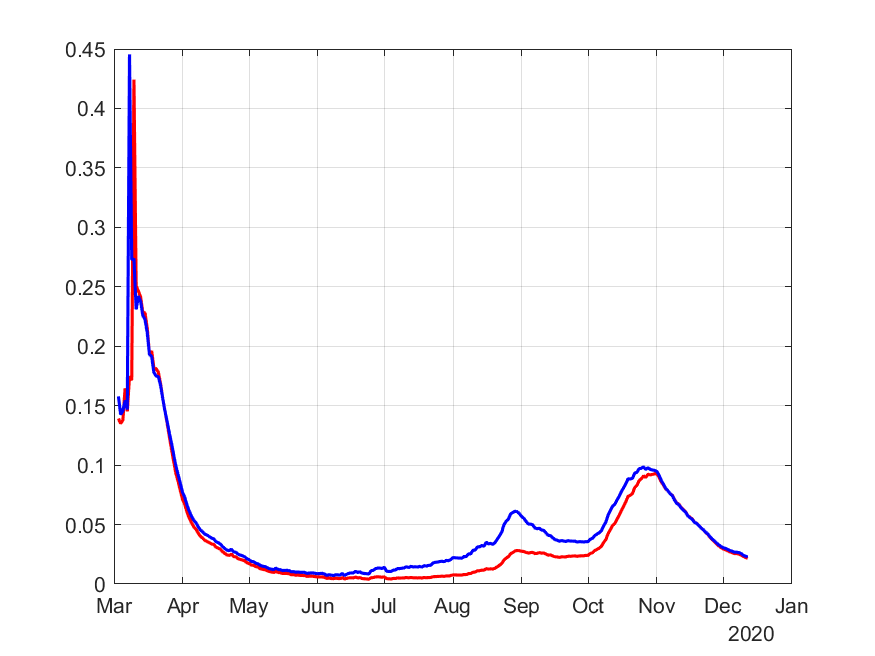}   &
  \includegraphics[trim = 10mm 6mm 10mm 0mm, clip,width=0.22\textwidth]{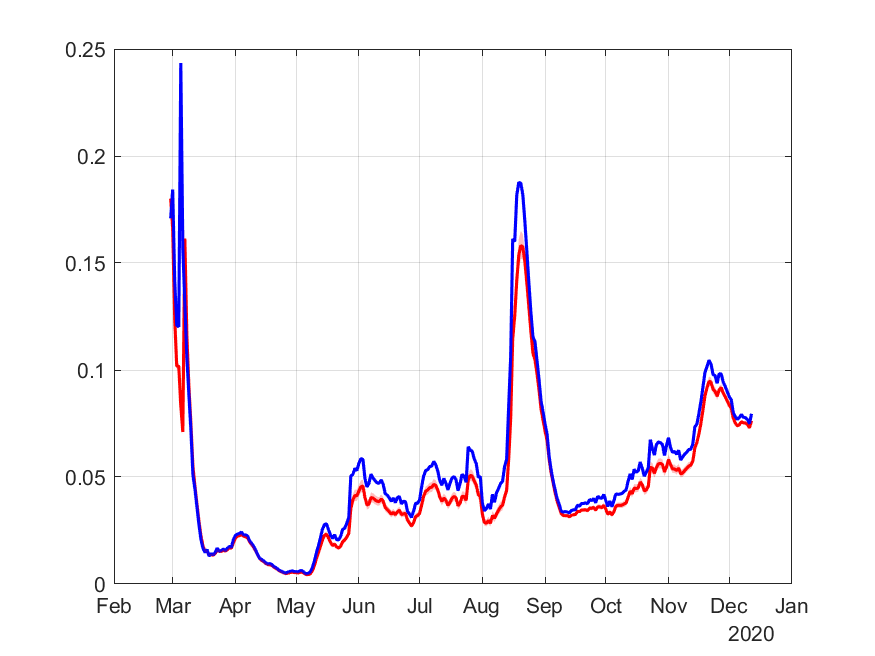}   & 
  \includegraphics[trim = 10mm 6mm 10mm 0mm, clip,width=0.22\textwidth]{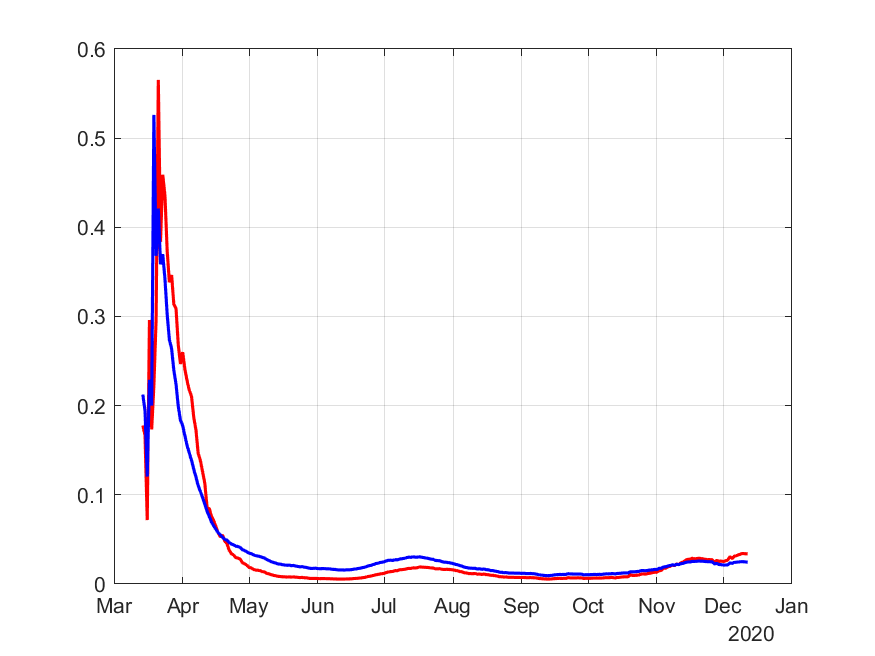} \\[-0.3em] 
  $\gamma_t$ \hspace{-0.4cm} &
  \includegraphics[trim = 10mm 6mm 10mm 0mm, clip,width=0.22\textwidth]{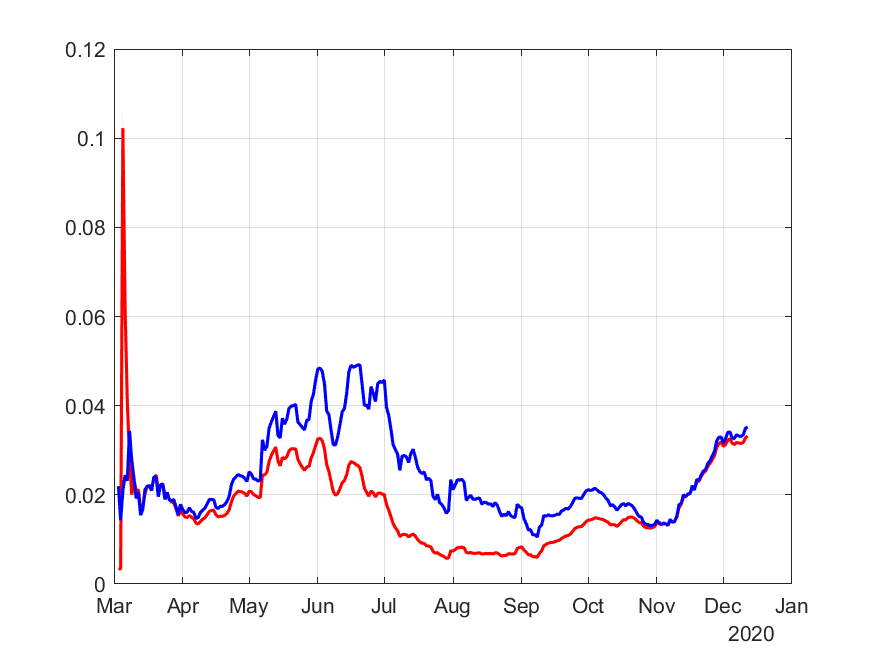}&
  \includegraphics[trim = 10mm 6mm 10mm 0mm, clip,width=0.22\textwidth]{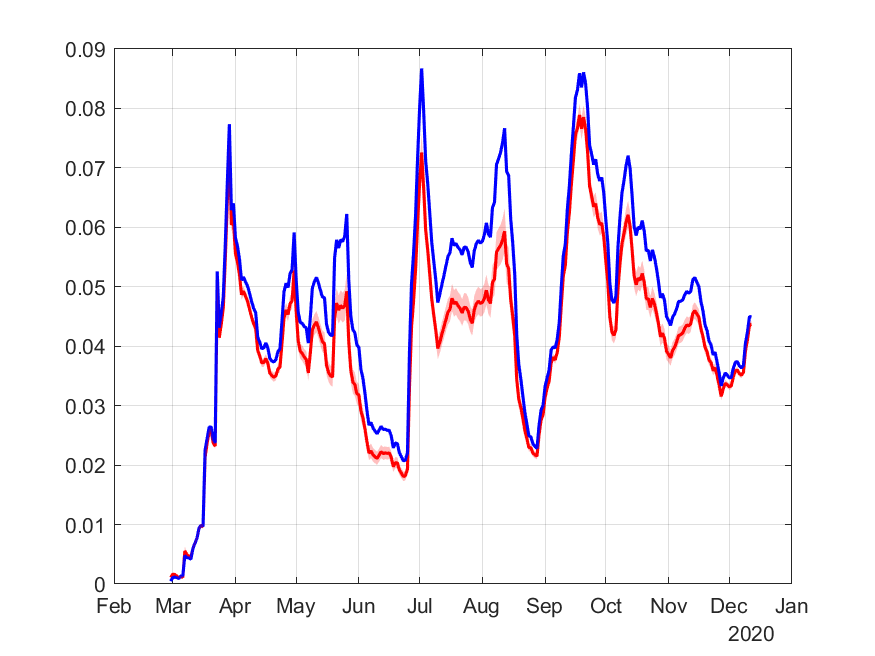}   &   
  \includegraphics[trim = 10mm 6mm 10mm 0mm, clip,width=0.22\textwidth]{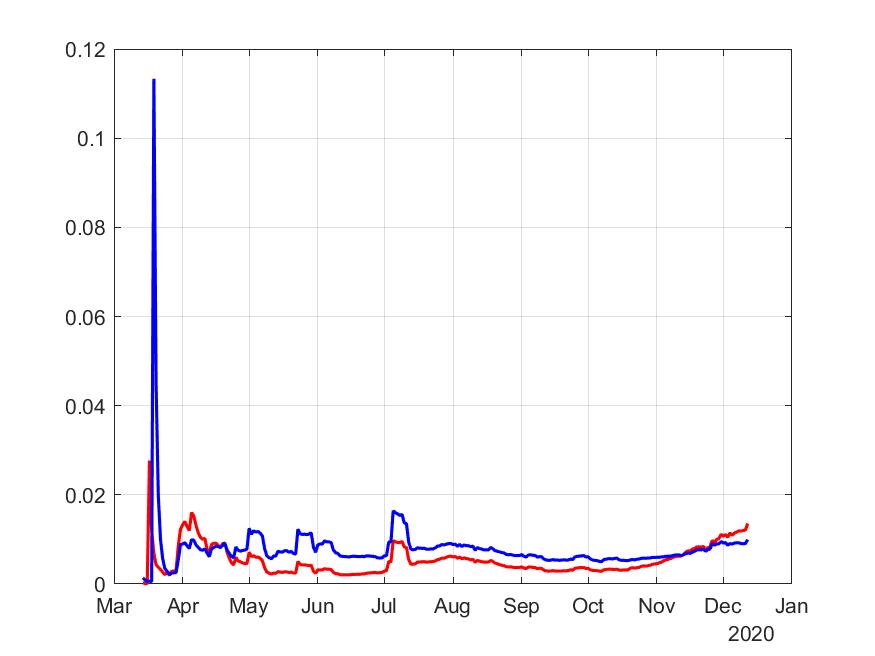}\\[-0.3em] 
  $\nu_t$ \hspace{-0.4cm} &
  \includegraphics[trim = 10mm 6mm 10mm 0mm, clip,width=0.22\textwidth]{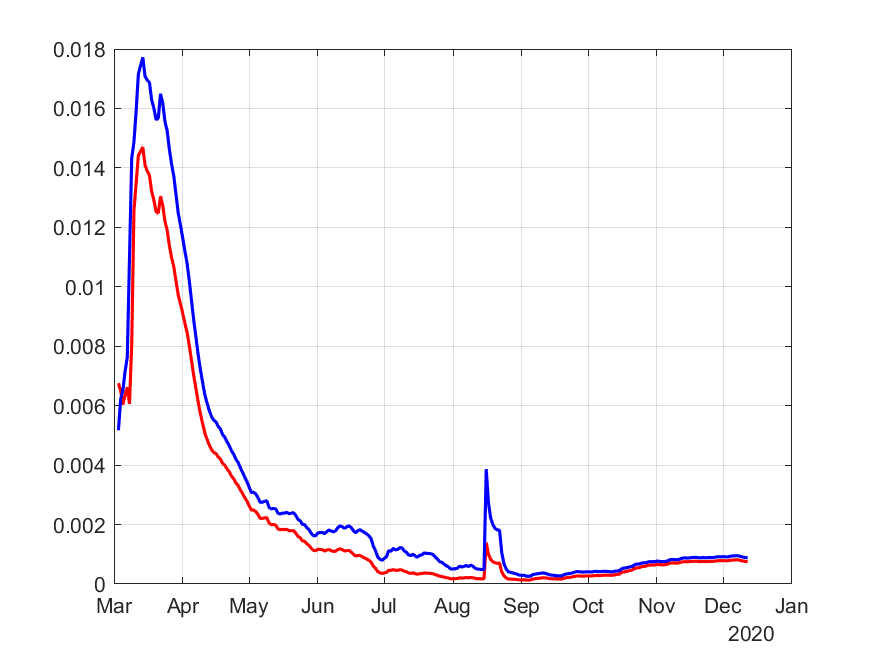} &
  \includegraphics[trim = 10mm 6mm 10mm 0mm, clip,width=0.22\textwidth]{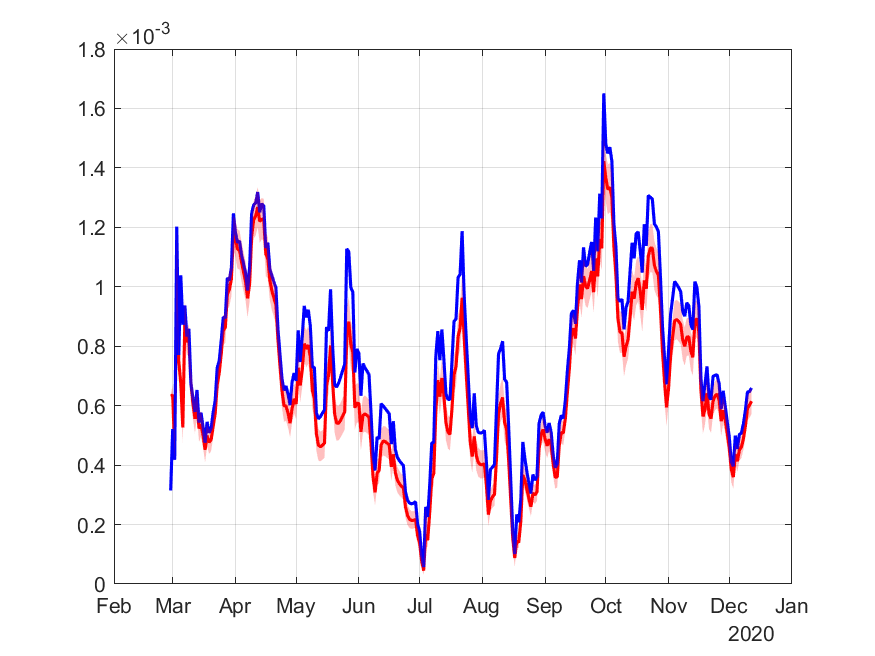} &
  \includegraphics[trim = 10mm 6mm 10mm 0mm, clip,width=0.22\textwidth]{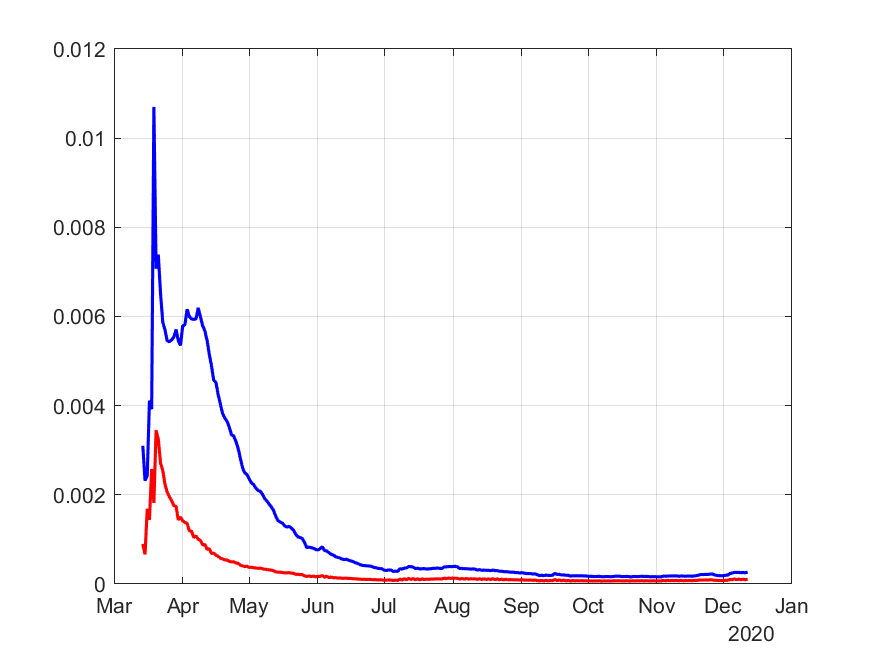}\\[-0.3em]
  $R_{0,t}$ \hspace{-0.4cm} &
  \includegraphics[trim = 10mm 6mm 10mm 0mm, clip,width=0.22\textwidth]{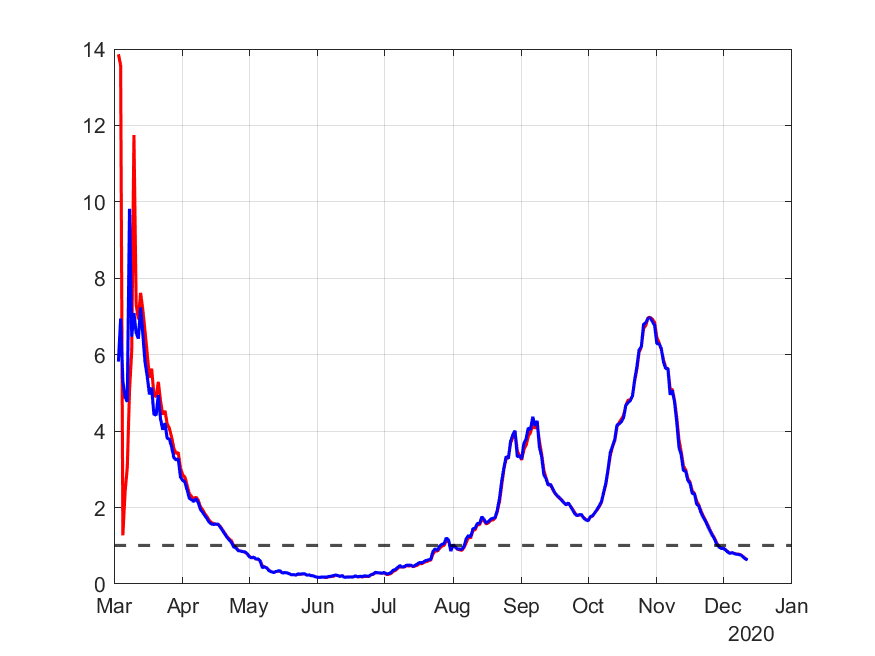} &
  \includegraphics[trim = 10mm 6mm 10mm 0mm, clip,width=0.22\textwidth]{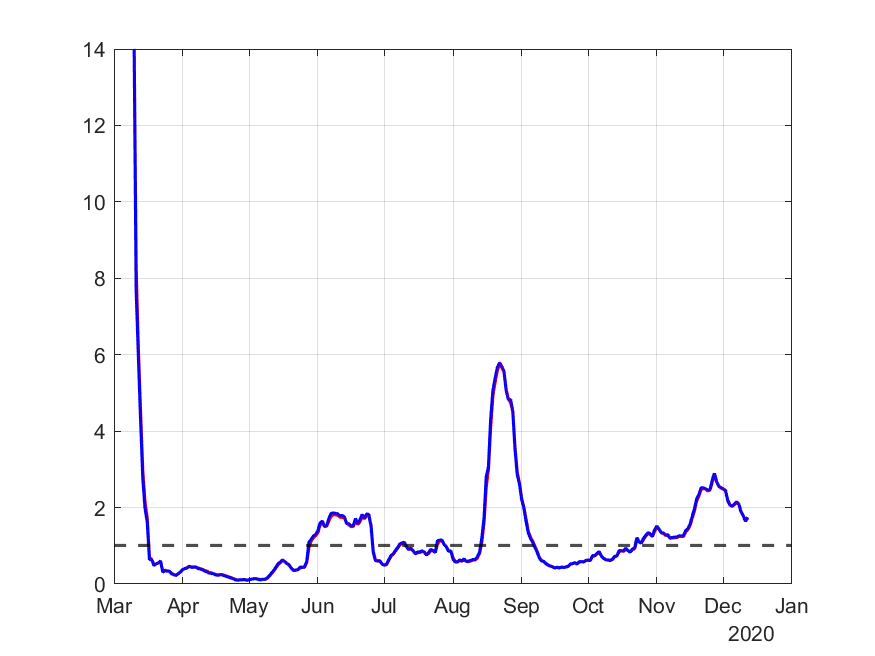} &
  \includegraphics[trim = 10mm 6mm 10mm 0mm, clip,width=0.22\textwidth]{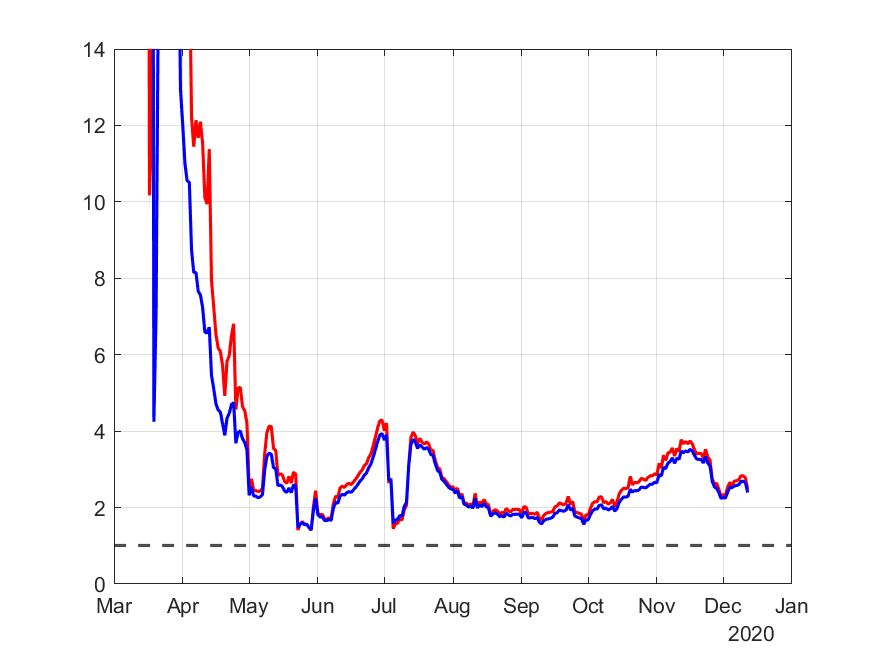}\\[-0.3em] 
\end{tabular}
    \label{fig:TVParameters-A}
\end{figure}

The results discussed in previous sections are computed using official statistics that include reported cases. In this section, we present our findings when we account for this selection bias using the information on the number of deaths and the number of tests together with these tests' positivity rate. We display the evolution of the model parameters estimated using \eqref{eq:TVP-SIRD-A} in Figure~\ref{fig:TVParameters-A} for selected countries. Compared to previous sections, countries, including Brazil, Germany, and India, are excluded from the sample due to limited testing data.\footnote{We use the database provided by \url{https://ourworldindata.org/coronavirus-testing} for the information on testing, see \cite{hasell2020cross} for details. For Germany and Brazil, data on Covid-19 testing is only available at a weekly frequency. We exclude India due to the high share of antigens tests among the total number of tests subject to discussions on these tests' low sensitivity. Results for India are available upon request.} 
\begin{center}
[Insert Figure~\ref{fig:TVParameters-A} about here]
\end{center} 
As can be seen from the graphs in the first row of Figure~\ref{fig:TVParameters-A}, we observe a sizable fraction of the infected individuals that do not show symptoms, $\delta_t$, at the onset of the pandemic for Italy and the US. The fraction is smallest for South Korea, starting from 5\% at the beginning of the sample, decreasing to lower values. This fraction is most considerable for the US, starting from well above 50\% of all infected individuals decreasing to  30\% in June. For Italy, the fraction is about 20\%, and it declines to about 1-5\% in June. Strikingly, the fraction of asymptomatic cases increases with the high number of infections during the pandemic waves. The number of tests might fall short of the number of total cases during these periods when the number of cases surges rapidly.

We can see the impact of the relatively sizable fraction of the US's asymptomatic individuals from the remaining rows of the last column of Figure~\ref{fig:TVParameters-A}. While the parameters' pattern remains unaffected, some level shifts in all rates seem to diminish towards the sample's end. The largest shifts are in the death rate, especially at the pandemic's onset when the fraction of the asymptomatic cases is highest. For Italy, the level shift is relatively more marked for the recovery rate compared to other countries. This downward shift is caused by the larger increase in the number of total infected cases compared to increase in the number of recovered cases. We observe similar level shifts of parameters for other countries, albeit limited compared to the US case. It turns out that the level shifts observed in some structural parameters are offsetting each other throughout the pandemic. Therefore, after the first wave, they do not lead to significant changes in the reproduction rate, $R_{0,t}$ which is the ratio of the infection rate over the resolution rate. In this case, for the US and Italy, $R_{0,t}$ differs considerably from the earlier estimates computed using the reported numbers at the onset of the pandemic. The parameter values converge as the cumulative figures mount, and they have a little effect after the first half of the samples. Therefore, we conclude that the $R_{0,t}$ computed using official statistics reflects progressively more and more the pandemic's actual stance.

	\section{Real-time performance of the models}
	\label{sec:Real-TimePerformance}
\begin{figure}[H]
    \centering
    \caption{Comparison of the real-time estimates of parameters for competing models}
\begin{tabular}{ccccc}
  &  $\beta_t$ & $\gamma_t$ & $\nu_t$ & $R_{0,t}$   \\[-0.2em]
  \begin{turn}{90} \hspace{0.3cm}  Brazil  \end{turn} \hspace{-0.7cm} 
  & \includegraphics[trim = 0mm 6mm 0mm 0mm, clip,width=0.22\textwidth]{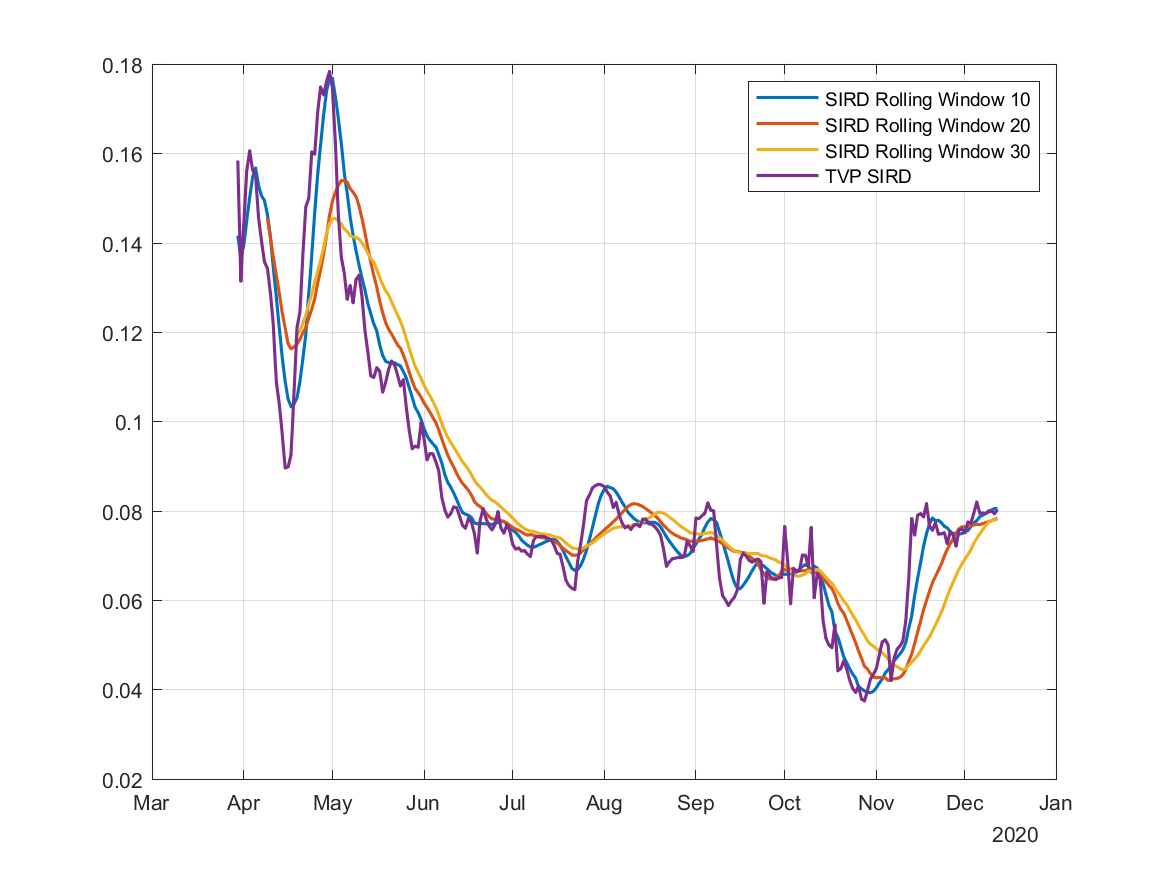}   & \includegraphics[trim = 0mm 6mm 0mm 0mm , clip, width=0.22\textwidth]{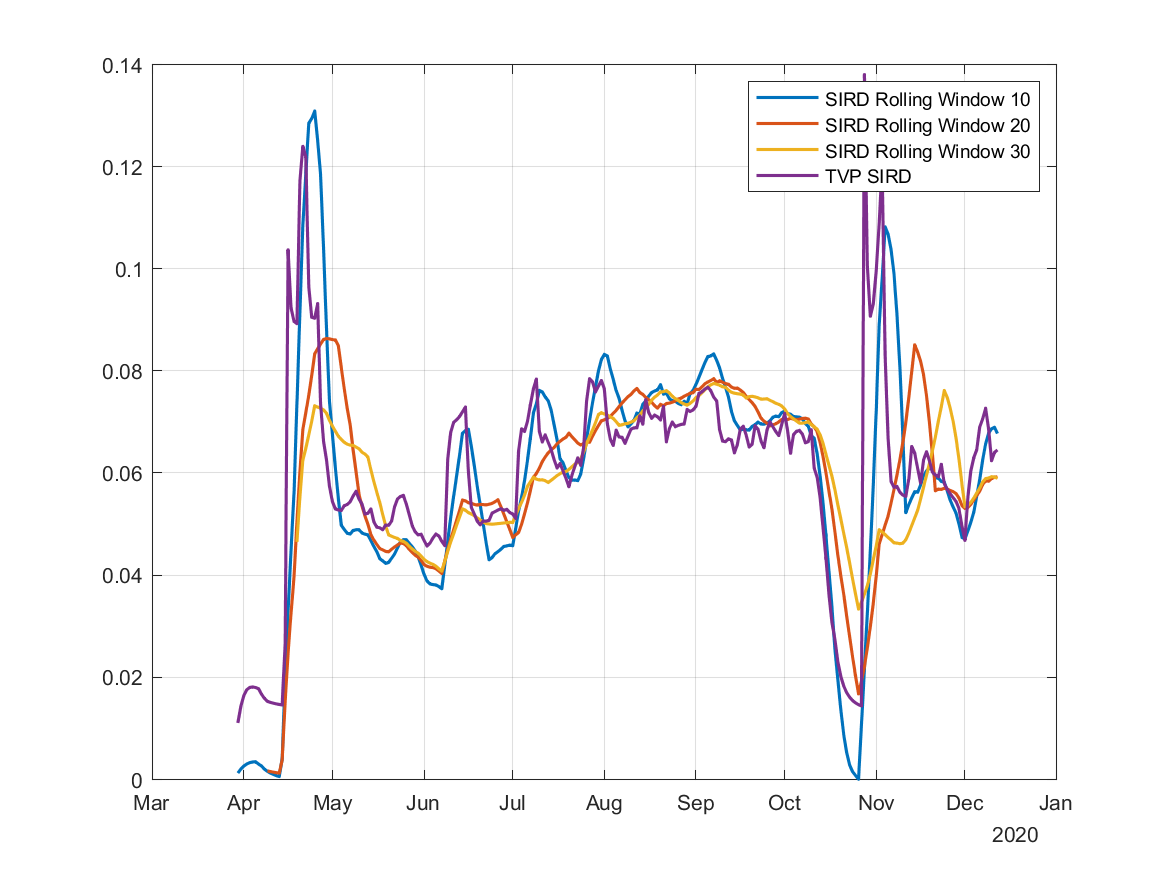} &\includegraphics[trim = 0mm 6mm 0mm 0mm, clip,width=0.22\textwidth]{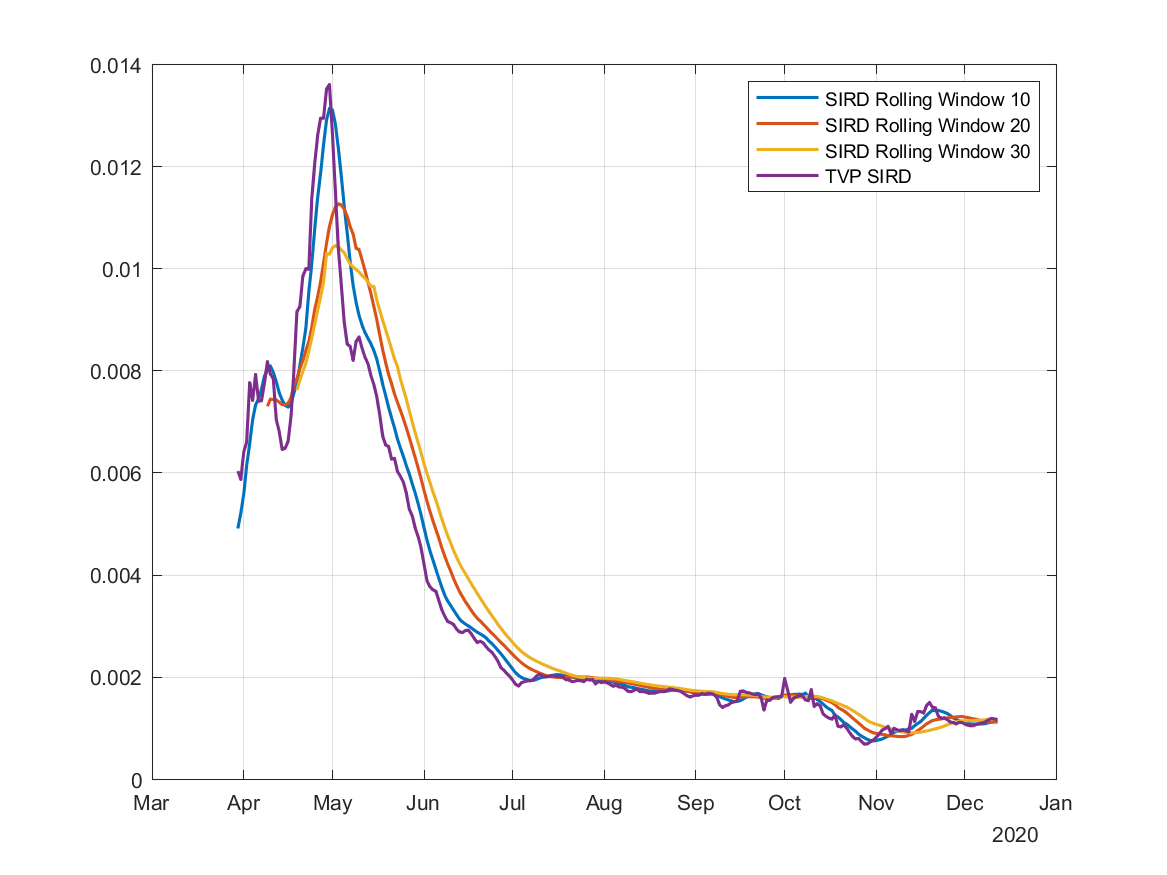}  &  \includegraphics[trim = 0mm 6mm 0mm 0mm, clip,width=0.22\textwidth]{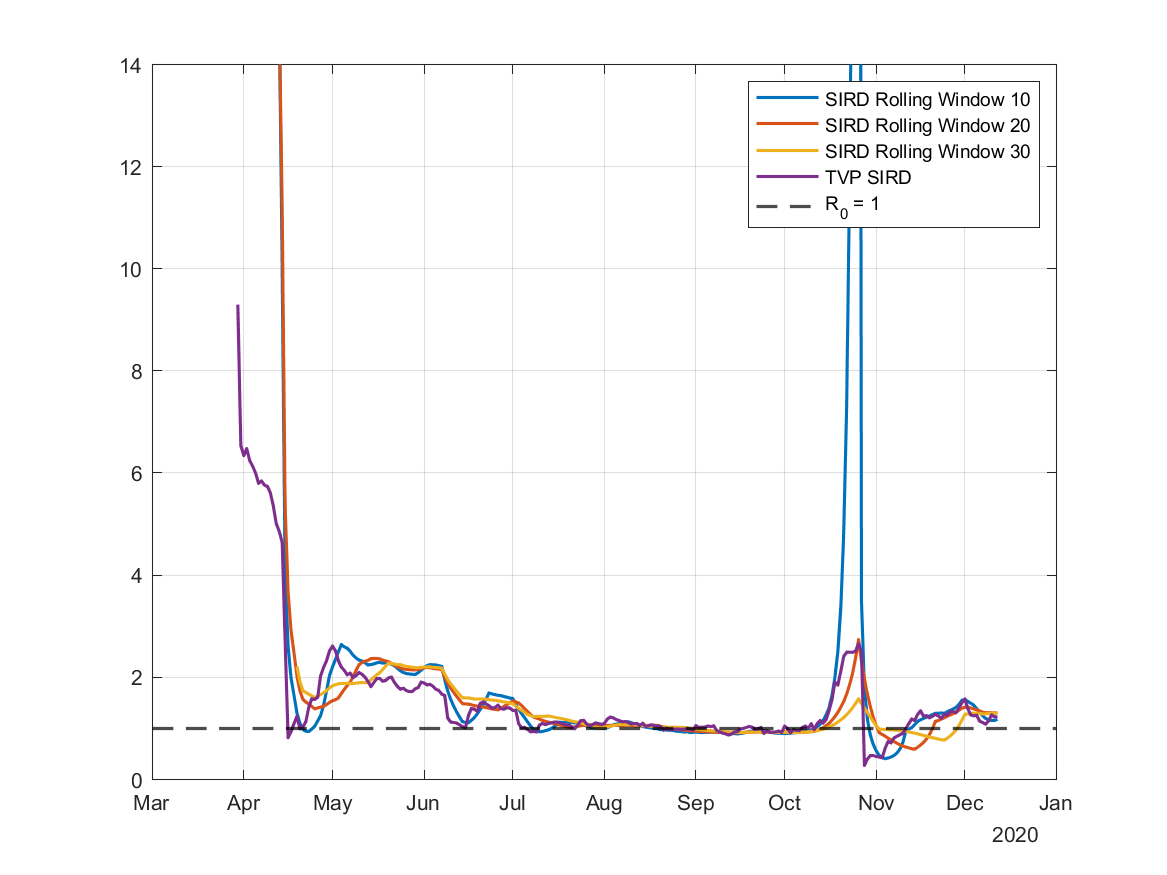} \\
    \begin{turn}{90} \hspace{0.3cm}  Germany  \end{turn} \hspace{-0.7cm} 
  & \includegraphics[trim = 0mm 6mm 0mm 0mm, clip,width=0.22\textwidth]{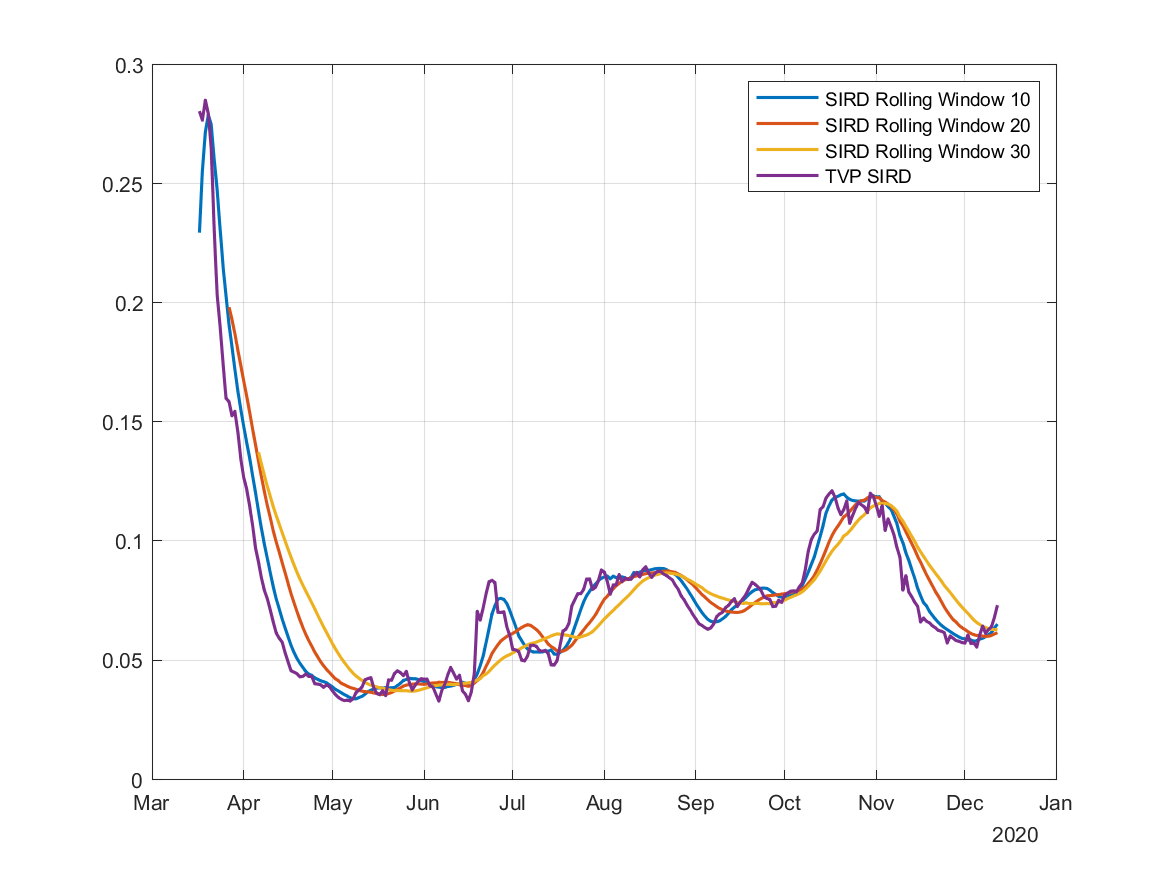}   & \includegraphics[trim = 0mm 6mm 0mm 0mm , clip, width=0.22\textwidth]{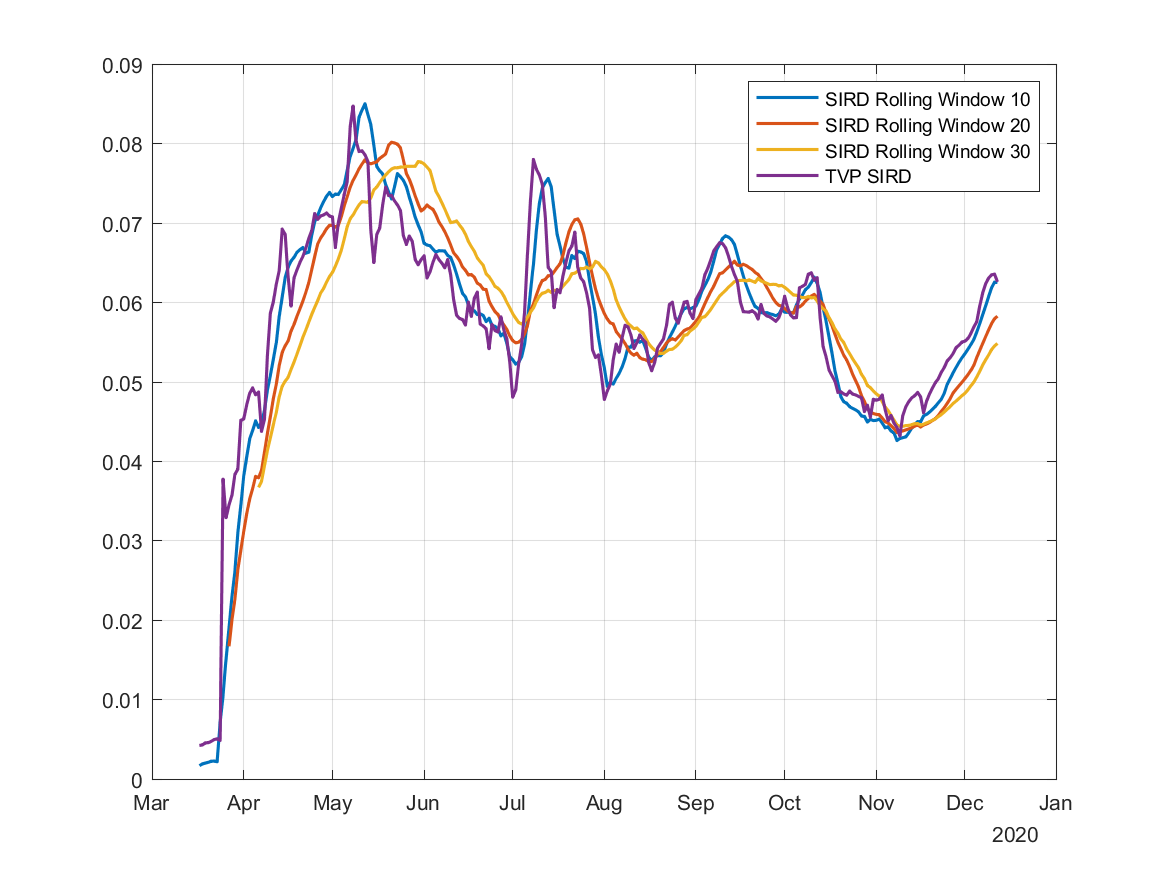} &\includegraphics[trim = 0mm 6mm 0mm 0mm, clip,width=0.22\textwidth]{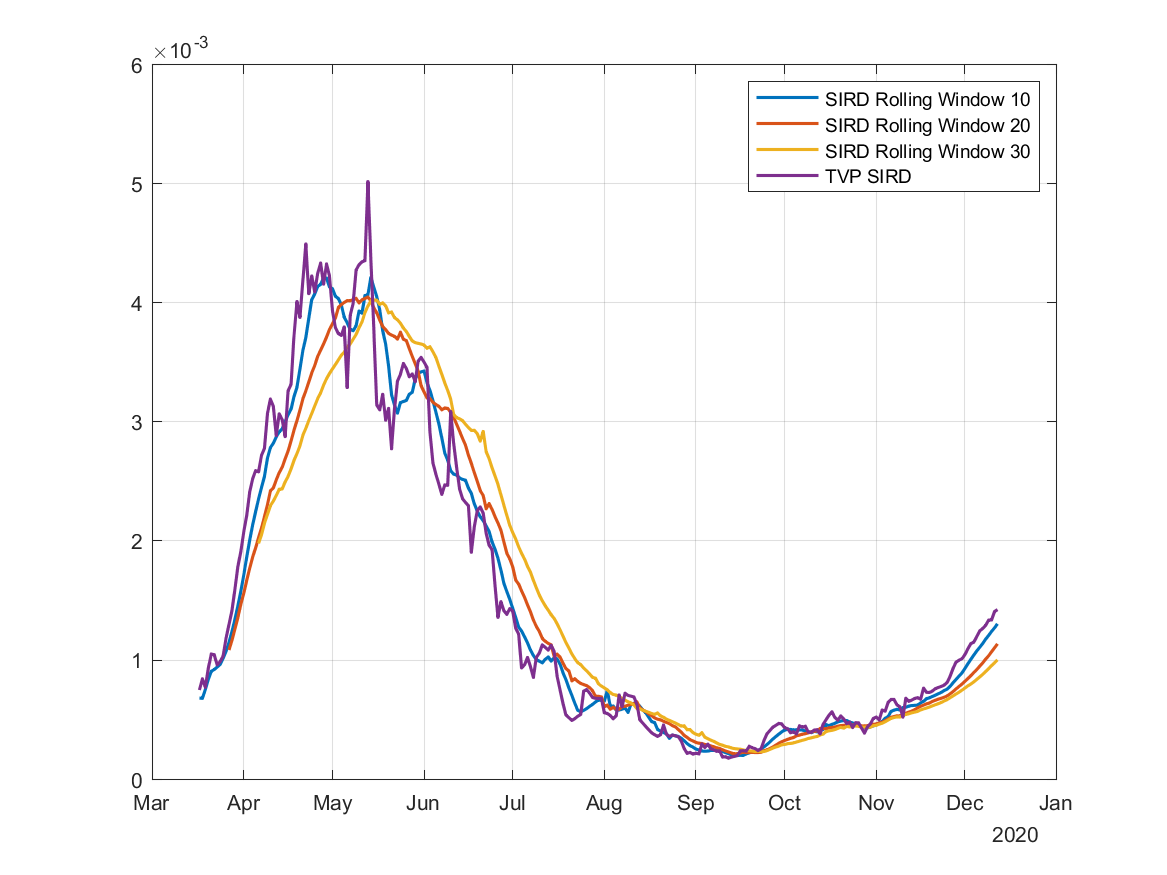}  &  \includegraphics[trim = 0mm 6mm 0mm 0mm, clip,width=0.22\textwidth]{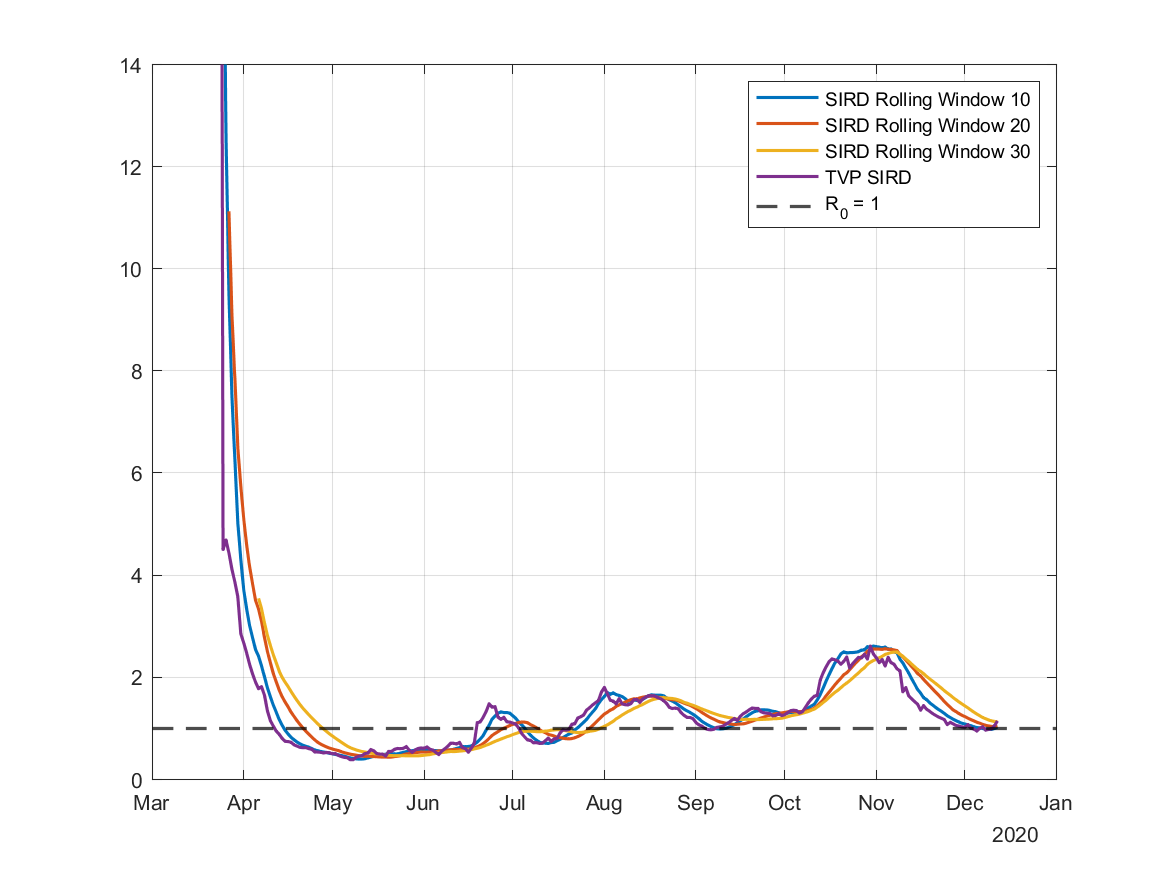} \\
    \begin{turn}{90} \hspace{0.3cm}  India  \end{turn} \hspace{-0.7cm} 
  & \includegraphics[trim = 0mm 6mm 0mm 0mm, clip,width=0.22\textwidth]{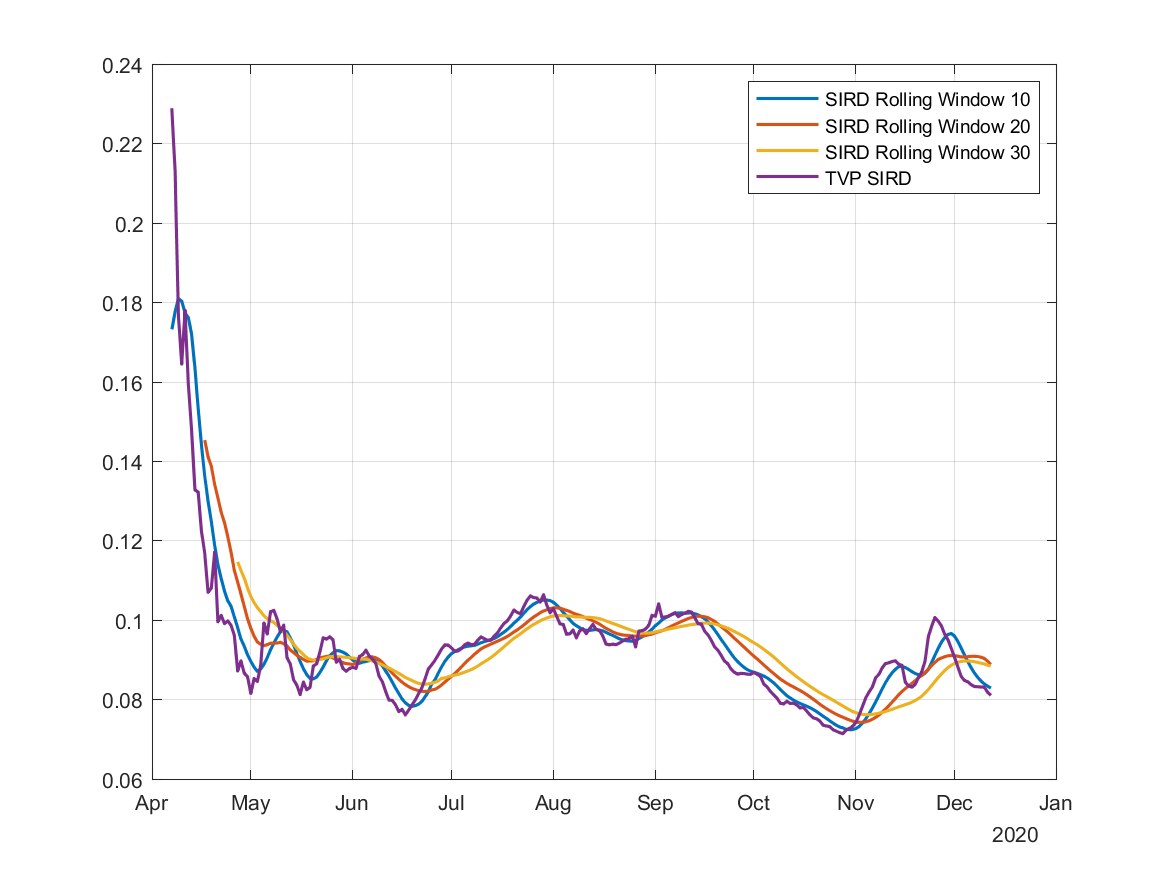}   & \includegraphics[trim = 0mm 6mm 0mm 0mm , clip, width=0.22\textwidth]{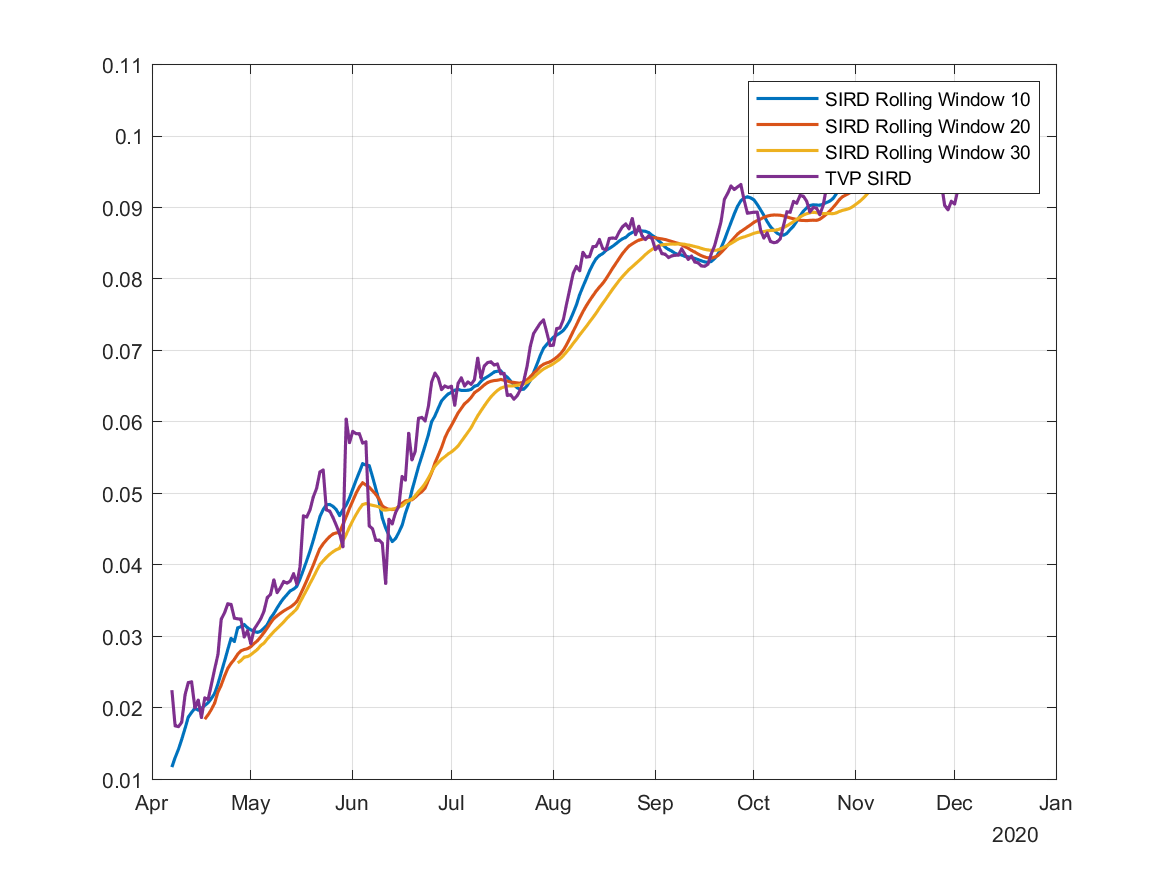} &\includegraphics[trim = 0mm 6mm 0mm 0mm, clip,width=0.22\textwidth]{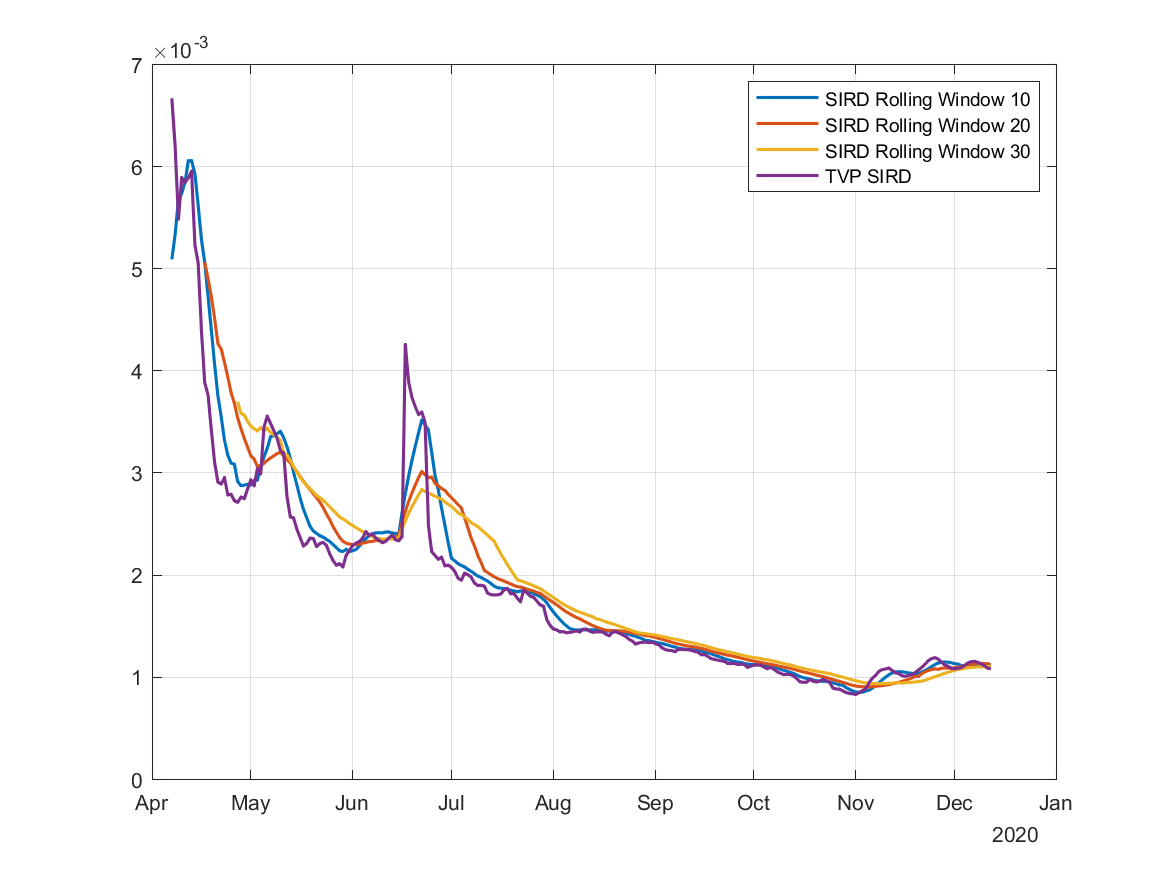}  &  \includegraphics[trim = 0mm 6mm 0mm 0mm, clip,width=0.22\textwidth]{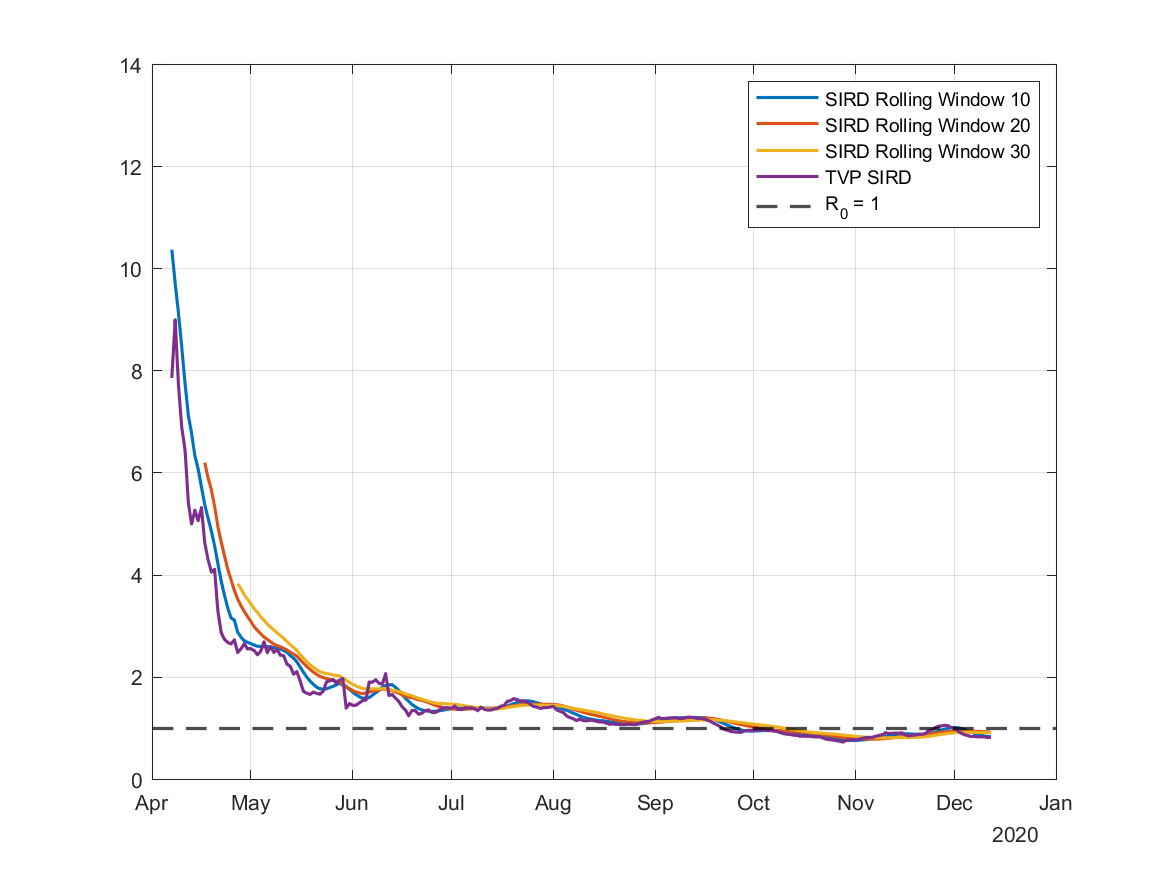} \\
  \begin{turn}{90} \hspace{0.3cm}  Italy  \end{turn} \hspace{-0.7cm} 
  & \includegraphics[trim = 0mm 6mm 0mm 0mm, clip,width=0.22\textwidth]{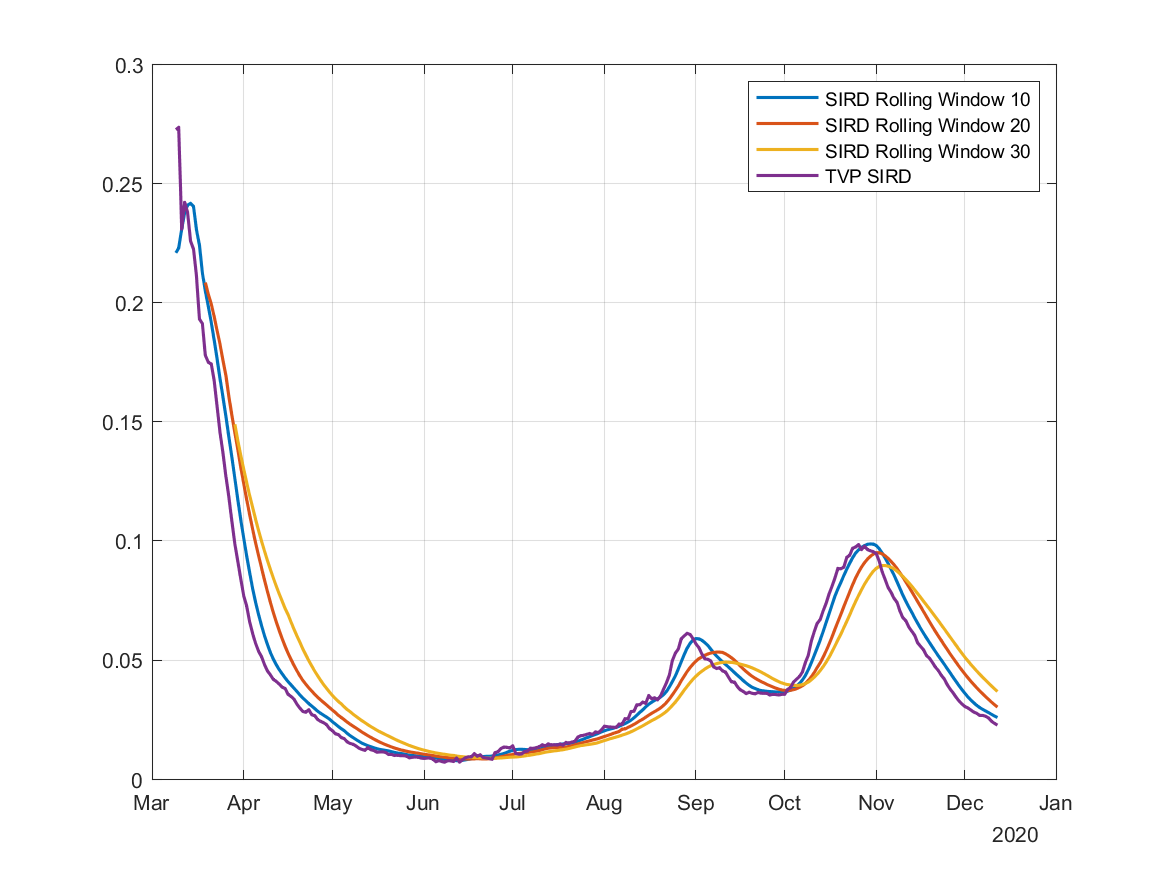}   & \includegraphics[trim = 0mm 6mm 0mm 0mm , clip, width=0.22\textwidth]{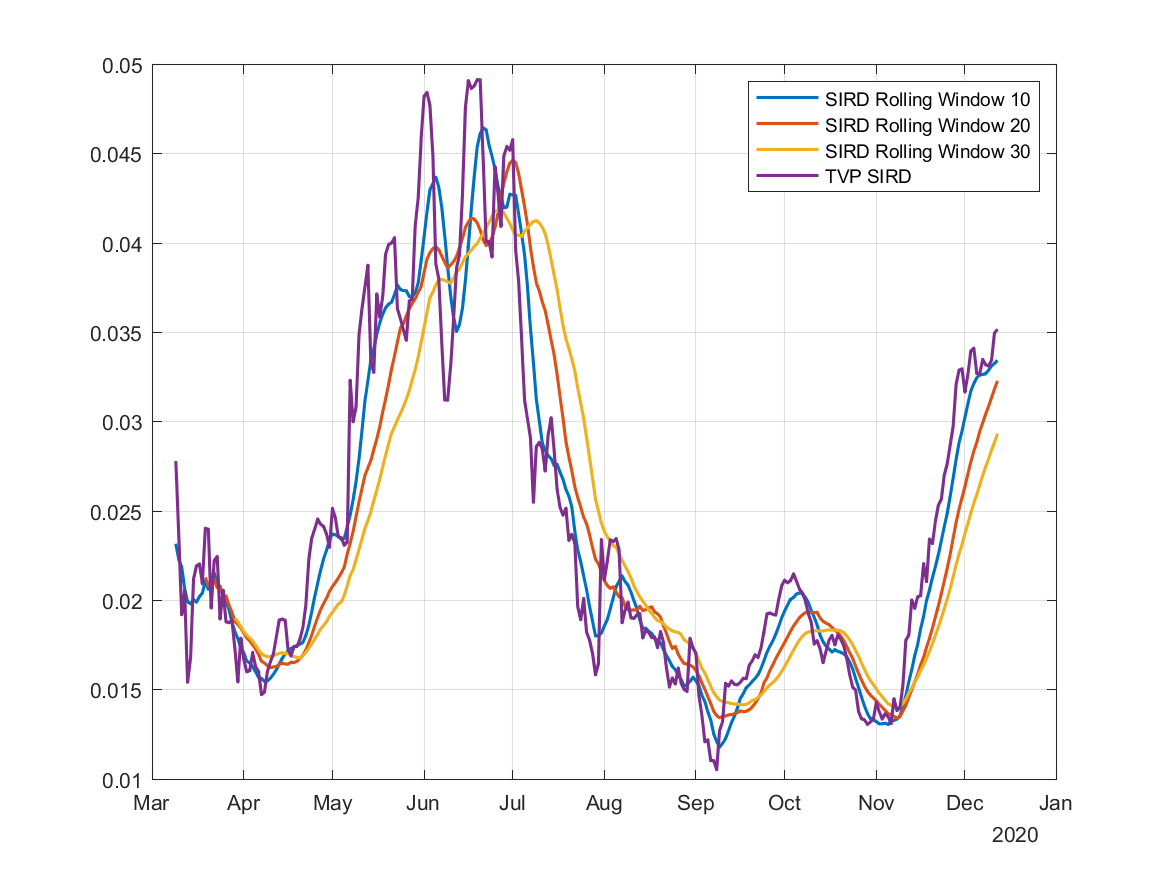} &\includegraphics[trim = 0mm 6mm 0mm 0mm, clip,width=0.22\textwidth]{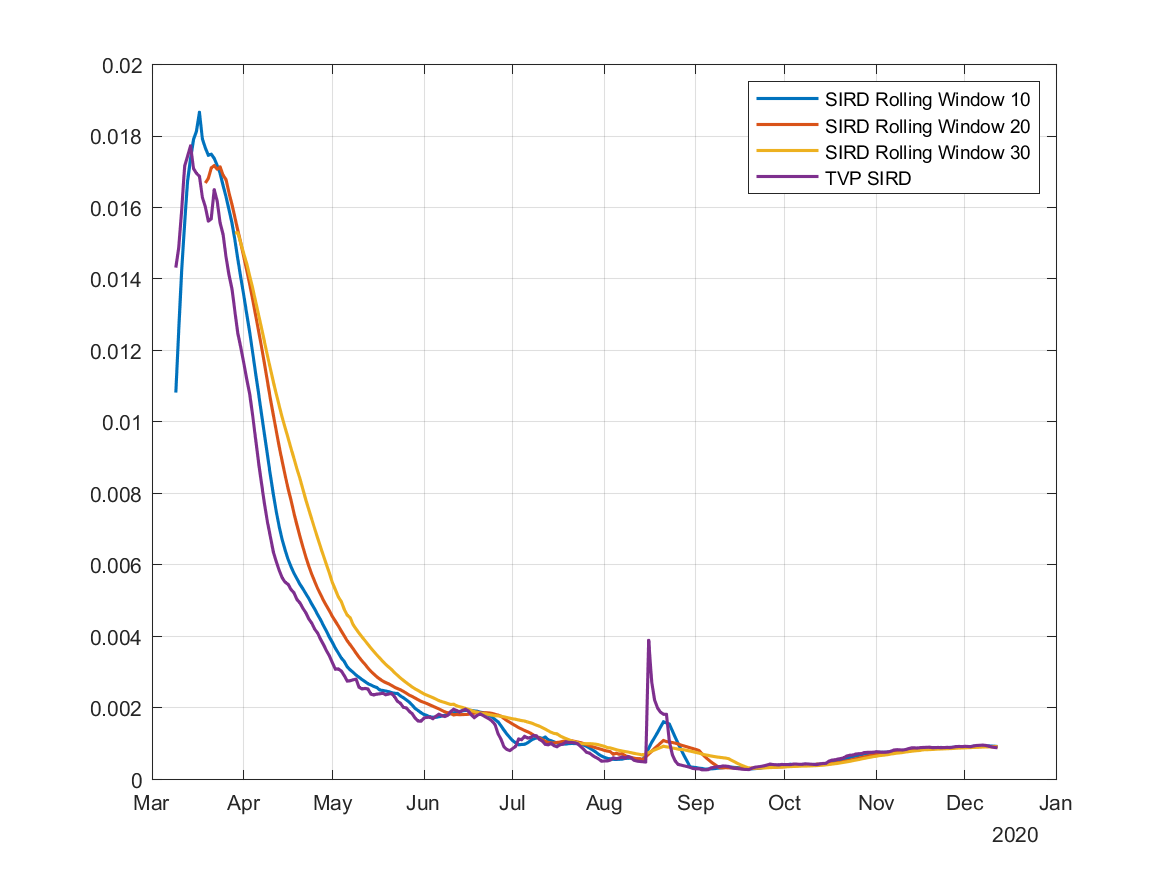}  &  \includegraphics[trim = 0mm 6mm 0mm 0mm, clip,width=0.22\textwidth]{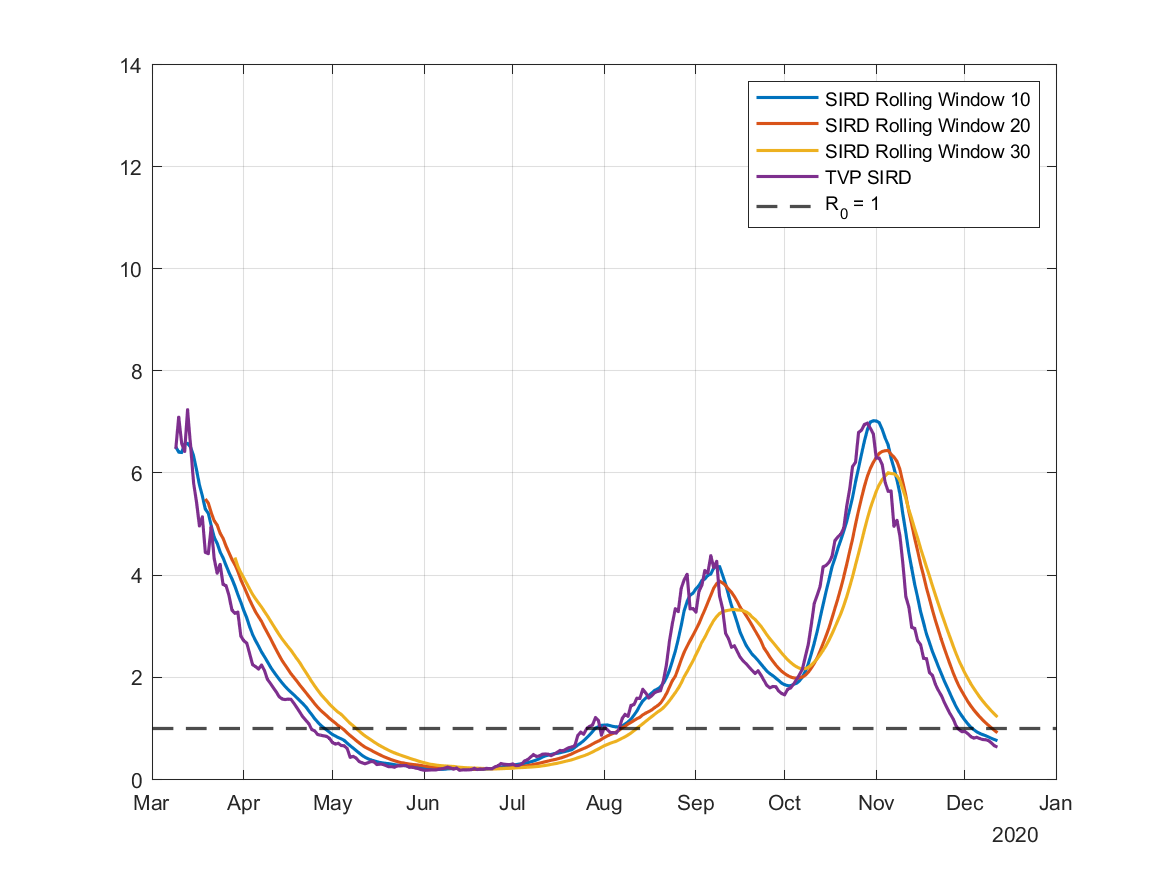} \\
  \begin{turn}{90} \hspace{0.0cm}  S. Korea  \end{turn} \hspace{-0.7cm} 
  & \includegraphics[trim = 0mm 6mm 0mm 0mm, clip,width=0.22\textwidth]{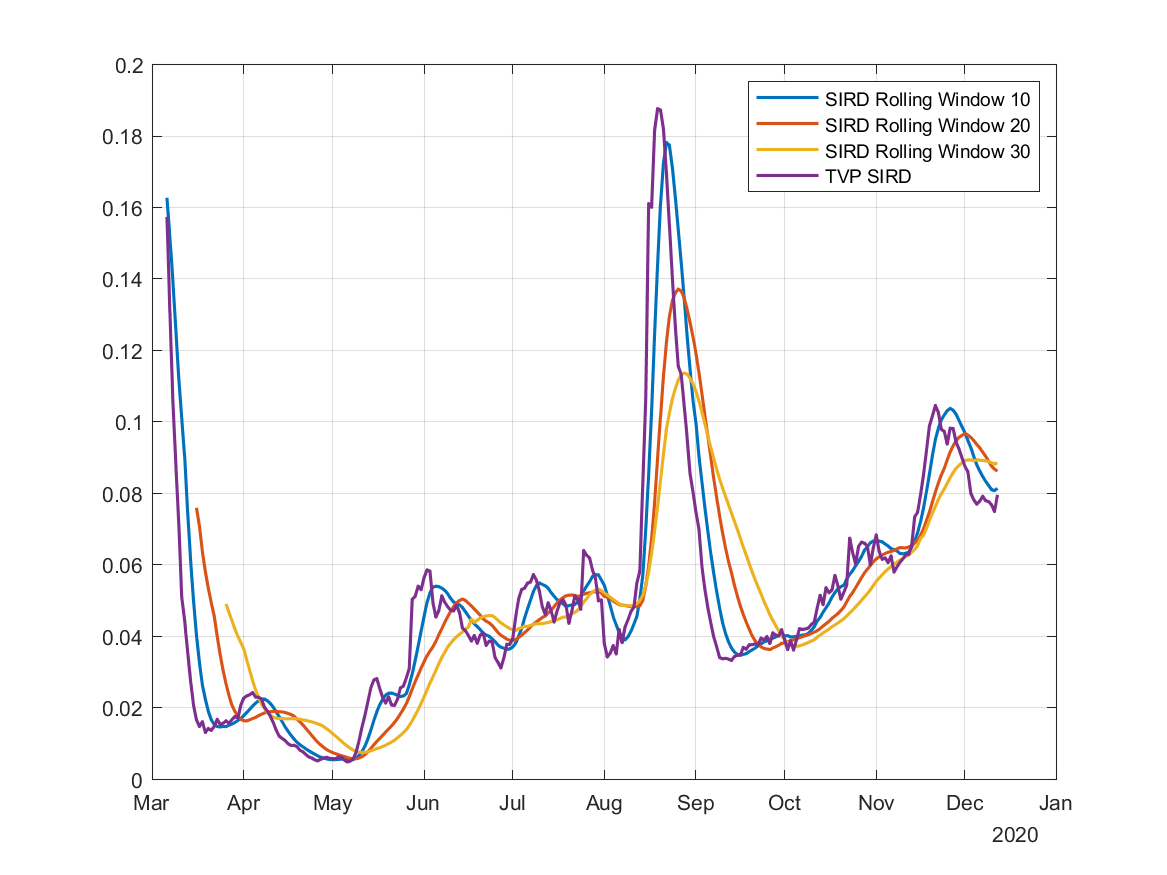}   & \includegraphics[trim = 0mm 6mm 0mm 0mm , clip, width=0.22\textwidth]{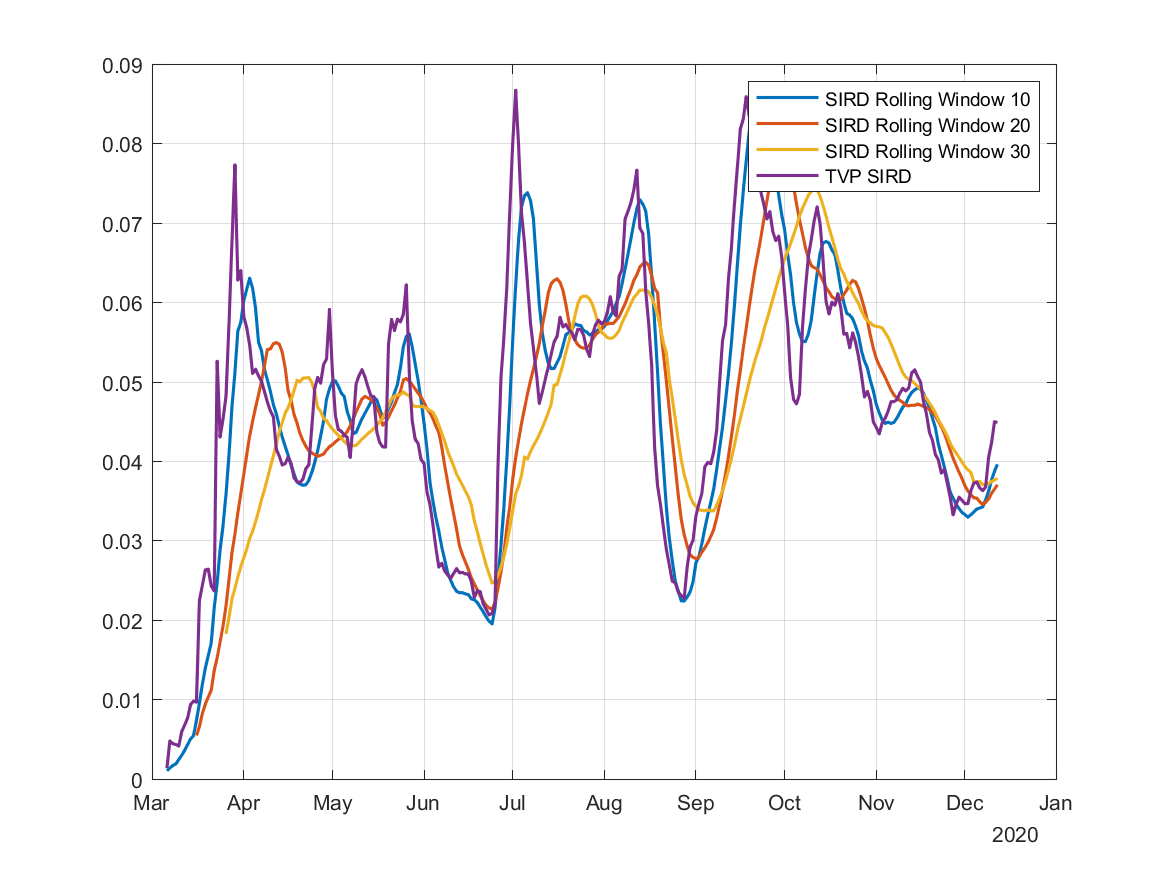} &\includegraphics[trim = 0mm 6mm 0mm 0mm, clip,width=0.22\textwidth]{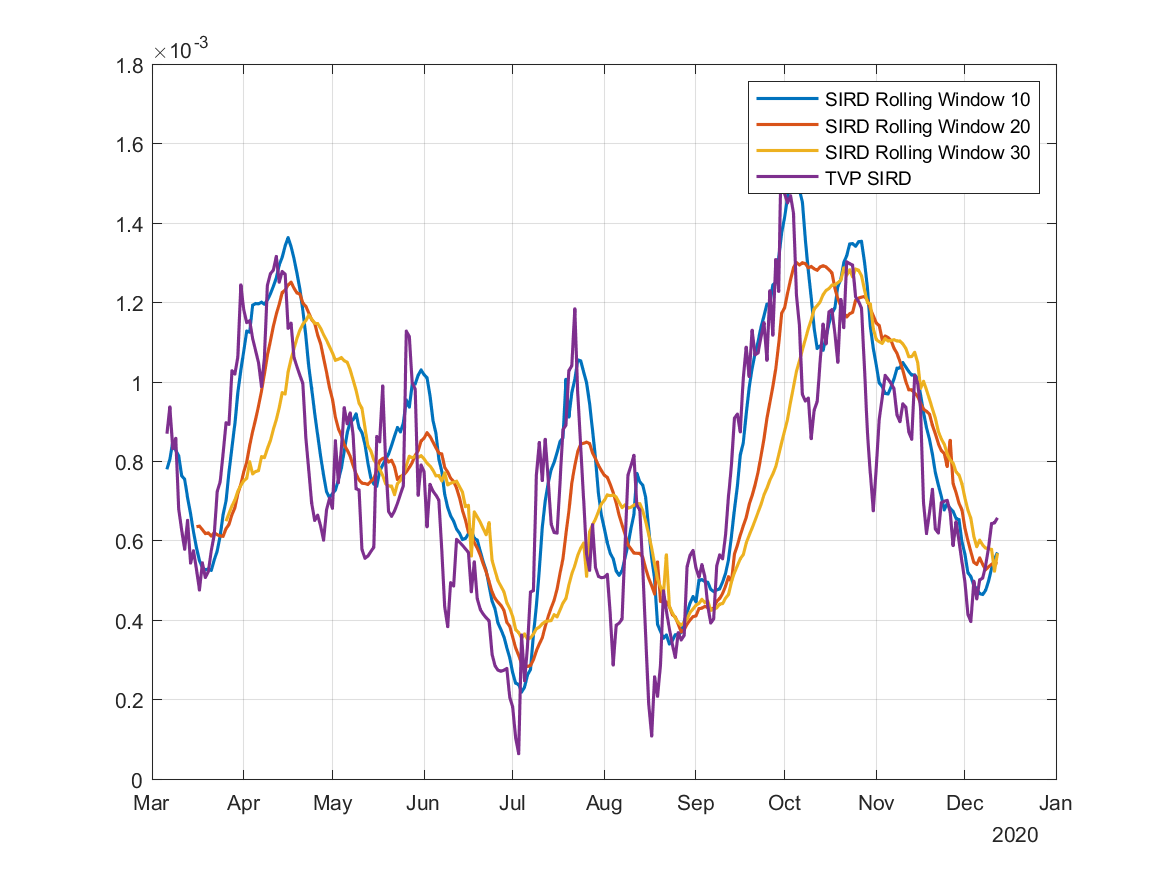}  &  \includegraphics[trim = 0mm 6mm 0mm 0mm, clip,width=0.22\textwidth]{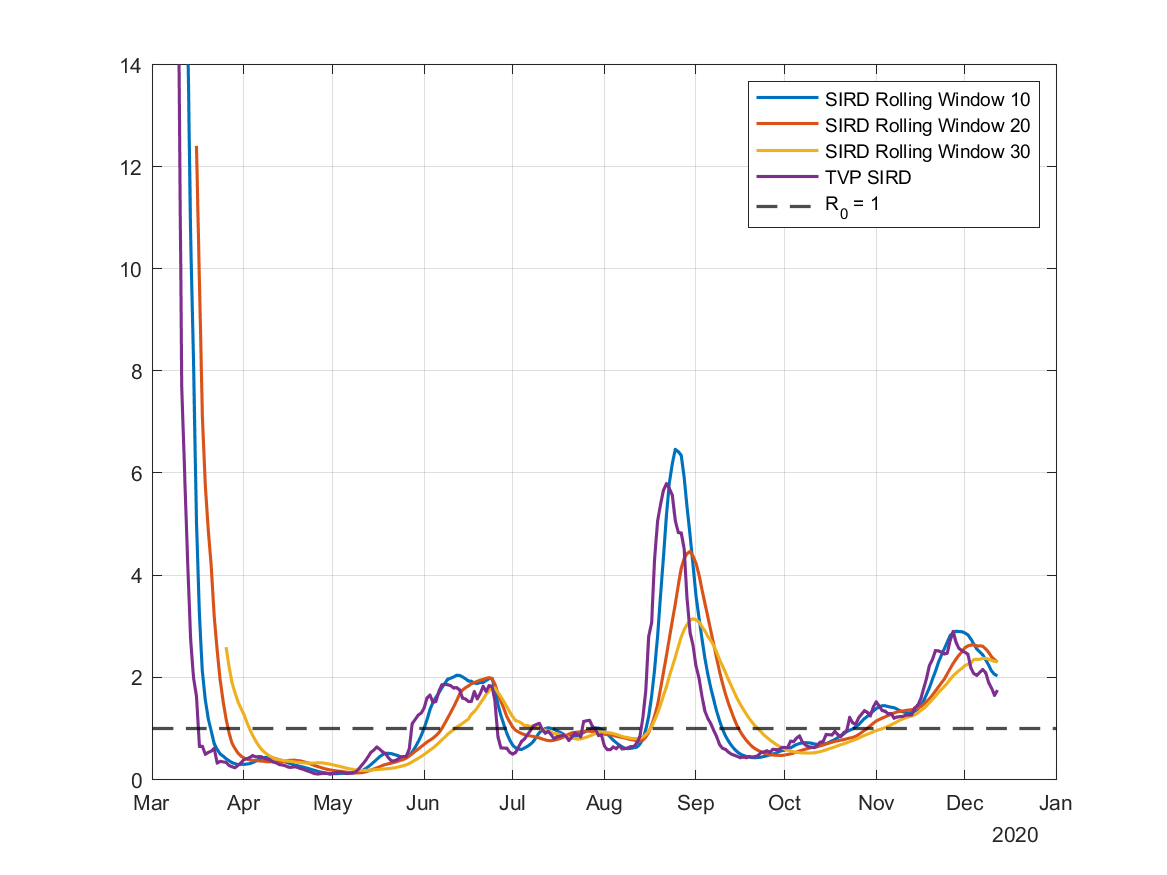} \\
  \begin{turn}{90} \hspace{0.3cm}  US  \end{turn} \hspace{-0.7cm} 
  & \includegraphics[trim = 0mm 6mm 0mm 0mm, clip,width=0.22\textwidth]{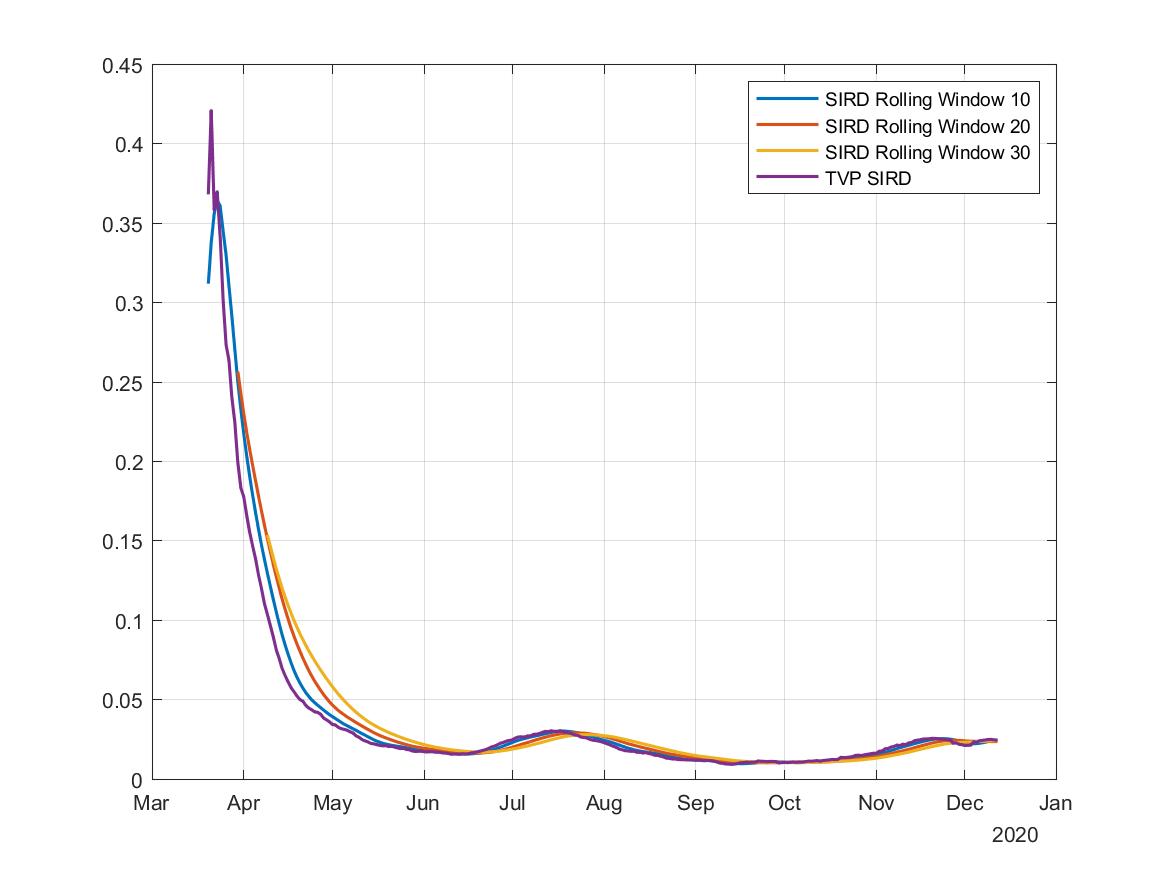}   & \includegraphics[trim = 0mm 6mm 0mm 0mm , clip, width=0.22\textwidth]{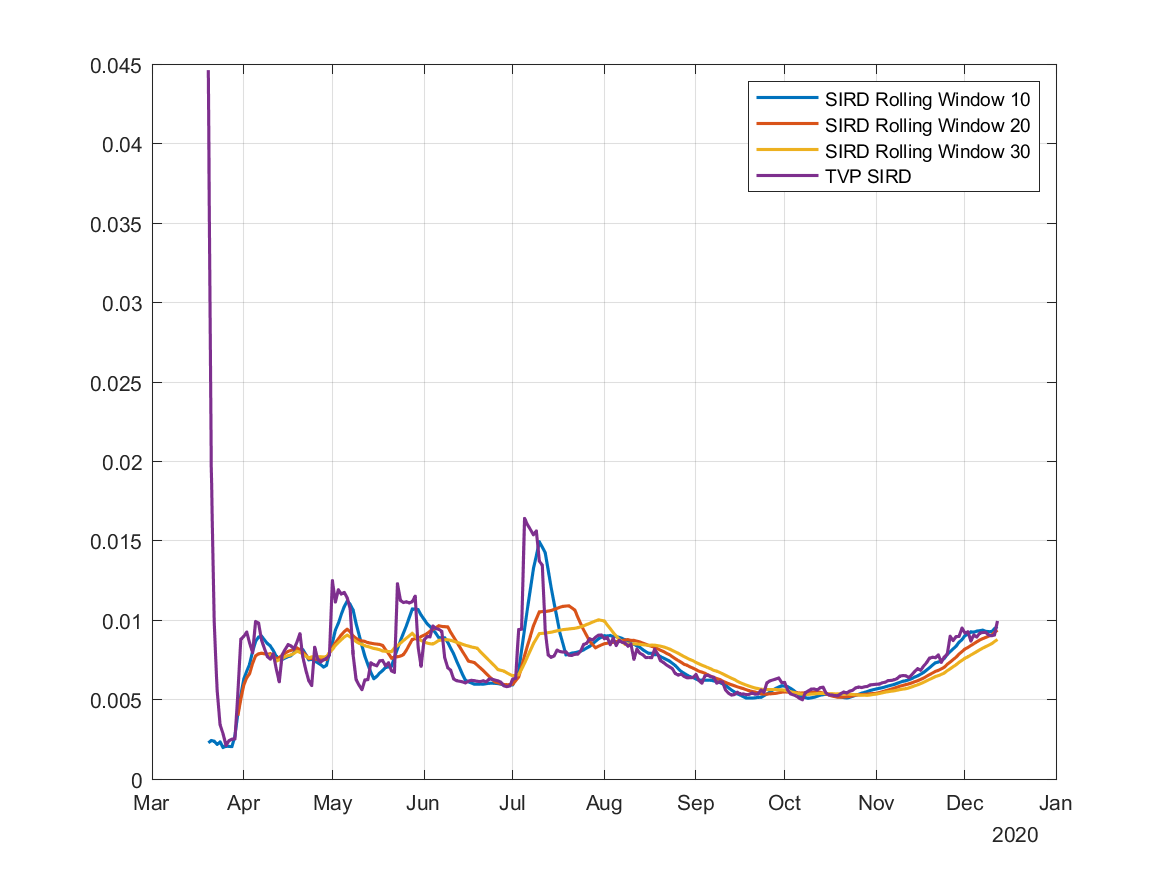} &\includegraphics[trim = 0mm 6mm 0mm 0mm, clip,width=0.22\textwidth]{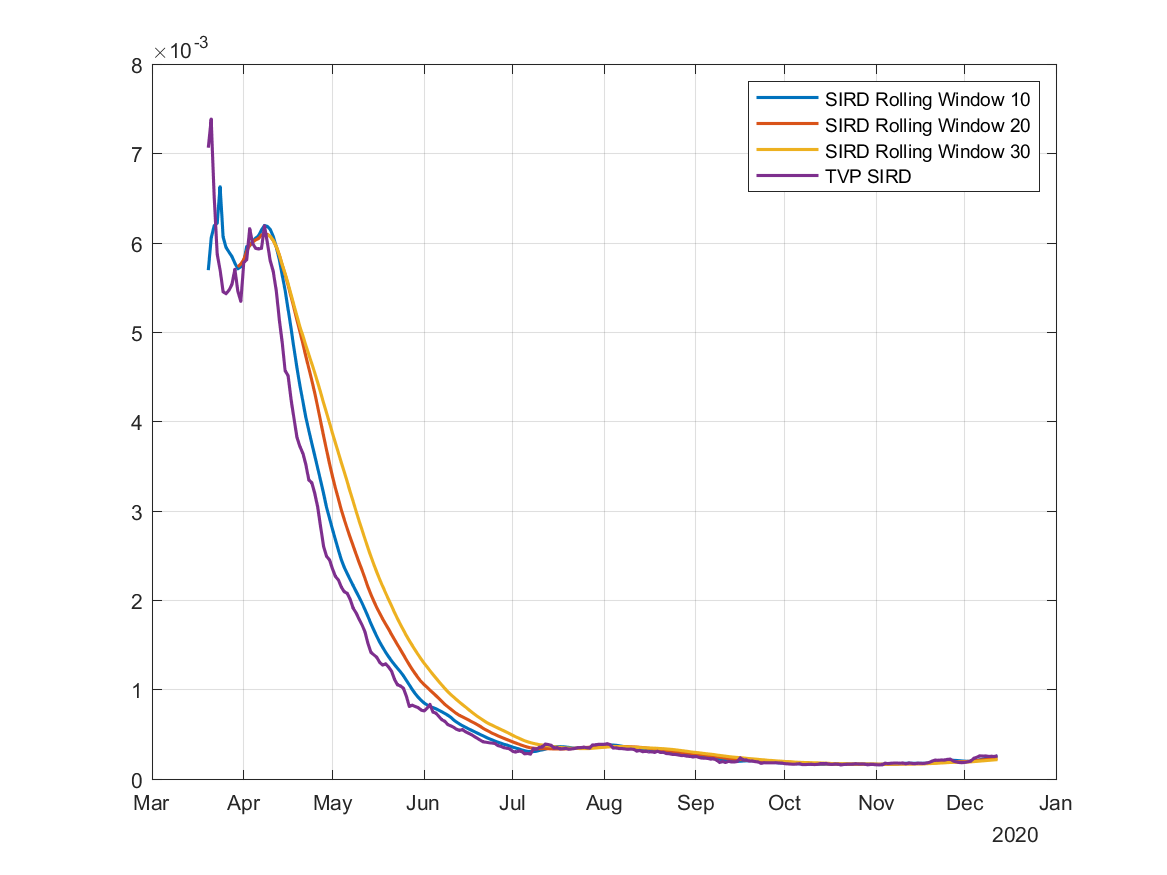}  &  
  \includegraphics[trim = 0mm 6mm 0mm 0mm, clip,width=0.22\textwidth]{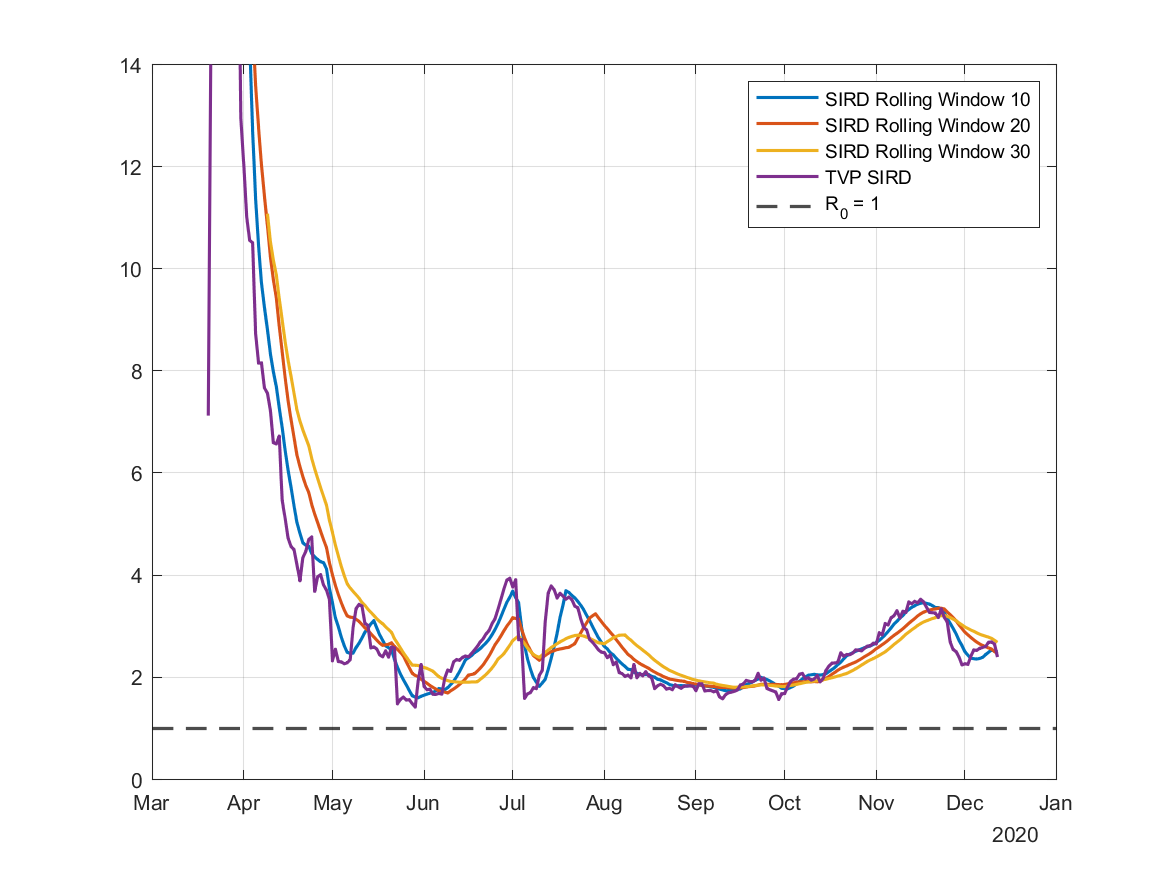} \\
\end{tabular}
    \label{fig:Comparison}
\end{figure}

The results in the previous section display our findings based on the estimates using the full sample dataset. These results indicate that our flexible modeling structure can accommodate various forms of parameter changes reflecting the pandemic's course. However, exploring the model's real-time performance would uncover whether this additional flexibility brought by the time-varying parameters could provide timely and accurate information on the pandemic's real-time stance. Therefore, in this section, we discuss the model parameters' estimation results in real-time using the model with fixed parameters and using the model with time-varying parameters. We use a moving window for performing the SIRD model's estimations with fixed parameters in addition to the expanding window, as the evidence in previous sections shows that the values at different stages of the pandemic could drive the findings intensively. Specifically, using the dataset from $t-M, t-M+1, \dots, t$, we estimate the SIRD model, and the resulting parameter estimates are those for the period $t$. We repeat this process by adding one more observation (and dropping one observation at the beginning of the sample for the moving window) recursively. We consider three cases by setting $M=10$, $20$, and $30$, i.e., starting from ten days of data up to one month of data. For the TVP-SIRD model, we use the data up to period $t$ using an expanding window rather than a rolling window as the parameters, in this case, are time-varying. We display the evolution of the structural parameters,  $\beta_t$, $\gamma_t$, $\nu_t$ and the resulting $R_{0, t}$ over time in Figure~\ref{fig:Comparison}.
\begin{center}
[Insert  Figure~\ref{fig:Comparison} about here]
\end{center} 
Considering the rolling window estimates using the SIRD model with fixed parameters, we observe a trade-off between the speed of reaction to the evolution of pandemic and the window size. The parameters evolve smoothly for the window size with 30 days but cannot promptly react to the rapid changes. On the contrary, using a window size of 10 days, parameters adjust to the new conditions more quickly. When we focus on the time-varying parameters SIRD model, we observe that the parameters can swiftly accommodate the newly changing conditions, ahead of the SIRD model with fixed parameters regardless of the window size. Besides, they can also react to the abrupt changes in the data. Specifically, the TVP model's parameter estimates lead the fixed-parameter model's estimates with 10 days of window size by almost two weeks ahead. This lead time increases when we consider wider rolling window sizes. While the lead time is more extensive when the pandemic waves hit fiercely, it decreases when the pandemic is stabilized, and thus, the variation in parameters is limited. This finding indicates that the TVP model is most useful when there are abrupt changes in the pattern of the data and the uncertainty about the course of the pandemic is at the highest level. For example, consider Brazil. In this case, the relatively mild increases in the number of active cases after $10^{th}$ of April is instantly reflected in the rate of infection with the TVP model, whereas this process takes longer for the SIRD model for 10 days of window size for estimation. We could observe a similar pattern at the end of July and August during the peak of the first wave for Brazil. In this case, the downturn is captured by the TVP-SIRD model earlier than the competing models. Finally, the pattern observed in April is repeated at the end of October, when the number of active cases starts to soar once again. Therefore,  we conclude that the TVP-SIRD model can anticipate the turning points earlier than the competing models in real-time.

In addition, the jump in recoveries' number on April 14 is instantly reflected in the recovery rate with the release of the first set of data on recoveries.\footnote{ Here, we do not take a stance on the official statistics' quality as all competing models are estimated using the same data.} For the SIRD model with 10 days of window size in inference, this process takes almost two weeks to reflect these changes. A reverse pattern is observed at the end of October when the number of recoveries was low for a couple of days. For these periods, while the fixed-parameter model estimates with 10 days of rolling window shrink towards minuscule values, the TVP-SIRD model seems to keep a balance between these unusual realizations and earlier observations. The fixed parameter model with 10 days of moving window is overly sensitive to extreme data realization, unlike the TVP-SIRD model.  These unusual realizations often stem from the misreporting of the cases. This finding shows that the TVP-SIRD is robust because it can rapidly accommodate the changes, but it is not overly sensitive to extreme observations. This robustness stems from the flexible modeling structure that exploits all available information. As a combination of these two opposing forces of infection and resolution rates, the reproduction rate immediately fell to the levels around 1 in mid-April, which bounced back again to values around 2 in about two weeks. It exceeds the values of 2 at the end of October due to low recovery rates, but it quickly reverted to values slightly above 1. Here, we do not discuss the fixed-parameter SIRD model results with the rolling window size of 20 and 30 days intensively as these perform worse than the model with 10 days of window size used for inference. 

During a severe pandemic like the current one, these lead times might be crucial in predicting the pandemic's short-term trajectory. Therefore, we explore whether this capability of the TVP-SIRD model in reflecting the pandemic's stance promptly proved to be useful in forecasting the number of daily confirmed cases. We, therefore, perform a real-time forecasting exercise where using our recursive estimations of the models in \eqref{eq:FPSIR} and \eqref{eq:TVP-SIRD} based on the information available in time period $t$, we perform $h=1,3,5,7,9,11,14-$day ahead predictions of daily confirmed cases, i.e. $\Delta C_{t+h}$. We use the first one-third of the full sample as the estimation sample, and we expand the window by adding one more observation and repeat the procedure. This ratio roughly provides us at least 180 days of evaluation period for each country. We display the results involving RMSFEs of the competing models relative to the TVP-SIRD model in Table~\ref{tab:OOS}. 
\begin{center}
[Insert Table~\ref{tab:OOS} about here]
\end{center} 
Table~\ref{tab:OOS} reveals that the TVP-SIRD model outperforms the SIRD models with fixed parameters estimated using an expanding window (EW) or using a rolling window regardless of the size of the window. First, fixed parameter models estimated using an expanding window or rolling windows with more than 10 days of observations perform inferior relative to the SIRD model with fixed parameters using 10 days of observations in the sense that the relative RMSFE increases monotonically with the use of more data. Second, the TVP-SIRD model that uses all available data outperforms competing models. Therefore, the closest competitor to the TVP-SIRD model is the SIRD model using 10 days of rolling window (RW-10). In this case, for 1-day ahead predictions of confirmed cases, the TVP-SIRD model provides much better predictions than the RW-10. However, for all the models, the forecasting performance deteriorates with the increasing prediction horizon. Therefore, the models' relative RMSFEs monotonically approach values closer to 1 for most models when the forecast horizon is as long as two weeks. These results show that while for the short horizons, the correction in the bias overcomes the reduction in the variance for the SIRD models, the reduction in the variance is more effective in the longer horizons. Still, the TVP-SIRD model outperforms the competing models for longer horizons for all countries except South Korea, where the two models perform quite close. TVP-SIRD model surpasses all competing models reflecting the flexible structure's ability to accommodate the pandemic's changing behavior rapidly.

\section{Conclusion}
\label{sec:Conclusion}
Countries have been imposing various measures to fight the pandemic ranging from partial curfew to full lockdown to lower the pandemic's transmission. These measures presumably pave the way for the normalization of economies and reopening policies. Health systems with overloaded intensive care units lead to substantial variation in the number of recoveries and daily death tolls throughout the pandemic. Additionally, mutations of the virus seem to alter the pandemic's essential features, such as infection rate. Therefore, the workhorse epidemiological SIRD model's parameters change over time due to these changes.

This paper extends the SIRD model, allowing for time-varying structural parameters for timely and accurate measurement of the pandemic's current stance and an accurate prediction of its future trajectory. Our modeling framework falls into the class of `generalized autoregressive score models'. These models involve parameters evolving deterministically according to an autoregressive process in the direction implied by the score function. Therefore, the resulting approach permits a flexible yet parsimonious and statistically coherent framework to efficiently operate in scarce data environments due to low computational cost. We demonstrate the proposed model's potential using daily data from a set of countries ranging from the US to South Korea with distinct pandemic dynamics since the pandemic outbreak. 

Our results show that the proposed framework can nicely track the stance of the pandemic in real-time. Our findings suggest that there is considerable fluctuation in the rate of infection and recovery and death rates. For most countries, the reproduction rate is above the critical level of 1, implying that they cannot contain the pandemic yet. We further extend the model to include the infected individuals who do not show symptoms and are therefore not diagnosed. This sample selection might have a sizable impact on the estimated level of reproduction rate at the onset of the sample, but the rates converge to similar levels towards the end of the sample.         

\newpage

\singlespacing
	
\bibliographystyle{jae}
\bibliography{ref}

\newpage

\section*{Tables and Figures}

\vspace{-0.8cm}

\begin{table}[H]
\singlespacing
\setlength{\tabcolsep}{18pt}
    \centering
\begin{threeparttable}
    \caption{The dataset}
\label{tab:dataset}
    \begin{tabular}{l|l}\hline \hline
    Country	    & \multicolumn{1}{c}{Time span} \\ \hline
 	Brazil	    & March 08 - Dec 13\\
 	Germany	    & March 08 - Dec 13\\
	India	    & March 23 - Dec 13 \\
	Italy	    & March 29 - Dec 13\\
	S. Korea	& Feb. ~~26 - Dec 13 \\
	US          & March 11 - Dec 13 \\ \hline \hline
    \end{tabular}
   \begin{tablenotes}
         \item \footnotesize {\it Note:} The data is obtained from GitHub, Covid-19 Data Repository by the Center for Systems Science and Engineering (CSSE) at Johns Hopkins University.
      \end{tablenotes}
\end{threeparttable}   
\end{table}	

\vspace{-0.4cm}

	\begin{table}[H]
	\singlespacing
	\setlength{\tabcolsep}{9pt}
		\centering
\begin{threeparttable}		
		\caption{Estimation results of the SIRD model with fixed parameters}
				\label{tab:FP-SIR-Parameters}
			\begin{tabular}{lcccc}\hline \hline
			& $\beta$ &  $\gamma$  & $\nu$ & $R_0$   \\ \hline
	Brazil	    & 0.072 (0.000) & 0.063 (0.000) & 0.002 (0.000) & 1.111 (0.001) \\
	Germany	    & 0.074 (0.000) & 0.054 (0.000) & 0.001 (0.000) & 1.340 (0.001) \\
	India 	    & 0.090 (0.000) & 0.085 (0.000) & 0.001 (0.000) & 1.044 (0.000) \\
	Italy	    & 0.046 (0.000) & 0.025 (0.000) & 0.002 (0.000) & 1.707 (0.002) \\	
	S. Korea    & 0.053 (0.000) & 0.042 (0.000) & 0.001 (0.000) & 1.253 (0.009) \\
	US          & 0.020 (0.001) & 0.007 (0.000) & 0.000 (0.000) & 2.555 (0.001) \\ \hline
			\hline 
		\end{tabular}
   \begin{tablenotes}
         \item \footnotesize {\it Note:} The table displays the estimation results of the model in \eqref{eq:FPSIR}. We display the posterior means and posterior standard deviations (in parenthesis) of the corresponding parameter shown in the first row and the country shown in the first column. 
      \end{tablenotes}
\end{threeparttable}  		
	\end{table} 
	
	\vspace{-0.4cm}

	\begin{table}[H]
\setlength{\tabcolsep}{6pt}	
	\singlespacing
	    \small
		\centering
\begin{threeparttable}				
		\caption{Estimation results of the TVP-SIRD model parameters}
			\label{tab:TVP-SIRD-par}
			\begin{tabular}{lrrrrrr}\hline \hline
				& \multicolumn{1}{c}{Brazil} & \multicolumn{1}{c}{Germany} & \multicolumn{1}{c}{India} & \multicolumn{1}{c}{Italy} & \multicolumn{1}{c}{S. Korea}  & \multicolumn{1}{c}{US} \\ \hline
$\alpha_0$ &-0.093 (0.004) &-0.101 (0.006) & -0.042 (0.009) &-0.031 (0.005) &-0.145 (0.024) &-0.102 (0.002) \\
$\alpha_1$ & 0.965 (0.001) & 0.966 (0.002) &  0.983 (0.004) & 0.999 (0.002) & 0.961 (0.009) & 0.974 (0.001) \\
$\alpha_2$ & 0.823 (0.006) & 0.848 (0.011) &  1.218 (0.016) & 1.204 (0.017) & 1.111 (0.036) & 1.220 (0.006) \\[0.2em]

$\phi_0$   &-0.703 (0.003) &-0.293 (0.012) & -0.056 (0.006) &-0.057 (0.011) &-0.281 (0.031) &-0.461 (0.005) \\
$\phi_1$   & 0.745 (0.001) & 0.897 (0.004) &  0.976 (0.003) & 0.980 (0.003) & 0.904 (0.010) & 0.905 (0.001) \\
$\phi_2$   & 0.392 (0.001) & 0.453 (0.007) &  1.102 (0.016) & 0.921 (0.016) & 0.684 (0.025) & 0.698 (0.003) \\[0.2em]

$\psi_0$   &-0.106 (0.020) &-0.051 (0.053) & -0.156 (0.046) &-0.117 (0.015) &-0.870 (0.671) &-0.107 (0.014) \\
$\psi_1$   & 0.985 (0.003) & 0.990 (0.008) &  0.977 (0.007) & 0.983 (0.002) & 0.878 (0.094) & 0.988 (0.002) \\
$\psi_2$   & 0.810 (0.037) & 0.834 (0.077) &  0.846 (0.039) & 0.662 (0.030) & 0.760 (0.138) & 1.223 (0.045) \\ \hline	\hline 
		\end{tabular}
   \begin{tablenotes}
         \item \footnotesize {\it Note:} The table displays the estimation results of the model in \eqref{eq:TVP-SIRD}. We display the posterior means and posterior standard deviations (in parenthesis) of the corresponding parameter shown in the first column and the country shown in the first row. \end{tablenotes}
\end{threeparttable}  		
	\end{table}

	\begin{table}[H]
	\setlength{\tabcolsep}{9pt}	
	\singlespacing
		\centering
\begin{threeparttable}			
		\caption{Relative RMSFEs of the competing models relative to the TVP-SIRD model}
			\label{tab:OOS}
			\begin{tabular}{ll rrrrrrr}  \hline \hline
			&	& $h=1$ & $h=3$ & $h=5$ & $h=7$ & $h=9$ & $h=11$ & $h=14$ \\  \hline 
\multirow{4}{*}{Brazil}   & EW    & 2.24  & 1.72  & 1.43 & 1.32 & 1.36 & 1.38 & 1.41 \\
                          & RW-10 & 1.74  & 1.42  & 1.23 & 1.17 & 1.24 & 1.28 & 1.33 \\
                          & RW-20 & 2.24  & 1.72  & 1.43 & 1.32 & 1.36 & 1.38 & 1.41 \\
                          & RW-30 & 2.76  & 2.03  & 1.64 & 1.48 & 1.51 & 1.52 & 1.54 \\[0.2em]
\multirow{4}{*}{Germany}  & EW    & 4.56  & 3.46  & 2.86 & 2.41 & 2.22 & 2.04 & 1.87 \\
                          & RW-10 & 2.30  & 2.25  & 2.24 & 2.21 & 2.34 & 2.44 & 2.70 \\
                          & RW-20 & 3.83  & 3.36  & 3.16 & 2.97 & 3.05 & 3.11 & 3.33 \\
                          & RW-30 & 4.87  & 4.11  & 3.74 & 3.44 & 3.45 & 3.45 & 3.59 \\[0.2em]
\multirow{4}{*}{India}    & EW    & 9.48  & 4.43  & 3.30 & 2.83 & 2.59 & 2.44 & 2.29 \\
                          & RW-10 & 3.71  & 2.02  & 1.63 & 1.47 & 1.40 & 1.36 & 1.33 \\
                          & RW-20 & 5.63  & 2.76  & 2.11 & 1.84 & 1.71 & 1.62 & 1.54 \\
                          & RW-30 & 7.07  & 3.30  & 2.44 & 2.07 & 1.89 & 1.77 & 1.65 \\[0.2em]
\multirow{4}{*}{Italy}    & EW    & 22.53 & 8.48  & 5.47 & 4.12 & 3.34 & 2.83 & 2.30 \\
                          & RW-10 & 9.95  & 4.99  & 3.98 & 3.57 & 3.38 & 3.30 & 3.29 \\
                          & RW-20 & 18.17 & 8.03  & 5.94 & 5.05 & 4.59 & 4.32 & 4.15 \\
                          & RW-30 & 24.16 & 10.13 & 7.20 & 5.93 & 5.25 & 4.84 & 4.49 \\[0.2em]
\multirow{4}{*}{S. Korea} & EW    & 7.19  & 2.69  & 1.83 & 1.45 & 1.26 & 1.14 & 1.02 \\
                          & RW-10 & 3.53  & 2.02  & 1.90 & 1.98 & 2.19 & 2.49 & 3.08 \\
                          & RW-20 & 5.39  & 2.62  & 2.20 & 2.09 & 2.12 & 2.24 & 2.51 \\
                          & RW-30 & 5.55  & 2.51  & 1.99 & 1.80 & 1.75 & 1.77 & 1.87 \\[0.2em]
\multirow{4}{*}{US}       & EW    & 13.09 & 5.36  & 3.52 & 2.75 & 2.38 & 2.17 & 1.99 \\
                          & RW-10 & 3.41  & 1.81  & 1.42 & 1.26 & 1.19 & 1.17 & 1.18 \\
                          & RW-20 & 5.40  & 2.55  & 1.86 & 1.58 & 1.46 & 1.41 & 1.38 \\
                          & RW-30 & 7.16  & 3.21  & 2.27 & 1.88 & 1.71 & 1.62 & 1.56 \\  
			\hline  \hline 
		\end{tabular}
   \begin{tablenotes}
         \item \footnotesize {\it Note:} The table displays the Root Mean Squared Forecast Errors (RMSFE) of the competing models relative to the TVP-SIRD model introduced in \eqref{eq:TVP-SIRD} for the country shown in the first column. EW stands for the Expanding Window. RW-M stands Rolling Window with $M$ observations as the sample size for $M=10,20,30$. 
\end{tablenotes}
\end{threeparttable}  		
	\end{table} 
	
\newpage

 \vspace{-0.3cm}
\footnotesize {\it Note:} The graphs show the evolution of the cumulative active infection cases in Brazil, Germany, India, Italy, S. Korea, and the US.

\newpage

\begin{figure}[H]
    \centering
    \caption{The evolution of $\beta_t$, $\gamma_t$, $\nu_t$ and $R_{0,t}$ estimated using the TVP-SIRD model}
\begin{tabular}{ccccc}
  &  $\beta_t$ & $\gamma_t$ & $\nu_t$ & $R_{0,t}$   \\[-0.2em]
  \begin{turn}{90} \hspace{0.3cm}  Brazil  \end{turn} \hspace{-0.7cm} 
  & \includegraphics[trim = 0mm 6mm 0mm 0mm, clip,width=0.22\textwidth]{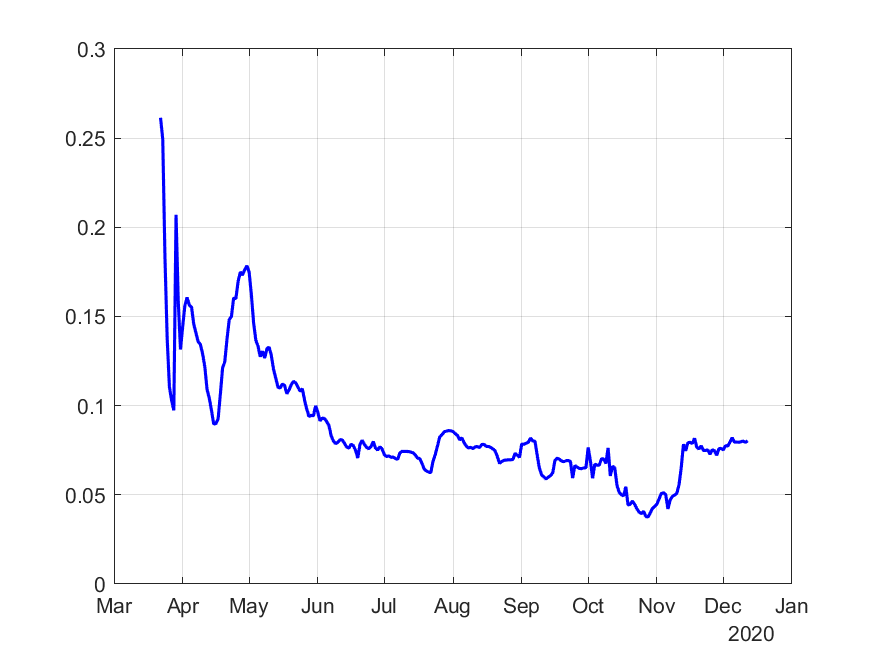}   & 
  \includegraphics[trim = 0mm 6mm 0mm 0mm , clip, width=0.22\textwidth]{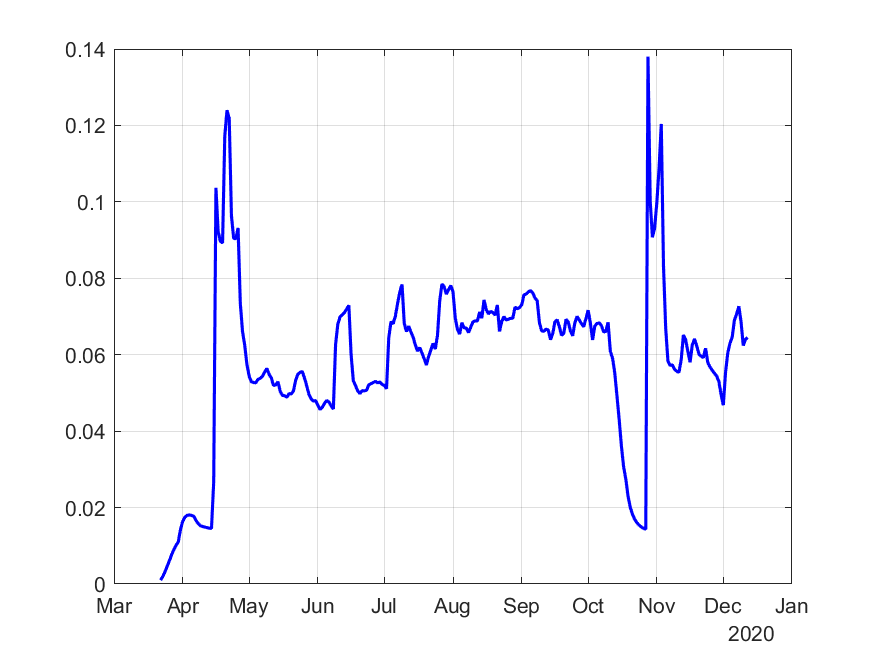} &\includegraphics[trim = 0mm 6mm 0mm 0mm, clip,width=0.22\textwidth]{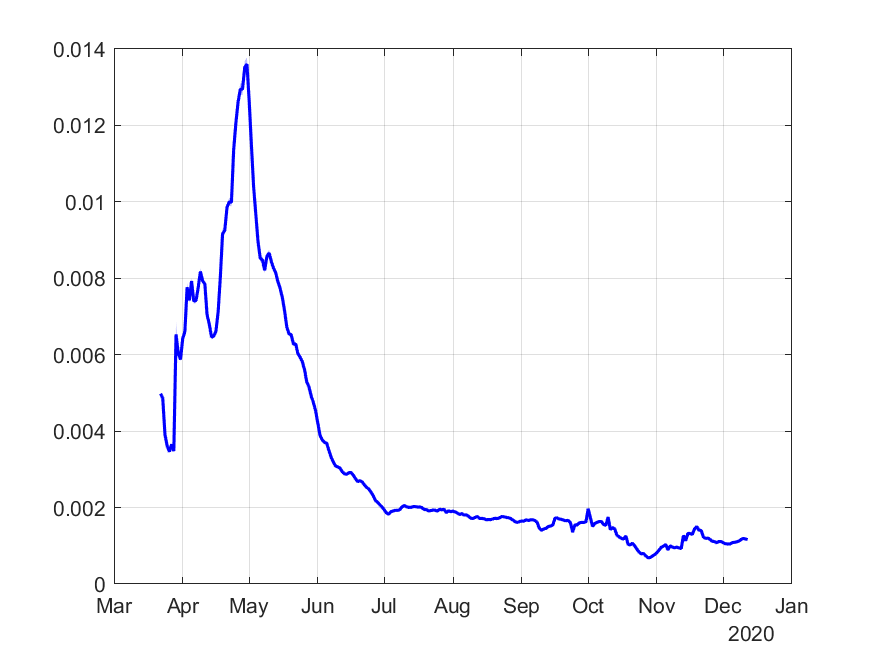}  &  \includegraphics[trim = 0mm 6mm 0mm 0mm, clip,width=0.22\textwidth]{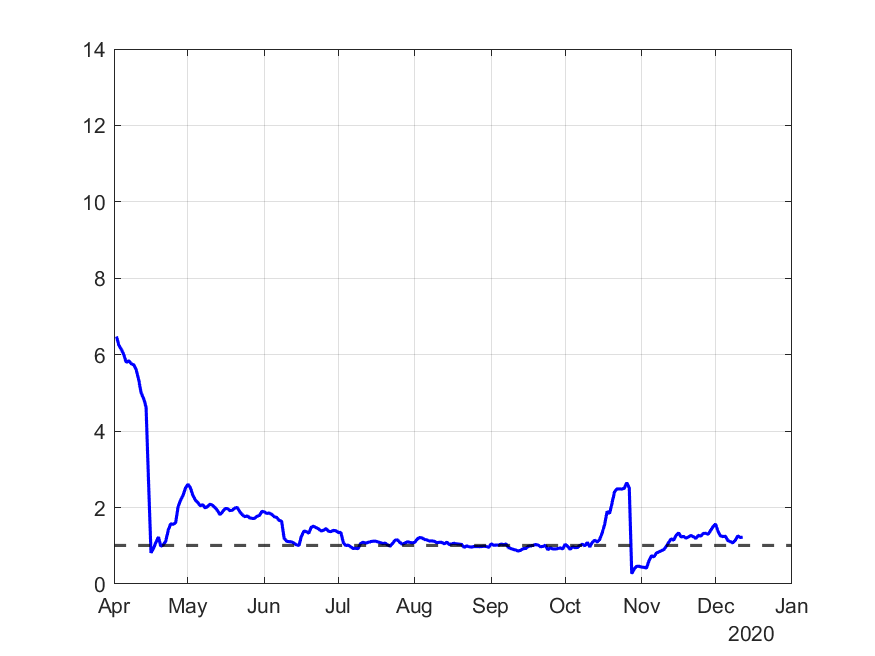} \\
 
  \begin{turn}{90} \hspace{0.3cm}  Germany  \end{turn} \hspace{-0.7cm} 
  & \includegraphics[trim = 0mm 6mm 0mm 0mm, clip,width=0.22\textwidth]{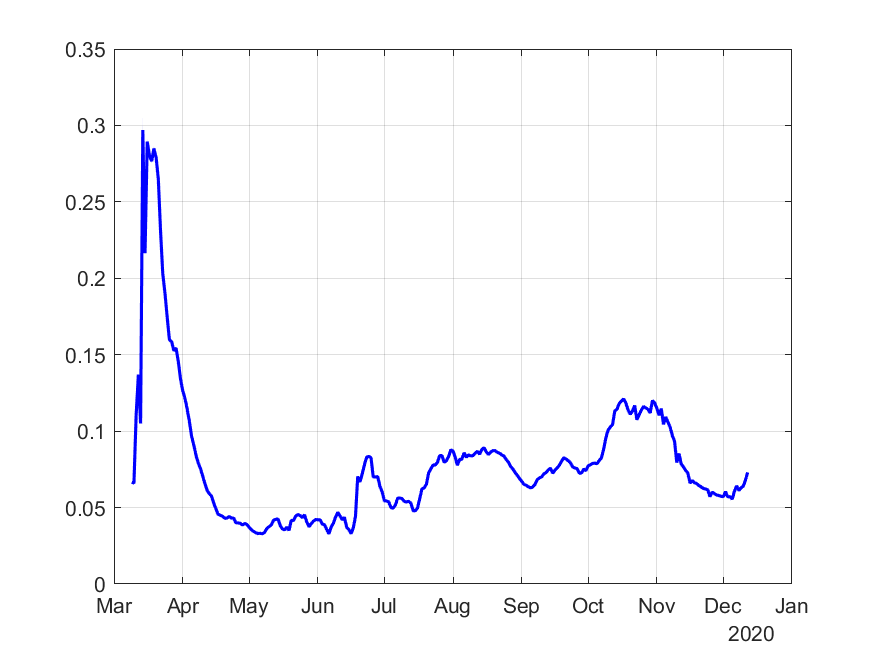}   & \includegraphics[trim = 0mm 6mm 0mm 0mm , clip, width=0.22\textwidth]{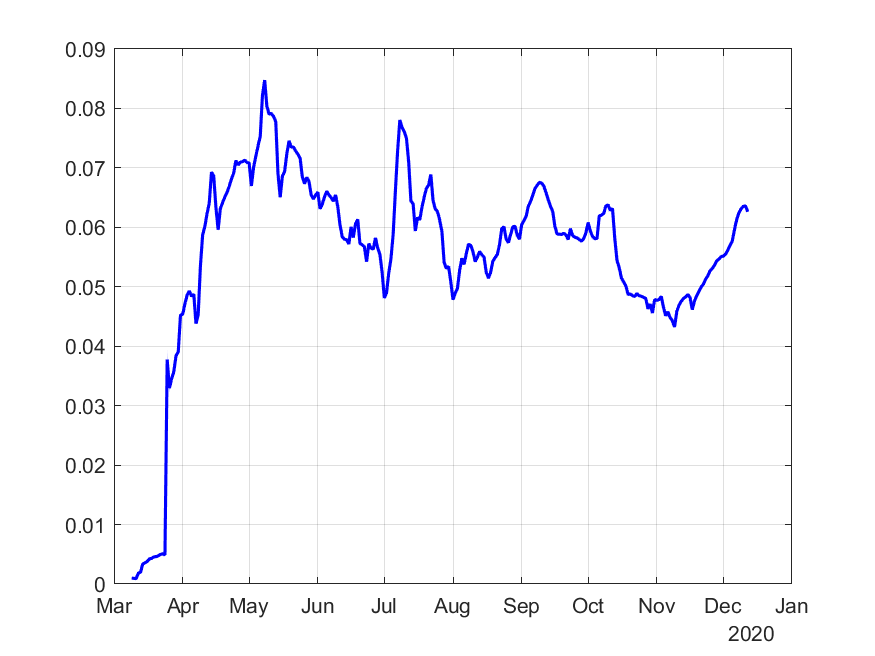} &\includegraphics[trim = 0mm 6mm 0mm 0mm, clip,width=0.22\textwidth]{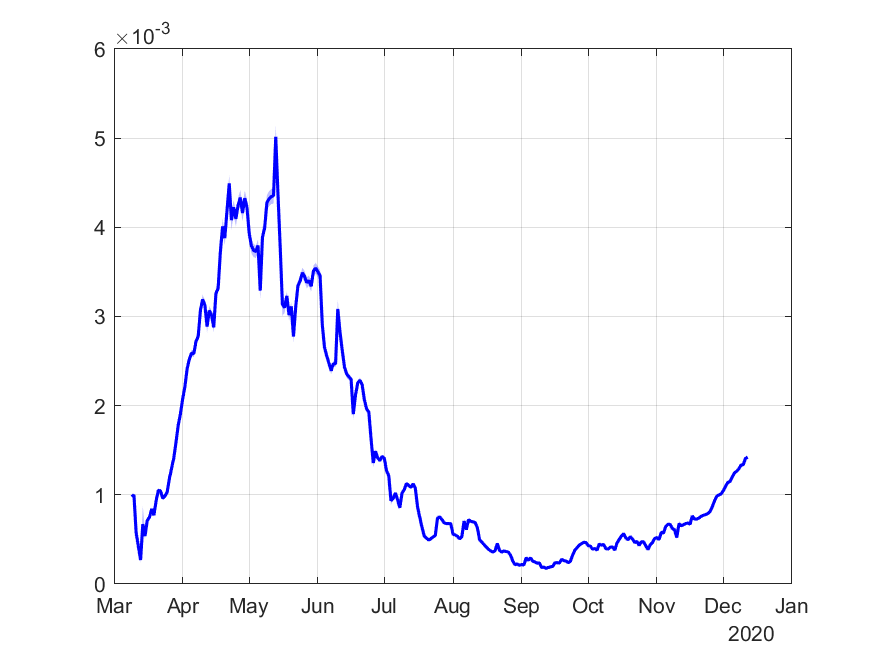}  &  \includegraphics[trim = 0mm 6mm 0mm 0mm, clip,width=0.22\textwidth]{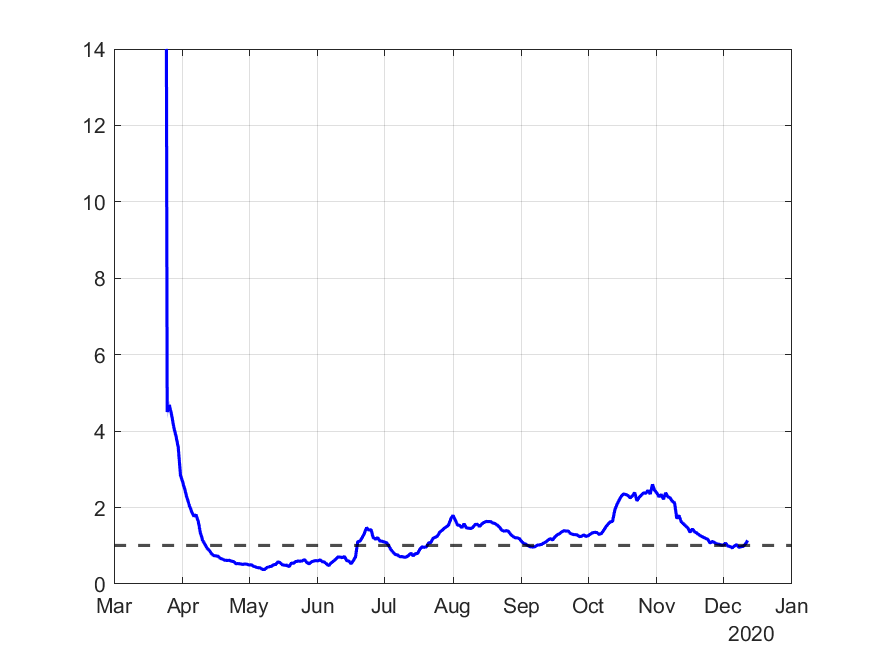} \\
  
    \begin{turn}{90} \hspace{0.3cm}  India  \end{turn} \hspace{-0.7cm} 
  & \includegraphics[trim = 0mm 6mm 0mm 0mm, clip,width=0.22\textwidth]{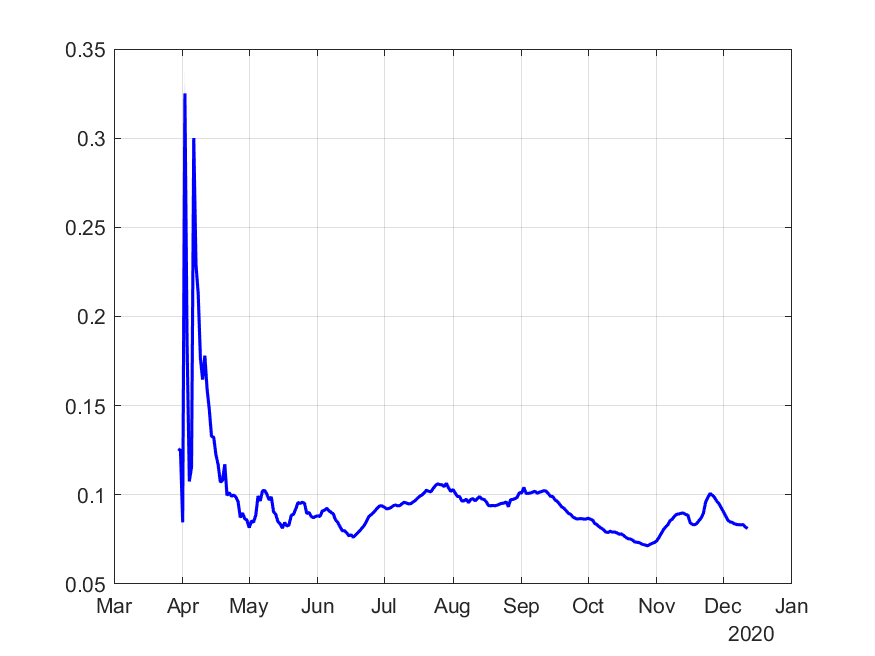}   & \includegraphics[trim = 0mm 6mm 0mm 0mm , clip, width=0.22\textwidth]{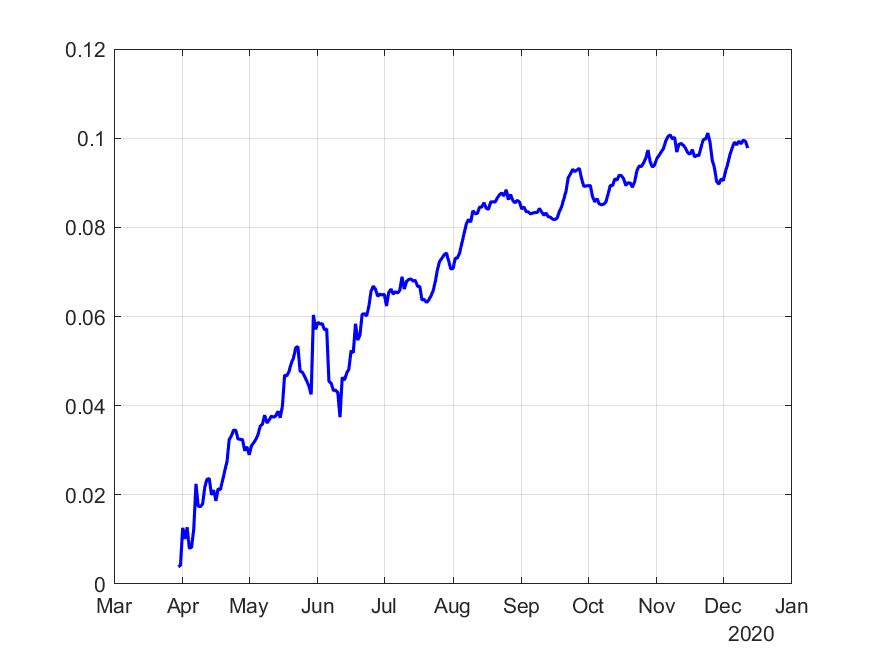} &\includegraphics[trim = 0mm 6mm 0mm 0mm, clip,width=0.22\textwidth]{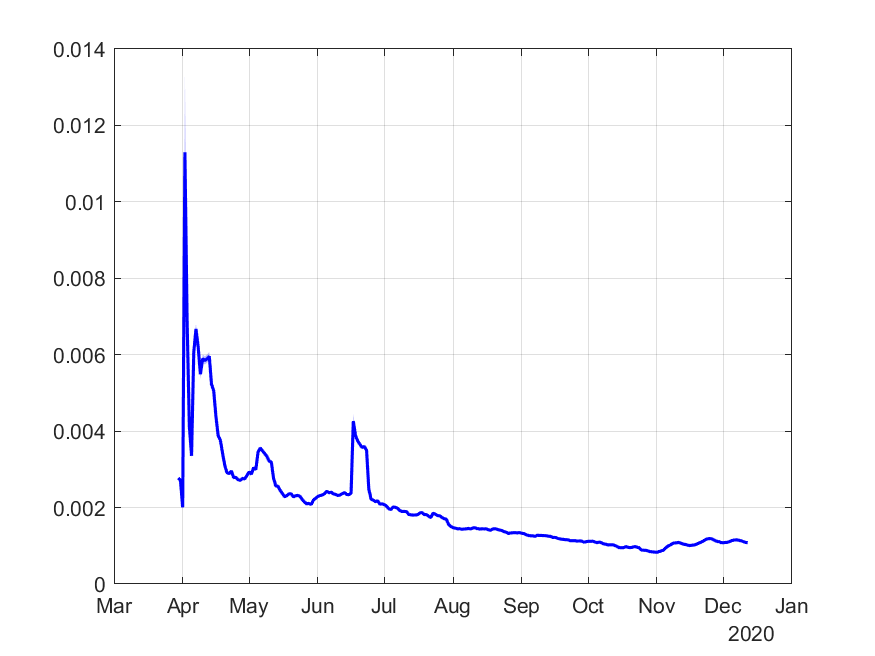}  &  \includegraphics[trim = 0mm 6mm 0mm 0mm, clip,width=0.22\textwidth]{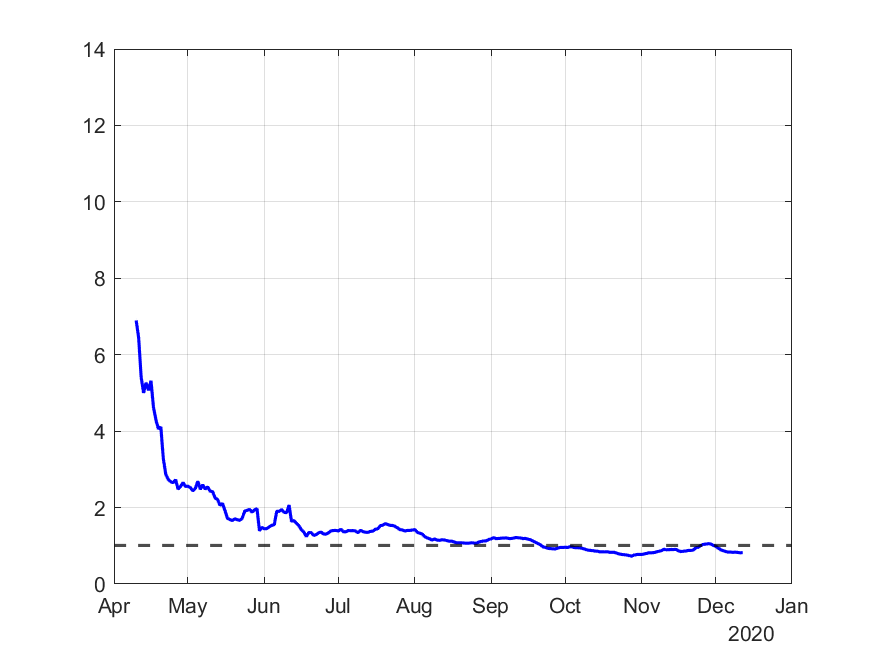} \\
 
  \begin{turn}{90} \hspace{0.3cm}  Italy  \end{turn} \hspace{-0.7cm} 
  & \includegraphics[trim = 0mm 6mm 0mm 0mm, clip,width=0.22\textwidth]{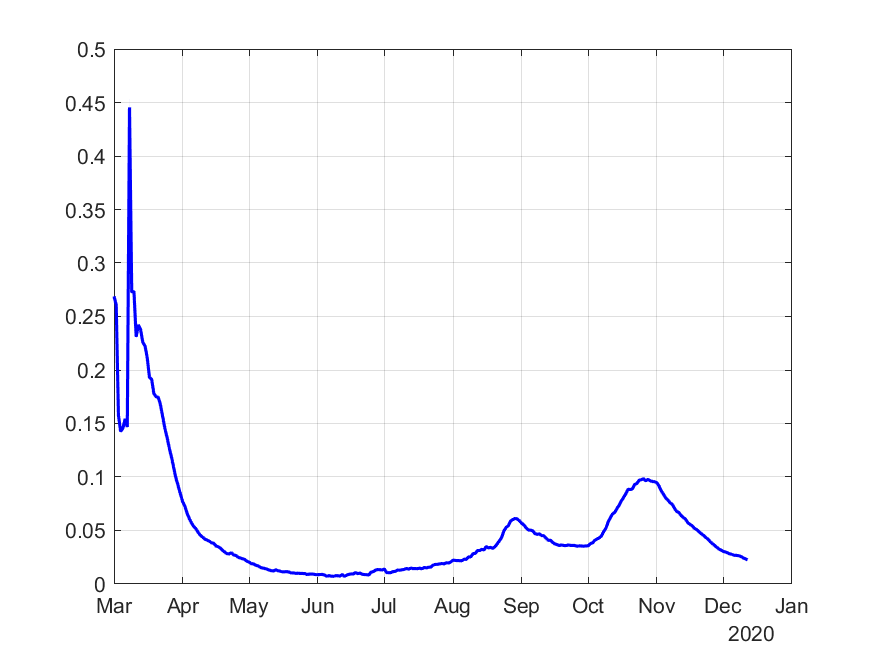}   & \includegraphics[trim = 0mm 6mm 0mm 0mm , clip, width=0.22\textwidth]{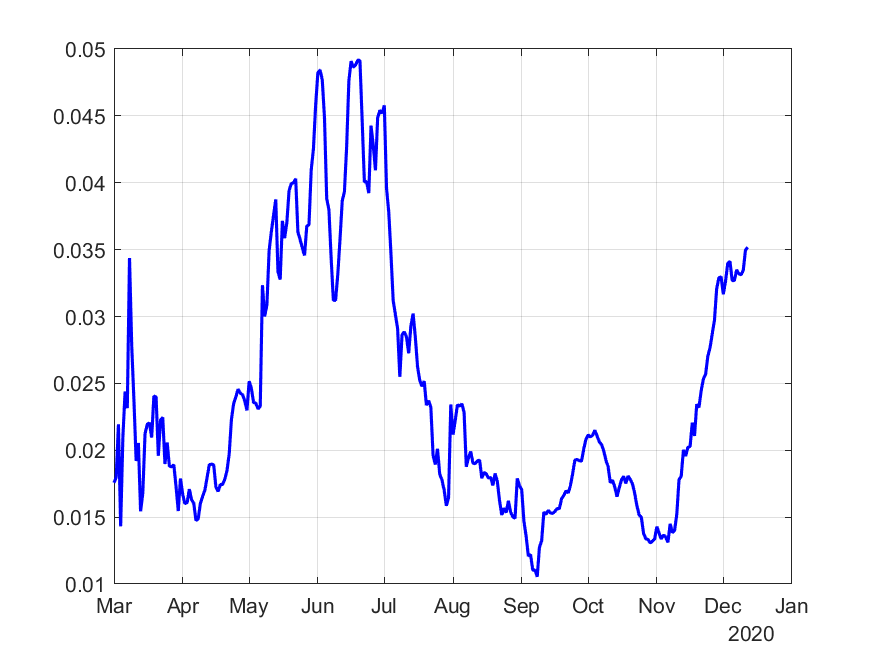} &\includegraphics[trim = 0mm 6mm 0mm 0mm, clip,width=0.22\textwidth]{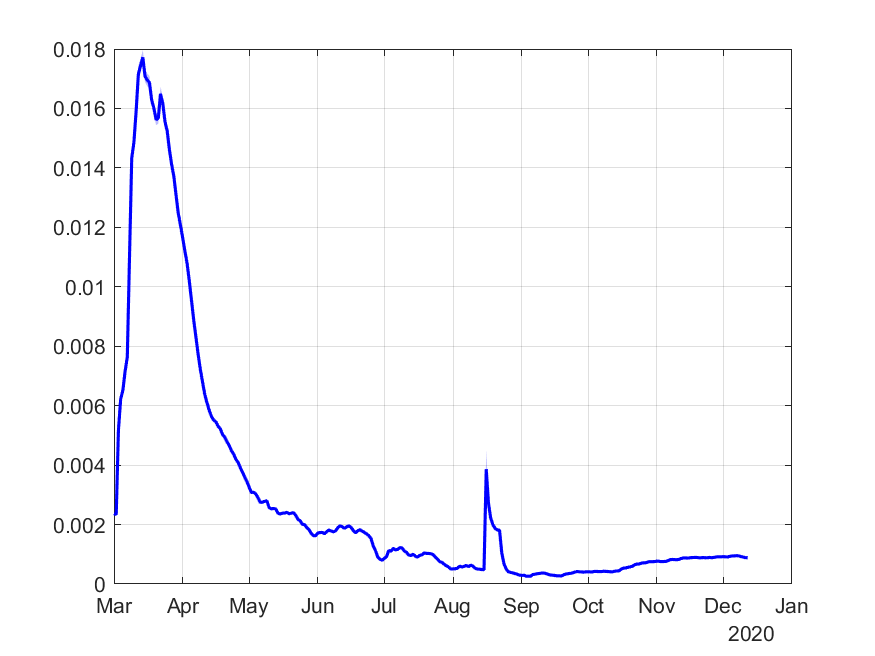}  &  \includegraphics[trim = 0mm 6mm 0mm 0mm, clip,width=0.22\textwidth]{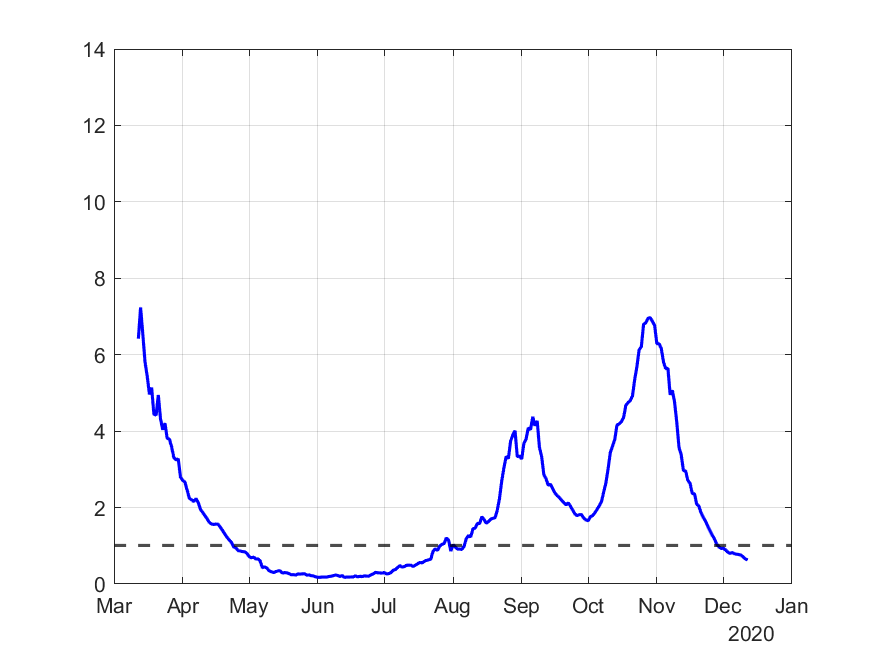} \\
 
  \begin{turn}{90} \hspace{0.0cm}  S. Korea  \end{turn} \hspace{-0.7cm} 
  & \includegraphics[trim = 0mm 6mm 0mm 0mm, clip,width=0.22\textwidth]{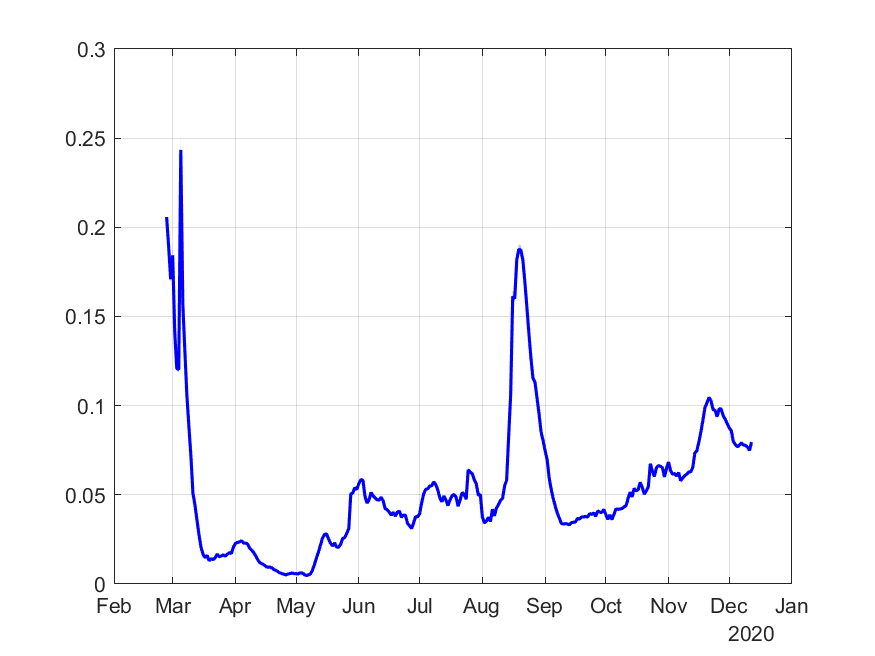}   & \includegraphics[trim = 0mm 6mm 0mm 0mm , clip, width=0.22\textwidth]{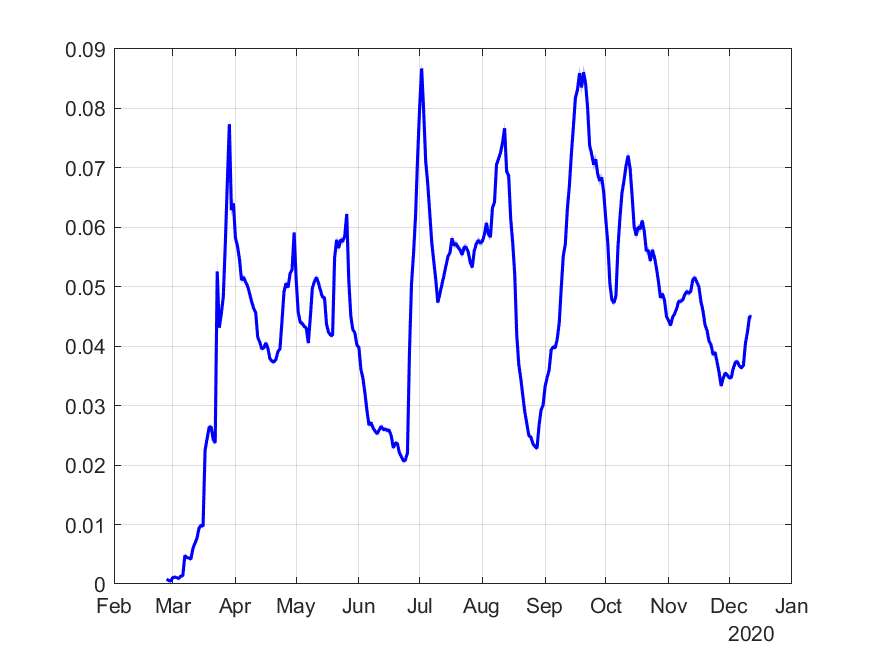} &\includegraphics[trim = 0mm 6mm 0mm 0mm, clip,width=0.22\textwidth]{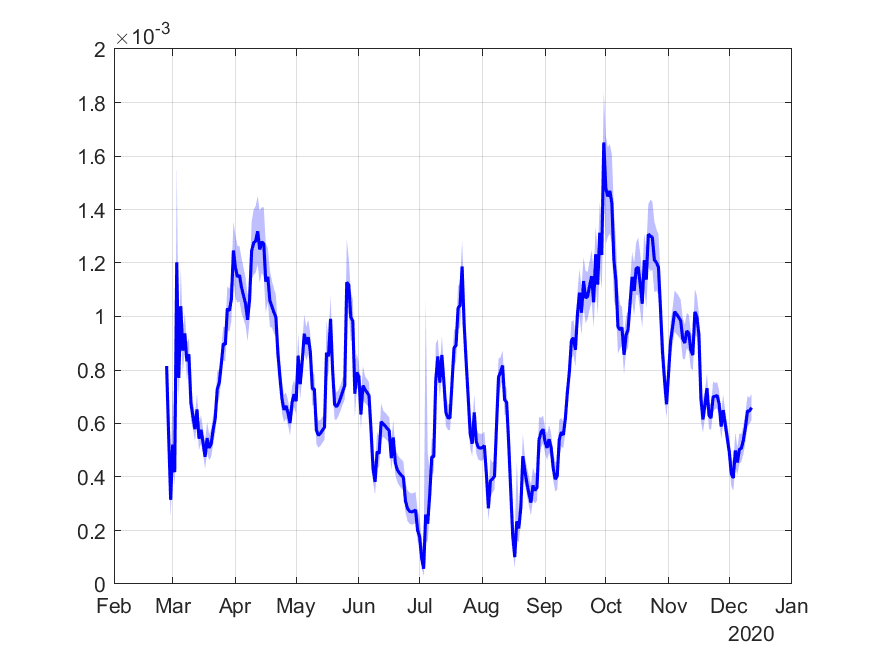}  &  \includegraphics[trim = 0mm 6mm 0mm 0mm, clip,width=0.22\textwidth]{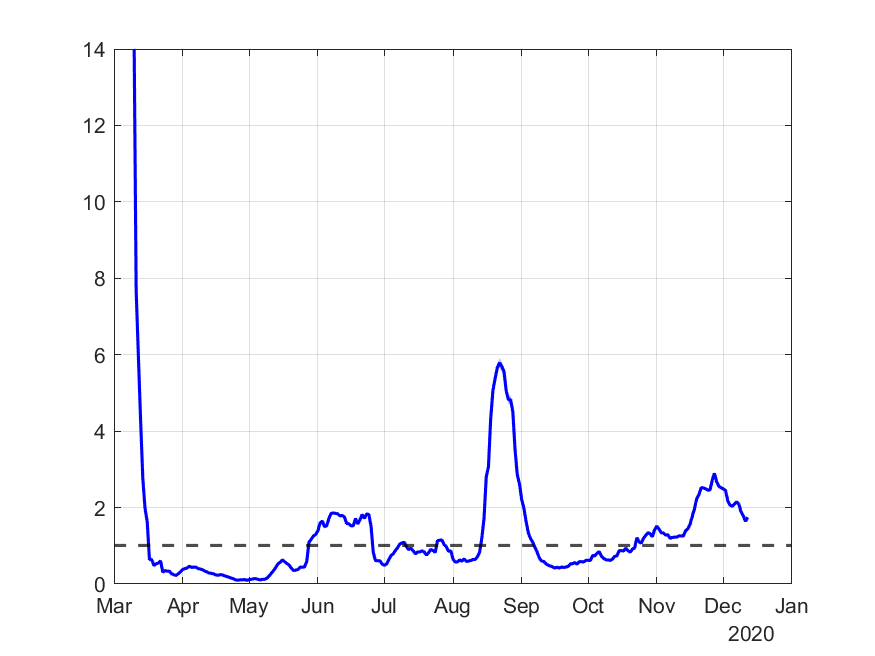} \\
  
  \begin{turn}{90} \hspace{0.3cm}  US  \end{turn} \hspace{-0.7cm} 
  & \includegraphics[trim = 0mm 6mm 0mm 0mm, clip,width=0.22\textwidth]{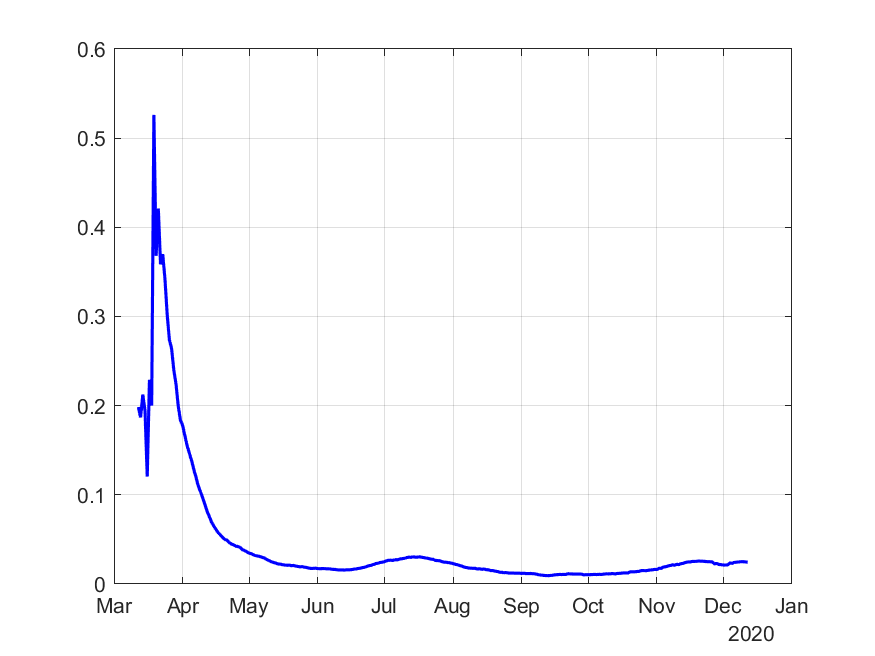}   & \includegraphics[trim = 0mm 6mm 0mm 0mm , clip, width=0.22\textwidth]{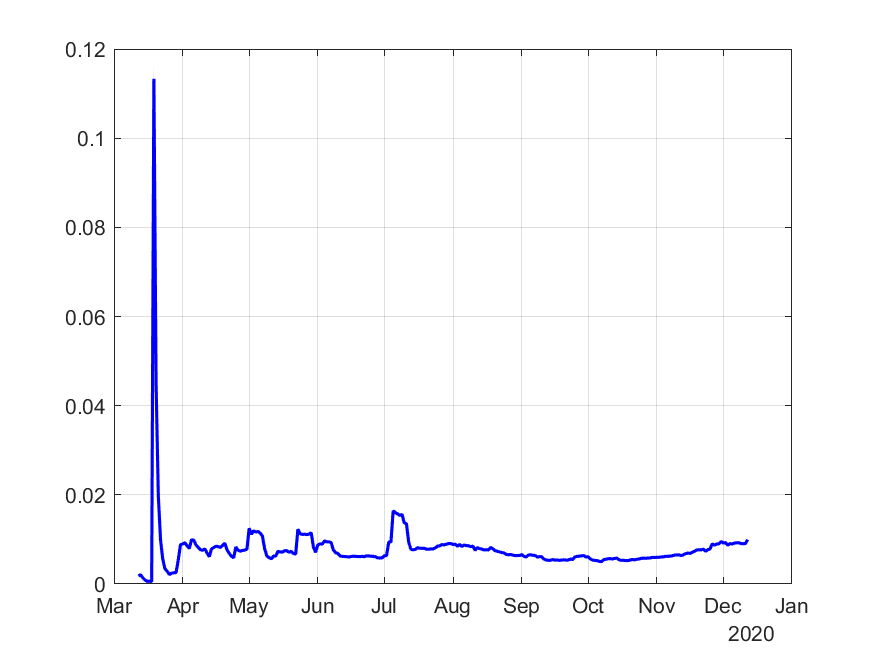} &\includegraphics[trim = 0mm 6mm 0mm 0mm, clip,width=0.22\textwidth]{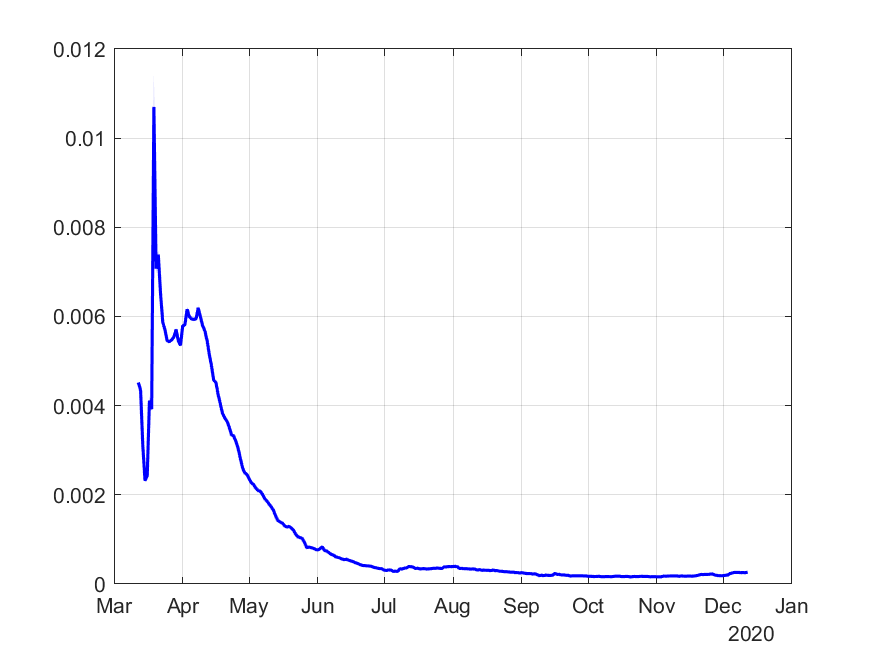}  &  \includegraphics[trim = 0mm 6mm 0mm 0mm, clip,width=0.22\textwidth]{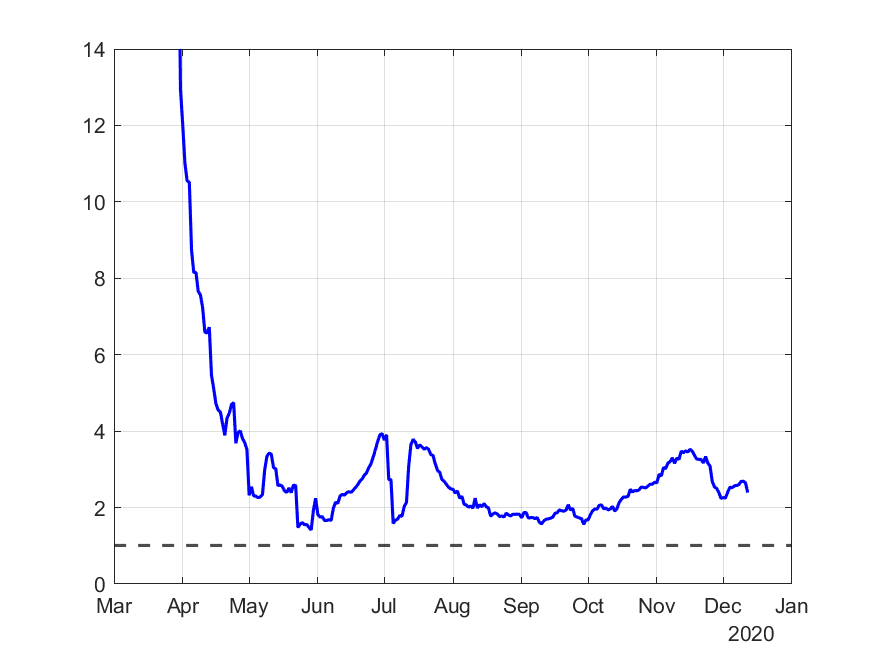} \\
\end{tabular}
    \label{fig:TVParameters}
\end{figure} \vspace{-0.3cm}
\footnotesize {\it Note:} The graphs show the evolution of the time-varying parameters, $\beta_t$, the rate of infection, $\gamma_t$, the rate of recovery, $\nu_t$, the death rate, and the resulting reproduction rate, $R_{0,t}$ estimated using the TVP-SIRD model introduced in \eqref{eq:TVP-SIRD} for the countries shown in the first column. We also include 95\% credible intervals around the posterior means with darker blue areas. 

\normalsize

\vspace{-0.3cm} \footnotesize {\it Note:} The graphs show the evolution of the time-varying parameters, $\delta_t$, the fraction of asymptomatic cases in total cases, $\beta_t$, the rate of infection, $\gamma_t$, the rate of recovery, $\nu_t$, the death rate, and the resulting reproduction rate, $R_{0,t}$. The parameters estimated using the TVP-SIRD model introduced in \eqref{eq:TVP-SIRD} are displayed with the blue line, and those for the TVP-SIRD model also takes the unreported cases into account introduced in \eqref{eq:TVP-SIRD-A} displayed with the red line.  

\clearpage

\normalsize

 \vspace{-0.3cm} \footnotesize {\it Note:} The graphs show the evolution of the time-varying parameters in real-time, i.e., estimated using the sample up to the period $t$ for the TVP-SIRD model (the blue line), estimated using the last $M$ observations up to the period $t$ for the SIRD model shown using the red, yellow and purple lines for $M=10,20$ and $30$ respectively. See Figure~\ref{fig:TVParameters} for further details.

\normalsize

\renewcommand\thesection{Appendix \Alph{section}}
\renewcommand\thesubsection{\Alph{section}.\arabic{subsection}}
\setcounter{section}{0}
\renewcommand\theequation{\Alph{section}.\arabic{equation}}
\setcounter{equation}{0}
\renewcommand\thetable{\Alph{section}.\arabic{table}}
\setcounter{table}{0}
\renewcommand\thefigure{\Alph{section}.\arabic{figure}}
\setcounter{figure}{0}

\newcommand\blfootnote[1]{%
  \begingroup
  \renewcommand\thefootnote{}\footnote{#1}%
  \addtocounter{footnote}{-1}%
  \endgroup}
\restoregeometry
\setstretch{1.5}
\justifying

\newpage

\section{Implied Moments of the SIRD model with fixed parameters}
\label{app:sec:UncMomSIRD}
We assume that states' initial values,  $S_0, I_0, R_0$ and $D_0$ are known. For the sake of simplicity, we assume that $S_t \approx N$. We focus on the general form of equations as $Y_t|\Omega_{t-1} \sim Poisson( \lambda I_{t-1})$ where $\lambda = \beta$, $\gamma $ and $\nu $ for $Y_t=\Delta C_t$, $\Delta R_t$ and $\Delta D_t$, respectively. The conditional mean and the variance for the Poisson distributed variables are
\begin{equation}
\begin{array}{rcl}
    E[Y_t | \Omega_{t-1}] &= &\lambda I_{t-1} \\
  Var(Y_t | \Omega_{t-1}) &= &\lambda I_{t-1}.
\end{array}    
\end{equation}
The resulting process is stationary if the underlying process for $I_t$ is stationary. $I_t$'s stationarity, in turn, depends on the basic reproduction rate, $R_0$. To see this, we first start with $\Delta I_t$ and use the fact that $E[\Delta I_t] = E[\Delta C_t]-E[\Delta R_t]-E[\Delta D_t]$. Therefore, the difference equation governing $I_t$ takes the form of
\begin{equation}
\label{app:eq:It process}
\begin{array}{rcl}
    I_t &=& I_{t-1} +  \Delta C_t - \Delta R_t - \Delta D_t \\
   E[I_t|\Omega_{t-1}] &=& I_{t-1} +  \beta I_{t-1} - \gamma I_{t-1} - \nu I_{t-1} \\
                        &=& (1+\beta - \gamma - \nu) I_{t-1} \\
    E[I_t] &=& (1 + \beta(1-R_0^{-1})) E[I_{t-1}]. 
\end{array}    
\end{equation}
where the last equation uses the definition of the basic reproduction rate, $R_0$, as the ratio of the infection rate to the resolution rate. If $R_0$ exceeds unity, as $\beta$ is positive, the process is explosive, i.e., the pandemic progress exponentially. On the other hand, if  $R_0$ falls below unity, the process becomes stationary. We can track down the process conditional on the starting value for $I_0$. Let $\pi=(1 + \beta(1-R_0^{-1}))$. \vspace{-0.2cm}
\begin{equation}
\label{app:eq:UnconMean}
 \begin{array}{rcl}
E[I_t] &=&  \pi^{t} I_0, 
\end{array}    \vspace{-0.2cm}
\end{equation} 
in case the initial condition is known; otherwise, it is replaced by $E[I_0]$. For the variance, we can use a similar recursion. We start with computing $Var(I_t|\Omega_{t-1})$.
\begin{equation}
\begin{array}{rcl}
     Var(I_t|\Omega_{t-1}) & = & Var[\Delta C_t - \Delta R_t - \Delta D_t|I_{t-1}]  \\
      & = & Var[\Delta C_t | I_{t-1}] + Var[\Delta R_t | I_{t-1}] + Var[\Delta D_t | I_{t-1}] \\
      & = & (\beta + \gamma + v) I_{t-1} \\
      & = & \beta (1 + R_0^{-1}) I_{t-1}
\end{array}
\end{equation}
By the law of total variance and using forward iteration, the unconditional variance can be computed as 
\begin{equation}
\label{app:eq:UnconVar}
\begin{array}{rcl}
Var(I_t) &=& \beta(1+R_0^{-1}) E[I_{t-1}] + \pi^2 Var(I_{t-1})  \\
& \vdots & \\
  &=& \beta (1+R_0^{-1})(\sum_{i=0}^{t-1} \pi^i )\pi^{t-1} E[I_0] + \pi^{2t} Var(I_{0}) \\
  &=& \beta (1+R_0^{-1})\frac{\pi^{t-1} (1-\pi^{t})}{1-\pi} E[I_0] + \pi^{2t} Var(I_{0}) \\
\end{array}    
\end{equation}
In case the initial condition is known the second term drops. 

\newpage

\section{Derivation of updating rules}
\label{app:sec:UpdatingRules}
Let $f_1(\Delta C_t|\Omega_{t-1})$, $f_2(\Delta R_t|\Omega_{t-1})$ and $f_3(\Delta D_t|\Omega_{t-1})$ denote the conditional probability density functions for $\Delta C_t$, $\Delta R_t$ and $\Delta D_t$ conditional on the information set at time period $t-1$, $\Omega_{t-1}$, respectively. Assuming (conditional) independence among these variables, the conditional joint probability density function could be written as $ f(\Delta C_t, \Delta R_t, \Delta D_t|\Omega_{t-1}) = f_1(\Delta C_t|\Omega_{t-1})f_2(\Delta R_t|\Omega_{t-1})f_3(\Delta D_t|\Omega_{t-1})$. We assume that these marginal distributions are Poisson distribution with arrival rates specified in SIRD model in equation \eqref{eq:FPSIR}. Thus, the joint density is as follows, 
 \begin{equation}
       \begin{array}{rcl}
    f(\Delta C_t, \Delta Rc_t, \Delta D_t|\Omega_{t-1}) = \frac{\lambda_{1,t}^{\Delta C_t} \exp(-\lambda_{1,t})}{\Gamma(\Delta C_t + 1)} ~ \frac{\lambda_{2,t}^{\Delta R_t} \exp(-\lambda_{2,t})}{\Gamma(\Delta Rc_t + 1)} ~\frac{\lambda_{3,t}^{\Delta D_t} \exp(-\lambda_{3,t})}{\Gamma(\Delta D_t + 1)},
 	\end{array}
 \end{equation}
 where $\lambda_{1,t} = \beta_t \frac{S_{t-1}I_{t-1}}{N}$, $\lambda_{2,t} = \gamma_t I_{t-1}$  and $\lambda_{3,t} = \nu_t I_{t-1}$. The score functions, denoted as $\nabla_{1,t}, \nabla_{2,t}$ and $\nabla_{3,t}$, can be written as
\begin{equation}
\label{app:eq:score}
       \begin{array}{rcl}
    \nabla_{1,t} &=& \left(\frac{\Delta C_t}{\lambda_{1,t}}- 1 \right)\left(\frac{I_{t-1}S_{t-1}}{N}\right) \\
    \nabla_{2,t} &=& \left(\frac{\Delta Rc_t}{\lambda_{2,t}}- 1 \right)I_{t-1} \\
    \nabla_{3,t} &=& \left(\frac{\Delta D_t}{\lambda_{3,t}}- 1 \right)I_{t-1} 
	\end{array}
\end{equation}
We use the variance of the score functions for the scaling parameter. Variance of the score function for $\nabla_{1,t}$, for example, can be computed as
   \begin{equation}
       \begin{array}{rcl}
    Var(\nabla_{1,t}|\Omega_{t-1}) &=& E[\nabla_{1,t}\nabla_{1,t}'|\Omega_{t-1}]\\
    &=& E\left[(\frac{\Delta C_t}{\lambda_{1,t}} - 1)^2\right|\Omega_{t-1}] \left(\frac{I_{t-1}^2 S_{t-1}^2 }{N^2}\right) \\
    &=& \frac{E[(\Delta C_t - \lambda_{1,t})^2|\Omega_{t-1}]}{\lambda_{1,t}^2} \left(\frac{I_{t-1}^2 S_{t-1}^2 }{N^2}\right).
	\end{array}
 \end{equation} 
As $E[(\Delta C_t - \lambda_{1,t})^2]$ refers to the variance of Poisson distributed random variable $\Delta C_t$, it is identical to $\lambda_{1,t}$. Hence, the resulting expression is,
 \begin{equation}
 \label{app:eq:scorevar1}
     Var(\nabla_t^1|\Omega_{t-1}) = \left(\frac{1}{\lambda_{1,t}}\right) \left(\frac{I_{t-1}^2 S_{t-1}^2}{N^2}\right)
 \end{equation}
Similar computations lead to the variances of score functions for $\Delta Rc_t$ and $\Delta D_t$ as
%
 \begin{equation}
 \label{app:eq:scorevar2_3}
    \begin{array}{rcl}
    Var(\nabla_{2,t}|\Omega_{t-1}) &= & \frac{I_{t-1}^2}{\lambda_{2,t}} \\
    Var(\nabla_{3,t}|\Omega_{t-1}) &= & \frac{I_{t-1}^2}{\lambda_{3,t}}
	\end{array}
 \end{equation}
Scaling \eqref{app:eq:score} together with \eqref{app:eq:scorevar1} and \eqref{app:eq:scorevar2_3}, the scaled score functions can be written as follows,
\begin{equation}
    \begin{array}{rcl}
    s_{1,t} &=& (\Delta C_t - \lambda_{1,t}) \left(\frac{N}{I_{t-1}S_{t-1}}\right) \\
    s_{2,t} &=& \frac{\Delta Rc_t-\lambda_{2,t}}{I_{t-1}} \\
    s_{3,t} &=& \frac{\Delta D_t-\lambda_{3,t}}{I_{t-1}}
	\end{array}
 \end{equation} 
The final step includes the division of the scaled score functions by $\beta_t$, $\gamma_t$ and $\nu_t$, respectively, to obtain the scaled score function in terms of parameters with logarithmic transformations applying the chain rule. The resulting time evolution for $\tilde{\beta}_t = log(\beta_t)$, $\tilde{\gamma}_t = log(\gamma_t)$ and $\tilde{\nu}_t = log(\nu_t)$ is
\begin{equation}
\label{app:eq:TVP-SIRD-stateeq}
    \begin{array}{rcl}
     \tilde{\beta}_t &=& \alpha_0 + \alpha_1 \tilde{\beta}_{t-1} + \alpha_2 \left(\frac{\Delta C_{t-1} - \lambda_{1,t-1}}{\lambda_{1,t-1}} \right) \\
     \tilde{\gamma}_t &=& \phi_0 + \phi_1 \tilde{\gamma}_{t-1} + \phi_2 \left(\frac{\Delta R_{t-1} - \lambda_{2,t-1}}{\lambda_{2,t-1}} \right) \\
    \tilde{\nu}_t &=& \psi_0 + \psi_1 \tilde{\nu}_{t-1} + \psi_2 \left( \frac{\Delta D_{t-1} - \lambda_{3,t-1}}{\lambda_{3,t-1}} \right) 
	\end{array}
 \end{equation} 
Combining \eqref{app:eq:TVP-SIRD-stateeq} together with the SIRD equations, the  final model becomes as follows
	\begin{equation}
\label{app:eq:TVP-SIRD}
    \begin{array}{rcl}
     \Delta C_{t}|\Omega_{t-1} &\sim&  Poisson( \beta_t \frac{S_{t-1}}{N} I_{t-1}) \\[-0.2em]
     \Delta R_{t}|\Omega_{t-1} &\sim&  Poisson( \gamma_t  I_{t-1} )    \\[-0.2em]
   	   \Delta D_{t}|\Omega_{t-1} &\sim&  Poisson( \nu_t I_{t-1})    \\[0.4em]
      \tilde{\beta_t}   & = &  \alpha_0 + \alpha_1 \tilde{\beta}_{t-1} + \alpha_2 s_{1,t}\\[-0.2em]
       \tilde{\gamma_t}  &= & \phi_0 + \phi_1 \tilde{\gamma}_{t-1} + \phi_2 s_{2,t} \\[-0.2em]
     \tilde{\nu_t}  &= & \psi_0 + \psi_1 \tilde{\nu}_{t-1} + \psi_2 s_{3,t} \\[0.4em]
    \Delta C_t = -\Delta S_t    &=&  \Delta I_{t} + \Delta R_{t} + \Delta D_{t} \\
    \end{array}
\end{equation}

\end{document}